\documentclass[particles,review,accept,moreauthors,pdftex]{Definitions/mdpi}  %For preprint

\pdfoutput=1

\firstpage{1} 
\pubvolume{1}
\issuenum{1}
\articlenumber{0}
\pubyear{2023}
\copyrightyear{2023}
\datereceived{\today} 
\dateaccepted{} 
\datepublished{} 
\hreflink{https://doi.org/} % If needed use \linebreak
\usepackage{amsfonts,amssymb,dsfont,epsfig}
\usepackage[english]{babel}
\usepackage{slashed}
\usepackage{xspace}
\newcommand{\be}{\begin{equation}}
\newcommand{\bea}{\begin{eqnarray}}
\newcommand{\ee}{\end{equation}}
\newcommand{\eea}{\end{eqnarray}}
\def\1eq#1{Eq.~(\ref{#1})}

\def\2eqs#1#2{Eqs.~(\ref{#1}) and~(\ref{#2})}
\def\3eqs#1#2#3{Eqs.~(\ref{#1}),~(\ref{#2}) and~(\ref{#3})}

\def\fig#1{Fig.~\ref{#1}}

\def\ie{{\it i.e.}, }
\def\eg{{\it e.g.}, }

\def\s#1{{\scriptscriptstyle #1}}

\def\tlambda{\mkern 2mu\widetilde{\mkern-4mu \lambda \mkern-2mu}\mkern 1.2mu}  %lambda tilde.

\newcommand{\pt}{{\widetilde \Pi}}

\newcommand{\Ls}{ \mathit{L}_{{sg}}}

\newcommand{\w}{{\cal W}}

\newcommand{\CB}{{\mathbb C}_{\star}}
\newcommand{\Cc}{{\mathcal C}_{\star}}

\def\g{\Gamma}
\newcommand{\gt}{\widetilde{\Gamma}\vphantom{\g}}
\def\gz{\Gamma_{\!0}}
\def\gbar{\overline{\g}\vphantom{\Gamma}}

\newcommand{\fatg}{{\rm{I}}\!\Gamma}
\newcommand{\fatgt}{\widetilde{{\rm{I}}\!\Gamma}\vphantom{\fatg}}

\newcommand{\Cfat}{{\mathbb C}}
\newcommand{\Cfattilde}{\widetilde{\mathbb C}}
\newcommand{\C}{{\mathcal C}}
\newcommand{\Cgh}{C}
\newcommand{\Ctilde}{{\widetilde{\mathcal C}}}

\newcommand{\IW}{{\cal I}_{\w}}

\Title{Gauge Sector Dynamics in QCD}

\TitleCitation{Gauge sector dynamics in QCD}

\Author{Mauricio Narciso Ferreira$^{1,\dagger}$\orcidA{} and Joannis Papavassiliou$^{1,\dagger}$\orcidB{}}

\AuthorNames{Mauricio N. Ferreira and Joannis Papavassiliou}

\AuthorCitation{Ferreira, M. N.; Papavassiliou, J.}

\address{%
$^{1}$ \quad \mbox{Department of Theoretical Physics and IFIC, 
University of Valencia and CSIC},
E-46100, Valencia, Spain.}

\corres{Correspondence: ansonar@uv.es (M. N. Ferreira); Joannis.Papavassiliou@uv.es (J. Papavassiliou)}

\firstnote{These authors contributed equally to this work.}

\abstract{ 
The dynamics of the gauge sector of QCD give rise 
to nonperturbative phenomena that 
are crucial for the 
internal consistency of the theory; most notably, 
they account for the generation of a 
gluon mass through the action of the Schwinger mechanism, the 
taming of the Landau pole and the ensuing stabilization of the 
gauge coupling, 
and the infrared suppression of the 
three-gluon vertex.
In the present work, we review some key advances  
in the ongoing investigation of this sector 
within the framework of the continuum Schwinger function methods, 
supplemented by results obtained from lattice 
simulations.
}

  %(List three to ten pertinent keywords specific to the article; yet reasonably common within the subject discipline.)

\begin{document}

\begin{footnotesize}

\tableofcontents

\end{footnotesize}

%%%%%%%%%%%%%%%%%%%%%%%%%%%%%%%%%%%%%%%%%%%%%%%%%%%%%%%%
\section{Introduction}\label{sec:intro}
%%%%%%%%%%%%%%%%%%%%%%%%%%%%%%%%%%%%%%%%%%%%%%%%%%%%%%%%

The systematic exploration of the 
Green's functions ($n$-point correlation functions) 
of Quantum Chromodynamics (QCD)~\cite{Marciano:1977su} 
by means of continuous Schwinger function  methods~\cite{Qin:2020rad,Roberts:2020hiw,Cui:2020tdf,Chang:2021utv,Cui:2022bxn,Lu:2022cjx,Ding:2022ows,Roberts:2022rxm}, such
as Schwinger-Dyson equations (SDEs)~\cite{Roberts:1994dr,Alkofer:2000wg,Fischer:2006ub,Roberts:2007ji,Binosi:2009qm,Bashir:2012fs,Binosi:2014aea,Cloet:2013jya,Aguilar:2015bud,Binosi:2016rxz,Binosi:2016nme,Huber:2018ned} and functional renormalization group~\cite{Pawlowski:2003hq,  Pawlowski:2005xe,Fischer:2008uz,Carrington:2012ea,Carrington:2014lba,Cyrol:2017ewj, Corell:2018yil,Huber:2020keu,Dupuis:2020fhh, Blaizot:2021ikl}, 
together with a plethora of gauge-fixed lattice simulations~\cite{Mandula:1987rh,Parrinello:1994wd,Alles:1996ka,Parrinello:1997wm,Boucaud:1998bq,Alexandrou:2001fh,Bowman:2002bm,Skullerud:2003qu,Bowman:2004jm,Cucchieri:2006tf,Ilgenfritz:2006he,Sternbeck:2006rd,Furui:2006ks,Bowman:2007du,Kamleh:2007ud,Cucchieri:2007md,Cucchieri:2007rg,Bogolubsky:2007ud,Cucchieri:2008qm,Cucchieri:2008fc,Cucchieri:2009zt,Cucchieri:2009xxr,Boucaud:2008gn,Cucchieri:2009kk,Bogolubsky:2009dc,Oliveira:2009eh,Cucchieri:2010mfr,Oliveira:2010xc,Blossier:2010ky,Cucchieri:2011pp,Maas:2011se,Boucaud:2011ug,Ayala:2012pb,Oliveira:2012eh,Sternbeck:2012mf,Bicudo:2015rma,Duarte:2016iko,Athenodorou:2016oyh,Duarte:2016ieu,Oliveira:2016muq,Boucaud:2017obn,Sternbeck:2017ntv,Boucaud:2018xup,Cucchieri:2018leo,Cucchieri:2018doy,Oliveira:2018lln,Dudal:2018cli,Vujinovic:2018nqc,Cui:2019dwv,Zafeiropoulos:2019flq,Aguilar:2019uob,Maas:2020zjp,Kizilersu:2021jen,Aguilar:2021lke,Aguilar:2021okw,Pinto-Gomez:2022brg,Pinto-Gomez:2022qjv,Pinto-Gomez:2022woq}, has afforded ample access to the dynamical mechanisms
responsible  for the 
nonperturbative properties of this remarkable theory. 
Particularly prominent in this quest is the notion of the
emergent hadron mass (EHM)~\cite{Roberts:2020udq, Roberts:2020hiw, Roberts:2021xnz, Roberts:2021nhw, Binosi:2022djx, Papavassiliou:2022wrb,Ding:2022ows,Roberts:2022rxm}, together with its  
three supporting pillars: first, the generation of a gluon mass~\cite{Cornwall:1979hz,Parisi:1980jy,Cornwall:1981zr,Bernard:1981pg,Bernard:1982my,Donoghue:1983fy,Mandula:1987rh,Cornwall:1989gv,Lavelle:1991ve,Halzen:1992vd,Wilson:1994fk,Mihara:2000wf,Philipsen:2001ip,Kondo:2001nq,Aguilar:2002tc,Aguilar:2004sw,Aguilar:2006gr,Epple:2007ut,Aguilar:2007fe,Aguilar:2008xm,Aguilar:2009ke,Campagnari:2010wc,Fagundes:2011zx,Aguilar:2011ux,Aguilar:2011xe,Aguilar:2011yb,Aguilar:2013hoa,Aguilar:2015bud,Glazek:2017rwe,Binosi:2017rwj,Aguilar:2019kxz,Eichmann:2021zuv,Aguilar:2021uwa,Horak:2022aqx,Papavassiliou:2022wrb,Aguilar:2022thg}
through the action of the Schwinger mechanism~\cite{Schwinger:1962tn,Schwinger:1962tp}; second, the  construction of the process-independent effective charge~\cite{Cornwall:1981zr,Watson:1996fg,Binosi:2002vk,Aguilar:2009nf,Binosi:2014aea,Binosi:2016nme,Cui:2019dwv,Roberts:2020hiw}, which arises as the 
QCD analogue of the Gell-Mann--Low charge known from Quantum Electrodynamics (QED)~\cite{Gell-Mann:1954yli,Itzykson:1980rh}, and has 
associated to it a renormalization-group invariant (RGI) scale of about half of the
proton mass~\cite{Binosi:2016nme,Cui:2019dwv}; and third,  the dynamical breaking of chiral symmetry and the
generation of constituent quark masses~\cite{Nambu:1961tp,Lane:1974he,Politzer:1976tv,Miransky:1981rt,Atkinson:1988mw,Brown:1988bm,Williams:1989tv,Papavassiliou:1991hx,Hawes:1993ef,Roberts:1994dr,Natale:1996eu,Fischer:2003rp,Maris:2003vk, Aguilar:2005sb,Bowman:2005vx,Sauli:2006ba,Cornwall:2008da,Alkofer:2008tt,Aguilar:2010cn,Cloet:2013jya,Rojas:2013tza,Mitter:2014wpa,Braun:2014ata,Heupel:2014ina,Binosi:2016wcx,Aguilar:2018epe,Gao:2021wun}.  

The dynamics of the gauge sector of QCD, which encompasses both gluonic and ghost interactions, is instrumental for the physical picture of the EHM outlined above. 
In fact, the basic concepts and pivotal mechanisms sustaining the first two pillars of the EHM have their 
original inception and most genuine realization in the realm of pure Yang-Mills theories~\cite{Eichten:1974et,Smit:1974je,Cornwall:1979hz,Cornwall:1981zr,Aguilar:2006gr,Aguilar:2008xm,Binosi:2012sj,Aguilar:2015bud,Aguilar:2011xe,Papavassiliou:2022wrb}. Therefore, in the present review, we focus 
precisely on the rich dynamical content 
of the gauge sector, especially  
in relation with the generation of a gluon mass scale out of the intricate gluon self-interactions.

%%%%%%%%%%%%%%%%%%%%%%%%%%%%%%%%%%%%%%%%%%%%%%%

The formulation of the nonperturbative QCD physics 
in terms of the Green's functions of the fundamental degrees of freedom, 
such as gluon and ghost propagators and vertices, 
provides an intuitive framework for unraveling a wide array of 
subtle mechanisms; in fact, 
certain distinctive features of these 
functions have been inextricably connected
with key phenomena such as gluon mass 
generation, violation of reflection positivity, 
and confinement, to name a few. 
Thus, the saturation of the gluon propagator in the deep infrared~\cite{Alexandrou:2001fh,Cucchieri:2007md,Cucchieri:2007rg,Bogolubsky:2007ud,Bogolubsky:2009dc,Oliveira:2009eh,Oliveira:2010xc,Cucchieri:2009zt,Cucchieri:2009kk,Cucchieri:2011pp,Sternbeck:2012mf,Bicudo:2015rma,Bowman:2007du,Kamleh:2007ud,Ayala:2012pb,Duarte:2016iko,Dudal:2018cli,Aguilar:2019uob}  
has been interpreted as the unequivocal  signal of a gluon mass~\cite{Smit:1974je,Cornwall:1981zr,Bernard:1981pg,Bernard:1982my,Donoghue:1983fy,Mandula:1987rh, Cornwall:1989gv,Wilson:1994fk,Philipsen:2001ip,Aguilar:2002tc,Aguilar:2004sw,Aguilar:2006gr,Aguilar:2008xm,Tissier:2010ts,Binosi:2012sj,Serreau:2012cg,Pelaez:2014mxa,Siringo:2015wtx,Aguilar:2016vin};  and the existence of an inflection point in the same function 
has been argued to lead to a non-positive gluon spectral density~\cite{Ding:2022ows}, 
and the ensuing loss of reflection positivity~\cite{Osterwalder:1973dx,Osterwalder:1974tc,Glimm:1981xz,Krein:1990sf,Alkofer:2000wg,Roberts:2007ji,Cornwall:2013zra,Binosi:2014aea,Ding:2022ows} for the dressed gluons. 
Similarly, the masslessness of the ghost induces~\cite{Aguilar:2013vaa} a maximum in the 
gluon propagator, and a zero crossing in the form factors of the three-gluon 
vertex~\cite{Cucchieri:2008qm,Aguilar:2013vaa,Pelaez:2013cpa,Blum:2014gna,Eichmann:2014xya,Williams:2015cvx,Blum:2015lsa,Cyrol:2016tym,Athenodorou:2016oyh,Duarte:2016ieu,Boucaud:2017obn,Sternbeck:2017ntv,Corell:2018yil,Aguilar:2019jsj,Aguilar:2019uob,Aguilar:2021lke,Barrios:2022hzr}, followed by an infrared divergence for vanishing momenta. The dynamical origin of these special traits will be the focal point of the analysis 
presented in the main body of this article.

The integral equations that govern the  full momentum evolution of the 
Green's functions, known as SDEs, 
constitute the indispensable formal and practical instrument for unraveling the special characteristics mentioned above.
In their primordial form, the SDEs are rigorously derived  
from the generating functional of the theory~\cite{Rivers:1987hi,Itzykson:1980rh}, and encode all 
dynamical information on the correlation functions, within the entire range of physical momenta. 
In practice, due to the enormous complexity 
of these equations, approximations and truncations need to be implemented; but, 
unlike perturbation theory, 
no expansion parameter is available 
in the strongly coupled regime of the theory for carrying out such a task. 
Despite this intrinsic shortcoming, in recent years 
the SDE predictions have become particularly  robust, in part due to various theoretical advances, and in part thanks to the   
intense synergy with gauge-fixed lattice simulations, 
as will be evidenced in the subsequent sections.

%%%%%%%%%%%%%%%%%%%%%%%%%%%%%%%%%%%%%%%%%%%%%%%%%%%%%%

Typically, the Green's function of QCD are 
defined within the quantization scheme 
obtained by implementing the linear covariant 
($R_{\xi}$) gauges~\cite{Fujikawa:1972fe}. The corresponding 
SDEs 
are derived and solved within this same 
quantization scheme, 
and in particular in the 
Landau gauge ($\xi=0$), where  lattice simulations 
are almost exclusively performed; for studies away from the Landau gauge, see \eg\cite{Epple:2007ut,Aguilar:2007nf,Cucchieri:2009kk,Campagnari:2010wc,Huber:2010tvj,Cucchieri:2011pp,Siringo:2014lva,Bicudo:2015rma,Aguilar:2015nqa,Huber:2015ria,Capri:2015ixa,Aguilar:2016ock,Glazek:2017rwe,Cucchieri:2018leo,Cucchieri:2018doy,DeMeerleer:2019kmh,Napetschnig:2021ria}. 
A great deal may be learned, however,  by 
considering the Green's functions and 
corresponding SDEs formulated within the 
``PT-BFM''scheme~\cite{Aguilar:2006gr,Binosi:2007pi}, namely the framework that arises from the fusion  
of the pinch technique (PT)~\cite{Cornwall:1981zr,Cornwall:1989gv,Pilaftsis:1996fh,Binosi:2002ft,Binosi:2003rr,Binosi:2009qm} with the
background field method (BFM)~\cite{DeWitt:1967ub,tHooft:1971qjg,Honerkamp:1972fd,Kallosh:1974yh,Kluberg-Stern:1974nmx,Arefeva:1974jv,Abbott:1980hw,Weinberg:1980wa,Abbott:1981ke,Shore:1981mj,Abbott:1983zw}.
The main advantage of the PT-BFM originates from 
the fact that certain appropriately chosen Green's 
functions satisfy Abelian Slavnov-Taylor identities (STIs), whose tree-level form does not 
get modified by quantum corrections.
This situation is to be contrasted to the standard STIs~\cite{Taylor:1971ff,Slavnov:1972fg} obtained in the conventional framework of the linear covariant gauges, 
which are deformed by non-trivial contributions 
stemming from the gauge sector of the theory.
In the present work, we will 
carry out computations and develop arguments 
within both frameworks ($R_{\xi}$ and PT-BFM), 
and will elaborate on their connection 
by means of the so-called Background-Quantum identities 
(BQIs)~\cite{Grassi:1999tp,Grassi:2001zz,Binosi:2002ez,Binosi:2009qm}.

The article is organized as follows: 

$\bullet$ In Section~\ref{sec:prel} we introduce some basic 
notation and review certain prominent 
features of the Green's functions 
within both the linear gauges and 
the PT-BFM formalism~\cite{Aguilar:2006gr,Binosi:2007pi}. We stress, in particular, 
the properties of the auxiliary function $G(q)$~\cite{Aguilar:2009pp,Aguilar:2009nf,Binosi:2013cea,Binosi:2014aea}, 
which relates the gluon propagators 
with quantum and background gluons, and 
is intimately connected with 
the definition of the process-independent 
and RGI interaction strength~\cite{Binosi:2014aea}, to be discussed in detail in Section~\ref{sec:dhat}.
In addition, we 
elucidate with a concrete example 
the important property of ``block-wise'' 
transversality, displayed by the background 
gluon self-energy~\cite{Aguilar:2006gr,Aguilar:2008xm,Aguilar:2015bud}. 

$\bullet$  In Section~\ref{sec:smg} we review the general principles associated with the Schwinger mechanism~\cite{Schwinger:1962tn,Schwinger:1962tp} that endows gauge bosons with an effective mass, focusing on the details associated with its realization in the context of Yang-Mills theories. We place particular emphasis on the pivotal requirement that must be satisfied by
the fundamental vertices of the theory, namely the appearance of massless poles in their form factors~\cite{Eichten:1974et,Aguilar:2006gr,Aguilar:2007fe,Aguilar:2008xm,Aguilar:2009ke,Aguilar:2011xe,Ibanez:2012zk,Aguilar:2015bud,Papavassiliou:2022wrb}.

$\bullet$  In Section~\ref{sec:CBSE} we 
examine the dynamical formation of 
{\it colored} composite excitations (bound states) 
of vanishing mass, which 
provide the required structures in the vertices, in order for the Schwinger mechanism to be activated~\cite{Eichten:1974et,Aguilar:2011xe,Ibanez:2012zk,Aguilar:2015bud}.
The formation of these states out of a pair of gluons or a ghost--antighost pair
is controlled by a set of coupled 
Bethe-Salpeter equations (BSEs)~\cite{Aguilar:2011xe,Ibanez:2012zk,Aguilar:2015bud,Aguilar:2017dco,Aguilar:2021uwa}, which are found to have nontrivial solutions for the corresponding Bethe-Salpeter (BS) amplitudes, to be denoted by $\Cfat(r)$ and $\C(r)$, respectively.  

$\bullet$  In Section~\ref{sec:massgen} we explain 
in detail how the presence of the massless poles in the dressed vertices
that enter in the SDE of the gluon propagator gives rise to a gluon mass. 
The demonstration is carried out separately for the $g_{\mu\nu}$ and $q_{\mu}q_{\nu}/q^2$ 
components of the gluon self-energy. The former case requires the evasion of the so-called ``seagull identity''~\cite{Aguilar:2009ke,Aguilar:2016vin}; 
this becomes possible by virtue of the 
crucial Ward identity (WI) displacement, to be further considered in Section~\ref{sec:widis3g}.

$\bullet$  In Section~\ref{sec:dhat} we go over 
the basic notions underpinning the PT~\cite{Cornwall:1981zr,Cornwall:1989gv,Pilaftsis:1996fh,Binosi:2002ft,Binosi:2009qm}, and show  
how their application  leads naturally to the definition of a dimensionful process-independent RGI interaction strength~\cite{Cornwall:1981zr,Watson:1996fg,Binosi:2002vk,Aguilar:2009nf,Binosi:2014aea,Binosi:2016nme,Cui:2019dwv,Roberts:2020hiw}, denoted by ${\widehat d}(q)$. The genuine process-independence of this quantity is concretely exemplified by demonstrating its appearance in two processes involving fundamentally different external fields. Next, ${\widehat d}(q)$ is computed by 
combining lattice data for the gluon propagator and 
SDE results for the function $G(q)$. Finally, the dimensionless 
quantity is derived that constitutes the physical definition 
of the one-gluon exchange interaction appearing in 
standard bound-state computations~\cite{Munczek:1994zz,Bender:1996bb,Maris:1997hd,Maris:1997tm,Chang:2009zb,Chang:2011vu,Bashir:2012fs,Cloet:2013jya,Binosi:2014aea,Qin:2020jig}.

$\bullet$  In Section~\ref{sec:planar_deg} we focus on the structure of
the ``transversely projected'' three-gluon vertex~\cite{Eichmann:2014xya,Blum:2014gna,Huber:2016tvc,Aguilar:2022thg}, and discuss briefly 
the property of planar degeneracy~\cite{Pinto-Gomez:2022brg}, 
satisfied, at a high level of accuracy~\cite{Eichmann:2014xya,Blum:2014gna,Huber:2016tvc,Pinto-Gomez:2022brg,Pinto-Gomez:2022qjv,Pinto-Gomez:2022woq}, by the vertex form factors. 
This special property induces a striking simplification to  
the structure of this vertex, 
captured by a particularly compact expression~\cite{Pinto-Gomez:2022brg}, 
which will be extensively used in some of the following sections.

$\bullet$  In Section~\ref{sec:ghost_dyn} we take a close look at the ghost sector of the theory, and 
solve the coupled system of SDEs governing the ghost propagator and ghost-gluon vertex~\cite{Schleifenbaum:2004id,Boucaud:2008ky,Huber:2012kd,Aguilar:2013xqa,Aguilar:2018csq,Aguilar:2021okw}; as is well-known, the ghost remains massless, but its dressing function saturates at the origin~\cite{Ilgenfritz:2006he,Cucchieri:2007md,Bogolubsky:2007ud,Cucchieri:2008fc,Aguilar:2008xm,Dudal:2008sp,Boucaud:2008ky,Boucaud:2008ji,Bogolubsky:2009dc,Kondo:2009gc,Boucaud:2011ug,Pennington:2011xs,Dudal:2012zx,Ayala:2012pb,Aguilar:2013xqa,Cyrol:2016tym,Huber:2018ned,Boucaud:2018xup,Aguilar:2018csq,Cui:2019dwv,Aguilar:2021okw}, because the infrared-finite gluon propagator used in the ghost SDE provides an effective infrared cutoff. 
In the SDE of the ghost-gluon vertex, we employ as 
central input the compact expression for the three-gluon vertex presented in the previous section. The results are in excellent agreement with the available lattice data for the ghost dressing function~\cite{Boucaud:2018xup,Aguilar:2021okw} and the form factor 
of the ghost-gluon vertex evaluated in the soft-gluon limit~\cite{Ilgenfritz:2006he,Sternbeck:2006rd}. 

$\bullet$ In Section~\ref{sec:ghost_loops} we discuss two important consequences of the masslessness of the ghost propagator, which manifest themselves at the level of both the gluon propagator and the three-gluon vertex. Specifically, the diagrams comprised by a ghost loop induce ``unprotected'' logarithms, \ie of the type $\ln q^2$; instead, gluonic loops give rise to ``protected'' logarithms, of the type $\ln (q^2+m^2)$, where $m$ is the effective gluon mass~\cite{Aguilar:2013vaa,Papavassiliou:2022umz}. As $q^2 \to 0$, the unprotected  contributions diverge, driving the appearance of a maximum in the gluon propagator and a divergence in its first derivative, as well as a zero-crossing and a corresponding divergence in the form factors of the three-gluon vertex. As we comment in this section, of particular phenomenological importance~\cite{Meyers:2012ka,Sanchis-Alepuz:2015hma,Xu:2018cor,Souza:2019ylx,Huber:2020ngt,Huber:2021yfy,Papavassiliou:2022umz} is the relative suppression that the above features induce to the dominant vertex form factors in the intermediate range of momenta. 

$\bullet$  In Section~\ref{sec:widis3g} 
we discuss an outstanding feature 
of the WI satisfied by the pole-free part of the three-gluon vertex, namely the displacement induced by the presence of the 
aforementioned massless poles~\cite{Aguilar:2021uwa,Papavassiliou:2022wrb}. In this context, we introduce  
the key quantity denominated ``displacement function'',  
whose appearance serves as a  smoking gun signal of the action of the Schwinger mechanism in QCD; 
quite interestingly, it coincides~\cite{Aguilar:2021uwa,Papavassiliou:2022wrb} with the BS amplitude $\Cfat(r)$ for the 
formation of a massless scalar out of a pair of gluons, introduced in Section~\ref{sec:CBSE}. 
In addition, we derive a crucial relation, which 
ultimately permits the indirect determination of $\Cfat(r)$ 
from lattice QCD~\cite{Aguilar:2021uwa,Papavassiliou:2022wrb,Aguilar:2022thg}; an important ingredient in this relation is 
a partial derivative~\cite{Aguilar:2020yni,Aguilar:2021uwa}, denoted by  $\w(r)$, of the ghost-gluon kernel~\cite{Aguilar:2018csq}, to be determined 
in the next section.

$\bullet$ In Section~\ref{sec:WSDE} we set up and solve the SDE  
that governs the evolution of $\w(r)$~\cite{Aguilar:2020yni,Aguilar:2020uqw,Aguilar:2021uwa,Aguilar:2022thg}; 
the main component of this SDE is a special projection of the three-gluon vertex, which is 
computed by appealing to formulas established 
in Section~\ref{sec:planar_deg}, and allows for the 
accurate determination of $\w(r)$ in the entire range of relevant momenta~\cite{Aguilar:2022thg}.

$\bullet$  In Section~\ref{sec:wilat} 
we substitute into the central relation derived in Section~\ref{sec:widis3g} the
solution for  
$\w(r)$ found in the previous section, together with the lattice data~\cite{Aguilar:2021okw,Aguilar:2021lke} for 
the gluon propagator, the ghost dressing function, and the form factor of the three-gluon 
vertex associated with the soft-gluon limit, in order to obtain  
the form of the displacement function $\Cfat(r)$~\cite{Aguilar:2021uwa,Aguilar:2022thg}. As we discuss, the results exclude with nearly absolute certainty  the null hypothesis (absence of Schwinger mechanism, $\Cfat(r)=0$), 
and corroborate the action of the Schwinger mechanism in QCD~\cite{Aguilar:2022thg}. 
In addition, we show that the form of $\Cfat(r)$ found  
is statistically completely compatible with that obtained from the BSE-based analysis 
presented in Section~\ref{sec:CBSE}.

$\bullet$  In Section~\ref{sec:conc} we present our conclusions. 

$\bullet$  Finally, in Appendix~\ref{app:poleBQI} we derive the BQIs relating the 
displacement functions of the conventional and background vertices. 

%%%%%%%%%%%%%%%%%%%%%%%%%%%%%%%%%%%%%%%%%%%%%%%%%%%%%%%%
\section{Basic  concepts and general theoretical framework} \label{sec:prel}
%%%%%%%%%%%%%%%%%%%%%%%%%%%%%%%%%%%%%%%%%%%%%%%%%%%%%%%%

We start by considering the Lagrangian density of an SU($N$)  
Yang-Mills theory, comprised of the classical part, ${\cal L}_{\mathrm{cl}}$, 
the contribution from the ghosts, ${\cal L}_{\mathrm{gh}}$, and  
the covariant gauge-fixing term, ${\cal L}_{\mathrm{gf}}$, namely 
\be
 {\cal L}_{\mathrm{YM}} = {\cal L}_{\mathrm{cl}} + {\cal L}_{\mathrm{gh}} + {\cal L}_{\mathrm{gf}} \,,
	\label{lagden}
\ee
where
\be 
{\cal L}_{\mathrm{cl}} = -\frac14F^a_{\mu\nu}F^{a \mu\nu} \,, \qquad  {\cal L}_{\mathrm{gh}} = - {\overline c}^a\partial^\mu D_\mu^{ab}c^b \,, \qquad {\cal L}_{\mathrm{gf}} = \frac{1}{2\xi} (\partial^\mu A^a_\mu)^2 \,.
\ee 
In the above formula, $A^a_\mu(x)$ denotes the gauge field, 
while $c^a(x)$ and ${\overline c}^a(x)$ represent the ghost and antighost 
fields, respectively, with  
$a=1,\dots,N^2-1$.

In addition,  
\be 
F^a_{\mu\nu}=\partial_\mu A^a_\nu-\partial_\nu A^a_\mu+gf^{abc}A^b_\mu A^c_\nu \,,
\ee
is the antisymmetric field tensor, where
$f^{abc}$ stands for the totally antisymmetric structure constants of the SU($N$) gauge group, and $g$ is the gauge coupling, while 
\be
D_\mu^{ab} = \partial_\mu \delta^{ac} + g f^{amb} A^m_\mu \,,
\ee
denotes the covariant derivative in the adjoint representation. 
Finally, 
$\xi$ represents the gauge-fixing parameter; the choice 
$\xi=0$ corresponds to the  Landau gauge, while 
$\xi=1$ specifies the Feynman -´t Hooft gauge.

The transition 
from the pure Yang-Mills theory 
of \1eq{lagden}
to QCD is implemented by 
supplementing 
the corresponding kinetic and interaction terms for the quark fields. However, since 
throughout this work 
we do not consider effects due 
to dynamical quarks, 
the aforementioned terms 
will be omitted entirely.  

The most fundamental correlation function 
is the gluon propagator, whose nonperturbative features are 
inextricably connected with key dynamical properties of the theory. 
In the {\it Landau gauge} that we will employ throughout,
the gluon propagator, \mbox{$\Delta^{ab}_{\mu\nu}(q)=-i\delta^{ab}\Delta_{\mu\nu}(q)$},
is completely transverse, \ie 
\be
\Delta_{\mu\nu}(q) = \Delta(q) {P}_{\mu\nu}(q)\,, \qquad {P}_{\mu\nu}(q) := g_{\mu\nu} - q_\mu q_\nu/{q^2}\,.
\label{defgl}
\ee

\begin{figure}[t]
  \centering
  \includegraphics[width=1.0\textwidth]{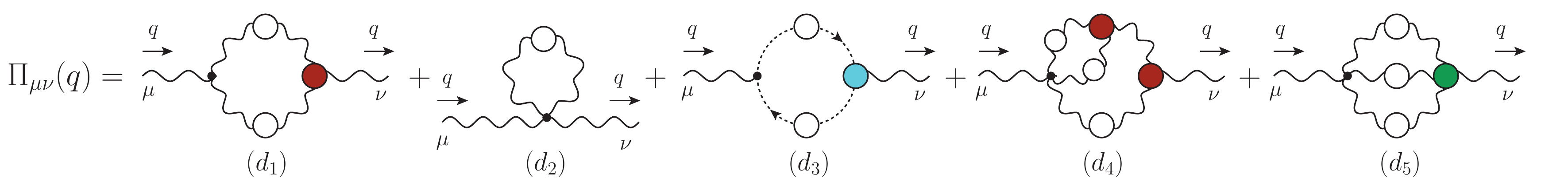}\\
  \includegraphics[width=1.0\textwidth]{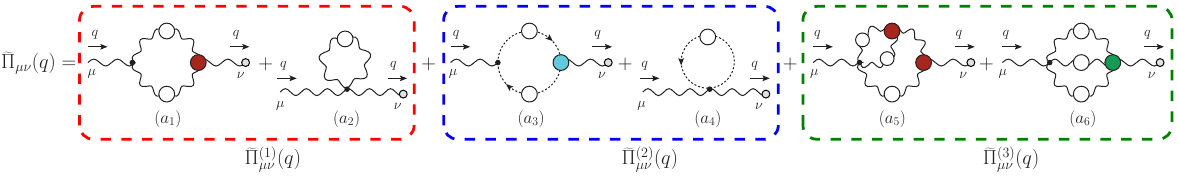}
\caption{ Upper panel: the diagrammatic representation of the conventional gluon self-energy, $\Pi_{\mu\nu}(q)$. Bottom panel: the diagrammatic representation of the $Q^a_\mu(q)B^b_\nu(-q)$, self-energy $\delta^{ab}\pt_{\mu\nu}(q)$; the grey circles at the end of the gluon lines indicate a background gluon. The corresponding Feynman rules are given in Appendix B of~\cite{Binosi:2009qm}.}
\label{fig:SDEs}
\end{figure}

In the continuum, the dynamical properties of the gluon propagator 
are encoded in the corresponding SDE, given by
\be
\Delta^{-1}(q)P_{\mu\nu}(q) = q^2P_{\mu\nu}(q)  + i {\Pi}_{\mu\nu}(q) \,,
\label{glSDE}
\ee
where  $\Pi_{\mu\nu}(q)$ is the gluon self-energy, shown diagrammatically in the first row of \fig{fig:SDEs}. 
The fully-dressed vertices entering the diagrams are determined by their own SDEs,  
obtaining finally a tower of coupled integral equations, which, for practical purposes, must be truncated or treated approximately.  

Given that, by virtue of the fundamental STI 
satisfied by the two-point function,  
the self-energy 
 $\Pi_{\mu\nu}(q)$ is transverse, 
\be
q^{\mu}\Pi_{\mu\nu}(q) = 0\,, 
\label{pitr}
\ee
we have that 
\be 
\Pi_{\mu\nu}(q) = \Pi (q) P_{\mu\nu}(q)\,,
\label{pitr2}
\ee
and 
from \1eq{glSDE} follows that  
\be
\Delta^{-1}(q) = q^2 + i \Pi (q)\,.
\label{DeltaPi}
\ee
Of particular importance is the 
exact way that  \1eq{pitr} is enforced 
at the level of the SDE given in \fig{glSDE}, which  
governs the gluon evolution. 
In particular, note that, if we were 
to contract the corresponding diagrams by $q^{\mu}$, 
the entire set of diagrams must be considered 
in order for \1eq{pitr} to emerge from the SDE. 
This pattern manifests itself already at the 
one-loop level, where it is known that the ghost-loop 
must be included in order to guarantee the 
transversality of the self-energy.
The main practical drawback stemming from this observation is that truncations, in the form of omission of 
certain subsets of graphs, are likely to distort this fundamental property.

%%%%%%%%%%%%%%%%%%%%%%%%%%%%%%%%%%
%Fig. 1 - Gluon SDE
%%%%%%%%%%%%%%%%%%%%%%%%%%%%%%%%%%
%%%%%%%%%%%%%%%%%%%%%%%%%%%%%%%%%%

Quite interestingly,  within the PT-BFM framework 
the transversality property of \1eq{pitr} is enforced 
in a very special way, which permits physically meaningful truncations.
In what follows we will employ predominantly the language of the BFM; for the basic principles of the PT and its 
connection with the BFM, the reader is referred to the extended literature on the subject~\cite{Cornwall:1981zr,Cornwall:1989gv,Pilaftsis:1996fh,Binosi:2002ft,Binosi:2002ez,Binosi:2009qm,Cornwall:2010upa}, as well as to Section~\ref{sec:dhat} of the present work.

%%%%%%%%%%%%%%%%%%%%%%%%%%%

The BFM is a powerful quantization procedure, where 
the gauge-fixing is implemented without compromising 
explicit gauge invariance. Within this framework 
the gauge field $A$ appearing in the classical action is decomposed as $A = B +Q$, where $B$ and $Q$ are the 
background and quantum (fluctuating) fields, respectively. 
Note that the variable of integration in the generating functional $Z(J)$ is the quantum field, which couples to the external sources, as $J\cdot Q$. The background field does not appear in loops. Instead, it couples externally to the Feynman diagrams, connecting them with the asymptotic states to form elements of the S-matrix.
Then, 
if the gauge-fixing term  
\be 
{\widehat {\cal L}}_{\mathrm{gf}} = \frac{1}{2\xi_{\s Q}}({\widehat D}_\mu^{ab}Q^{b \mu})^2 \,, \qquad {\widehat D}_\mu^{ab} = \partial_\mu \delta^{ab} + g f^{amb} B_\mu^m \,,
\ee 
is used, 
the resulting gauge-fixed action retains its 
invariance under gauge transformations of the background field.
As a result of this invariance, 
when the Green's functions are contracted 
by the momentum carried by a  background gluon,  
they satisfy Abelian 
(ghost-free) STIs, akin to 
the Takahashi identities known from QED.
In particular, the STIs of the BFM retain their tree-level form to all orders, in contradistinction to the STIs of the $R_\xi$ gauges, 
whose form is modified by contributions 
stemming from the ghost sector.

Within the BFM, one may 
consider three kinds of propagators, by choosing the type of incoming and outgoing gluons~\cite{Binosi:2008qk}. In particular, we have:

({\it i}) The propagator $\langle 0 \vert \,T \! [Q^a_\mu(q)Q^b_\nu(-q) ]\!\vert 0 \rangle$ that connects 
two quantum gluons. Notice that this propagator 
coincides with the conventional 
gluon propagator of the covariant gauges, defined in \1eq{defgl}, under the assumption that  
the corresponding gauge-fixing parameters, $\xi$ and 
$\xi_{\s Q}$, are identified, \ie $\xi=\xi_{\s Q}$.

({\it ii}) 
The propagator $\langle 0 \vert \,T \! [Q^a_\mu(q)B^b_\nu(-q) ]\!\vert 0 \rangle$ that connects a $Q^a_\mu(q)$ 
with a $B^b_\nu(-q)$, to be denoted by
$\widetilde{\Delta}^{ab}_{\mu\nu}(q)=-i\delta^{ab}\widetilde{\Delta}_{\mu\nu}(q)$.

({\it iii})
The propagator $\langle 0 \vert \,T \! [B^a_\mu(q)B^b_\nu(-q) ]\!\vert 0 \rangle$ that connects a $B^a_\mu(q)$ 
with a $B^b_\nu(-q)$, to be denoted by
$\widehat{\Delta}^{ab}_{\mu\nu}(q)=-i\delta^{ab}\widehat{\Delta}_{\mu\nu}(q)$. Note that its full definition 
requires an additional gauge-fixing term, with the associated ``classical'' 
gauge-fixing parameter, $\xi_{\s C}$~\cite{Abbott:1980hw,Abbott:1983zw,Binosi:2009qm}. 

Given that the relations captured by \2eqs{defgl}{glSDE} apply 
also in the cases  
of $\widetilde{\Delta}_{\mu\nu}(q)$ and 
$\widehat{\Delta}_{\mu\nu}(q)$, one may 
define 
the corresponding self-energies $\widetilde{\Pi}_{\mu\nu}(q)$ and  $\widehat{\Pi}_{\mu\nu}(q)$, as well as the functions 
$\widetilde\Delta(q)$ and $\widehat\Delta(q)$. 

%%%%%%%%%%%%%%%%%%%%%%%%%%%%%%%%%%%%%
%  Table I - Gluon propagator types
%%%%%%%%%%%%%%%%%%%%%%%%%%%%%%%%%%%%%
\begin{table}[htb]
	\caption{The different types of gluon propagators of the background field method (BFM), together with their diagrammatic representations, symbols, corresponding self-energies, and the background quantum identities (BQIs) that relate them to the conventional propagator.}
	\label{prop_types}
	\begin{center} 
	\begin{tabular}{|>{\centering\arraybackslash}m{2.2cm}|>{\centering\arraybackslash}m{2.3cm}|>{\centering\arraybackslash}m{1.9cm}|>{\centering\arraybackslash}m{2.cm}|>{\centering\arraybackslash}m{3cm}|}
		\hline 	
		\vspace{0.1cm}
		External legs & Diagrammatic representation & Symbol & Self-energy & BQI \\
		\hline 	
		\vspace{0.1cm}
		$Q^a_\mu(q)Q^b_\nu(-q)$ &
		\includegraphics[scale=0.6, trim = 0 0 41pt 0 ]{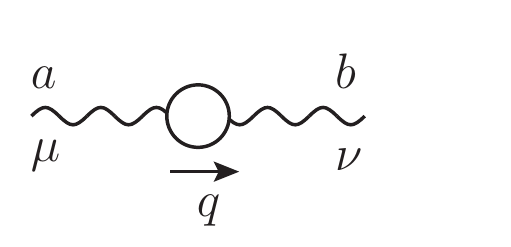} &
		$$-i\delta^{ab}\Delta_{\mu\nu}(q) $$ & $$ \Pi_{\mu\nu}(q) $$ &
		--- \\
		\hline
		\vspace{0.1cm}
		
		$Q^a_\mu(q)B^b_\nu(-q)$ &
		\includegraphics[scale=0.6, trim = 0 0 41pt 0 ]{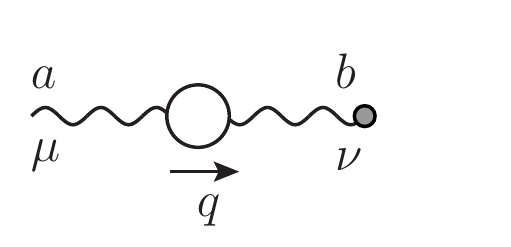} &
		$$-i\delta^{ab}{\widetilde \Delta}_{\mu\nu}(q) $$ & $${\widetilde \Pi}_{\mu\nu}(q)$$ & $${\widetilde \Delta}(q) = \frac{\Delta(q)}{1 + G(q)}$$ \\
		\hline
		\vspace{0.1cm}		
		
		$B^a_\mu(q)B^b_\nu(-q)$ &
		\includegraphics[scale=0.6, trim = 0 0 41pt 0 ]{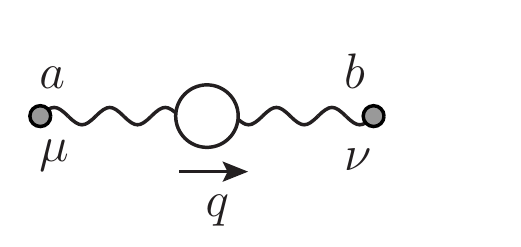} &
		$$-i\delta^{ab}{\widehat\Delta}_{\mu\nu}(q)$$ & $${\widehat \Pi}_{\mu\nu}(q)$$ & $${\widehat \Delta}(q) = \frac{\Delta(q)}{[1 + G(q)]^2}$$ \\
		\hline
	\end{tabular}
	\end{center}
\end{table}
%%%%%%%%%%%%%%%%%%%%%%%%%%%%%%%%%%%%%%

\begin{figure}[h]
  \centering
  \includegraphics[width=0.6\textwidth]{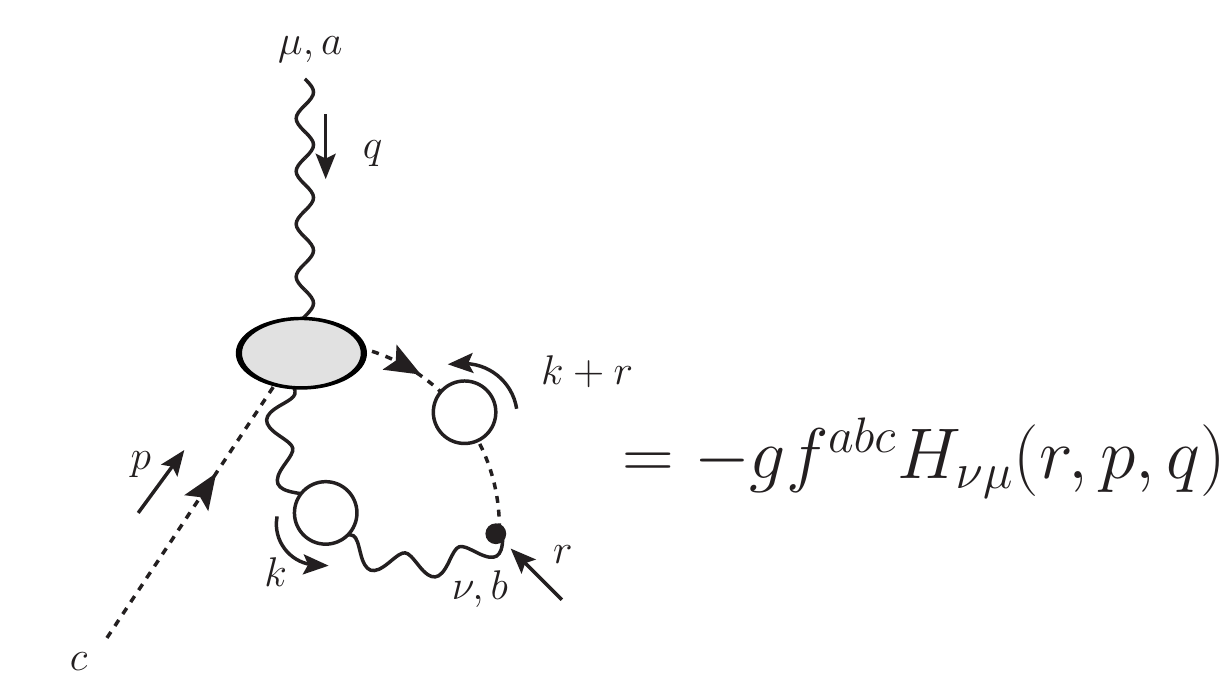}
\caption{ Diagrammatic definition of the ghost-gluon scattering kernel, $H_{\nu\mu}(r,p,q)$. At tree level, $H_{\nu\mu}^{0} = g_{\nu\mu}$. }
\label{fig:H_def}
\end{figure}

Quite interestingly,
the three propagators 
defined in ({\it i})-({\it iii})
are related by a set of exact 
identities, known as BQIs~\cite{Grassi:1999tp,Grassi:2001zz,Binosi:2002ez,Binosi:2009qm}. In particular, we have that (see also Table \ref{prop_types})
\be
\Delta(q) = [1 + G(q)] \widetilde\Delta(q) = [1 + G(q)]^2 \widehat\Delta(q)\,,
\label{propBQI}
\ee
where the function $G(q)$ is the $g_{\mu\nu}$ component of a particular two-point ghost function, $\Lambda_{\mu\nu}(q)$, given by~\cite{Grassi:1999tp,Binosi:2002ez,Grassi:2004yq,Binosi:2013cea}
\be 
\Lambda_{\mu\nu}(q) := i g^2 C_{\rm A}\int_k \Delta^\rho_\mu(k)D(k+q)H_{\nu\rho}(-q,k+q,-k) = g_{\mu\nu} G(q) + \frac{q_\mu q_\nu}{q^2}L(q) \,, \label{Lambda_GL}
\ee 
where $C_\mathrm{A}$ is the Casimir eigenvalue of the adjoint representation [$N$ for SU$(N)$], $D^{ab}(q) = i\delta^{ab}D(q)$ is the ghost propagator, and $H_{\nu\mu}(r,p,q)$ denotes the ghost-gluon kernel defined in \fig{fig:H_def}.

%%%%%%%%%%%%%%%%%%%%%%%%%%%%%%%%%%%%%%%%%%%%%%

In the Landau gauge, a special identity relates the form factors of 
$\Lambda_{\mu\nu}(q)$
to the ghost dressing function, $F(q)$, defined as 
$F(q)=q^2 D(q)$,
namely~\cite{Aguilar:2009nf,Binosi:2013cea,Binosi:2014aea}
\be 
F^{-1}(q) = 1 + G(q) + L(q) \,, \label{FGL_unren}
\ee 
which is valid before renormalization.
In fact, in this particular gauge, 
$G(q)$ coincides with the so-called Kugo-Ojima function~\cite{Kugo:1995km,Grassi:2004yq,Kondo:2009ug,Aguilar:2009pp}.

%%%%%%%%%%%%%%%%%%%%%%%%%%%%%%%%%%
%Fig. 2 - H_munu definition
%%%%%%%%%%%%%%%%%%%%%%%%%%%%%%%%%%
%%%%%%%%%%%%%%%%%%%%%%%%%%%%%%%%%%

To determine the renormalized form of \1eq{FGL_unren}, we introduce the renormalization constants of the conventional Green's functions
\begin{eqnarray}
&\Delta_{{\rm \s{R}}}(q) = Z_{A}^{-1} \Delta(q) \,, \qquad & F_{{\rm \s{R}}}(q) = Z_c^{-1}F(q) \,, \nonumber\\
&\fatg_\mu^{{\rm \s{R}}}(r,p,q) = Z_1 \fatg_\mu(r,p,q) \,, \qquad & \fatg_{\alpha\mu\nu}^{{\rm \s{R}}}(q,r,p) = Z_3 \fatg_{\alpha\mu\nu}(q,r,p) \,, \nonumber\\
& g_{{\rm \s{R}}} = Z_g^{-1} g \,, \qquad & \left[g_{\mu\nu} + \Lambda_{\mu\nu}^{{\rm \s{R}}}(q)\right] = Z_\Lambda \left[g_{\mu\nu} + \Lambda_{\mu\nu}(q)\right] \,, \nonumber\\
& Z_g^{-1} = Z_1^{-1}Z_A^{1/2}Z_c = Z_3^{-1}Z_A^{3/2} \,, & \label{Zs_def}
\end{eqnarray}
where we denote by $\fatg^{abc}_\mu(r,p,q) = - gf^{abc}\fatg_\mu(r,p,q)$ and $\fatg^{abc}_{\alpha\mu\nu}(q,r,p) = gf^{abc}\fatg_{\alpha\mu\nu}(q,r,p)$ the conventional ghost-gluon [$Q^a_\mu(q)c^c(p){\bar c}^b(r)$] and three-gluon [$Q^a_\alpha(q)Q^b_\mu(r)Q^c_\nu(p)$] vertices, respectively. Note that, by virtue of Taylor's theorem~\cite{Taylor:1971ff}, $Z_1$ is \emph{finite} in the Landau gauge; its precise value depends on the renormalization scheme adopted, see  Sec.~\ref{sec:ghost_dyn}.
Moreover, denoting by 
${\widehat Z}_A$ the (wave-function) 
renormalization constant of 
$\widehat \Delta(q)$, 
the Abelian STIs 
of the BFM impose the validity of the pivotal relation~\cite{Abbott:1980hw,Abbott:1983zw,Binosi:2009qm}
\be 
Z_g = {\widehat Z}_A^{-1/2} \,, 
\label{Zg_ZA}
\ee
which is the non-Abelian analogue 
of the textbook relation $Z_e = Z_A^{-1/2}$~\cite{Itzykson:1980rh}, relating 
the renormalization constants of the electric charge 
and the photon propagator in QED. 

Then, since the BQIs of \1eq{propBQI} are direct consequences of the Becchi-Rouet-Stora-Tyutin (BRST) symmetry~\cite{Becchi:1974md,Becchi:1975nq,Tyutin:1975qk} of the theory~\cite{Grassi:1999tp,Binosi:2002ez,Grassi:2004yq,Binosi:2013cea}, their form is preserved by renormalization. Hence, combining \3eqs{propBQI}{Zg_ZA}{Zs_def} we obtain
\be 
Z_{\Lambda} = Z_1^{-1} Z_c \,, \label{Z_Lambda}
\ee 
which yields\footnote{In the original and widely used~\cite{Aguilar:2009nf,Binosi:2014aea,Binosi:2016nme,Cui:2019dwv,Roberts:2020hiw,Ding:2022ows} version of \1eq{FGL} the renormalization is performed in the so-called Taylor scheme, where $Z_1 = 1$.}
\be 
Z^{-1}_1 F^{-1}(q) = 1 + G(q) + L(q) \,. \label{FGL}
\ee 

As has been shown in~\cite{Aguilar:2009nf}, the dynamical equation 
governing $L(q)$ yields $L(0) =0$, provided that 
the gluon propagator entering in it is finite at the origin. 
Thus, one obtains from \1eq{FGL}
the useful identity~\cite{Aguilar:2009pp}
\be 
Z^{-1}_1 F^{-1}(0)=1+G(0) \,. 
\label{F0_G0} \,
\ee 
According to numerous lattice simulations and studies in the
continuum (see,  
\eg\cite{Ilgenfritz:2006he,Cucchieri:2007md,Bogolubsky:2007ud,Cucchieri:2008fc,Aguilar:2008xm,Dudal:2008sp,Boucaud:2008ky,Boucaud:2008ji,Bogolubsky:2009dc,Kondo:2009gc,Boucaud:2011ug,Pennington:2011xs,Dudal:2012zx,Ayala:2012pb,Aguilar:2013xqa,Cyrol:2016tym,Huber:2018ned,Boucaud:2018xup,Aguilar:2018csq,Cui:2019dwv,Aguilar:2021okw}), the ghost dressing function 
reaches a finite
(nonvanishing) value at the origin, which, due to 
\1eq{F0_G0}, furnishes also the value of $G(0)$. 

%%%%%%%%%%%%%%%%%%%%%
The final upshot of the above considerations is that one may 
use the BQIs in \1eq{propBQI} to 
express 
the SDE given in \1eq{glSDE} in terms of the 
 $\widetilde{\Pi}_{\mu\nu}(q)$ or $\widehat{\Pi}_{\mu\nu}(q)$, 
 at the modest cost of introducing in the dynamics 
 the quantities $1+G(q)$ or $[1+G(q)]^2$. Focusing on the former possibility, \1eq{propBQI} becomes
\be
\Delta^{-1}(q)P_{\mu\nu}(q) = \frac{q^2P_{\mu\nu}(q)  + i \pt_{\mu\nu}(q)}{1 + G(q)} \,, 
\label{sdebq}
\ee
where the diagrammatic representation of the 
self-energy $\pt_{\mu\nu}(q)$ is shown in the lower panel of \fig{fig:SDEs}.

The principal advantage of this formulation
is that the self-energy $\pt_{\mu\nu}(q)$
contains fully-dressed vertices with a background gluon of momentum $q$ exiting from them, which satisfy Abelian STIs.
In particular, denoting by $\widetilde{\fatg}_{\mu\alpha\beta}(q,r,p)$, 
$\widetilde{\fatg}_\mu(r,p,q)$, and $\widetilde{\fatg}^{mnrs}_{\mu\alpha\beta\gamma}(q,r,p,t)$  
the BQQ, Bcc, and BQQQ vertices, respectively, we have that~\cite{Cornwall:1989gv,Aguilar:2006gr,Binosi:2009qm} 
\begin{eqnarray}
q^\mu \widetilde{\fatg}_{\mu\alpha\beta}(q,r,p) &=& \Delta_{\alpha\beta}^{-1}(r) - \Delta_{\alpha\beta}^{-1}(p)\,,
\label{st1} \\
q^\mu \widetilde{\fatg}_\mu(r,p,q) &=& {D}^{-1}(p) - {D}^{-1}(r) \,,
\label{st2}\\
q^\mu \widetilde{\fatg}^{mnrs}_{\mu\alpha\beta\gamma}(q,r,p,t) &=& f^{mse}f^{ern} {\fatg}_{\alpha\beta\gamma}(r,p,q+t) + f^{mne}f^{esr}{\fatg}_{\beta\gamma\alpha}(p,t,q+r)
\nonumber \\
&+& f^{mre}f^{ens} {\fatg}_{\gamma\alpha\beta}(t,r,q+p)\,.
\label{st3}
\end{eqnarray}

In contrast, the conventional three-gluon and ghost-gluon vertices, $\fatg_{\alpha\mu\nu}(q,r,p)$ and $\fatg_{\alpha}(r,p,q)$, respectively, satisfy the STIs~\cite{Marciano:1977su,Ball:1980ax,Davydychev:1996pb,vonSmekal:1997ern,Binosi:2011wi,Gracey:2019mix}
\begin{align}
& q^\alpha \fatg_{\alpha \mu \nu}(q,r,p) = F(q)
\left[\Delta^{-1}(p) P_\nu^\sigma(p) H_{\sigma\mu}(p,q,r) - \Delta^{-1}(r) P_\mu^\sigma(r) H_{\sigma\nu}(r,q,p)\right]\,,
\label{st1_conv} \\
& q^\mu F^{-1}(q) \fatg_{\mu}(r,p,q) + p^\mu F^{-1}(p) \fatg_{\mu}(r,q,p) = - r^2 F^{-1}(r) U(r,q,p)\,, \label{st2_conv}
\end{align}
where $U(r,q,p)$ is an interaction kernel containing only ghost fields; its tree level value is $U^0(r,q,p) = 1$.
The STI for the conventional four-gluon vertex is given in Eq.~(C.24) of~\cite{Binosi:2009qm}.

The special STIs listed in \3eqs{st1}{st2}{st3} are responsible 
for the remarkable property of ``block-wise'' transversality~\cite{Aguilar:2006gr,Binosi:2007pi,Binosi:2008qk}, displayed by $\pt_{\mu\nu}(q)$. 
To appreciate this point, notice 
that the diagrams comprising $\pt_{\mu\nu}(q)$ in \fig{fig:SDEs}
have been separated into three different subsets (blocks) comprised of:
({\it i}) one-loop dressed diagrams containing only gluons, ({\it ii}) one-loop dressed diagrams containing a ghost loop, and
({\it iii}) two-loop dressed diagrams containing only gluons. The corresponding contributions of each block
to $\pt_{\mu\nu}(q)$ are denoted by 
$\pt^{(i)}_{\mu\nu}(q)$, with $i=1,2,3$.

The block-wise transversality is a stronger version of the standard 
transversality relation \mbox{$q^{\mu} \pt_{\mu\nu}(q) =0$}; it states that 
each block of diagrams mentioned above is individually transverse,  
namely 
\be
q^{\mu} \pt^{(i)}_{\mu\nu}(q)= 0\,,\qquad i=1,2,3.
\label{blockwise}
\ee

In order to appreciate in detail 
the reason why the 
STIs in \3eqs{st1}{st2}{st3} 
are instrumental for 
the block-wise transversality,
we will consider the case of  $\pt^{(2)}_{\mu\nu}(q)$; 
the relevant 
diagrams are enclosed by the blue box of \fig{fig:SDEs}. 

The diagrams $(a_3)$ and $(a_4)$ are given by
\begin{align}
    (a_3)_{\mu\nu}(q) &= g^2C_{\rm A}  \int_k (k+q)_\mu D(k+q)D(k)\fatgt_\nu(-k,k+q,-q) \,,  \\
    (a_4)_{\mu\nu}(q) &= g^2C_{\rm A}\, g_{\mu\nu} \int_k  D(k) \,,
\label{a3a4}
\end{align}
where a color factor $\delta^{ab}$ has been suppressed 
in both expressions. In addition, for the formal manipulations of integrals, we employ 
dimensional regularization~\cite{Collins:1984xc}; to that end, we introduce the short-hand notation 
\be
\int_{k} :=\frac{\mu_{\s 0}^{\epsilon}}{(2\pi)^{d}}\!\int_{-\infty}^{+\infty}\!\!\mathrm{d}^d k\,,
\label{dqd}
\ee
where $d=4-\epsilon$ is the dimension of the space-time, and $\mu_{\s 0}$ denotes the 't Hooft mass.

The contraction of  graph $(a_3)_{\mu\nu}(q)$ by $q^\nu$ triggers  
the STI satisfied by $\gt_\nu(-k,k+q,-q)$ 
[given by \1eq{st2}], and we obtain 
\begin{eqnarray}
    q^\nu(a_3)_{\mu\nu}(q) &=&  g^2 C_{\rm A} \int_k (k+q)_\mu D(k+q)D(k) \left[D^{-1}(k) - D^{-1}(k+q)\right]
 \nonumber\\   
    &=& g^2 C_{\rm A} \int_k (k+q)_\mu \left[D(k+q) - D(k)\right]
    \label{qa3}
\nonumber\\ 
&=& - g^2 C_{\rm A} \,q_\mu \int_k D(k) \,,
\end{eqnarray}
which is precisely the negative of the contraction $q^\nu(a_4)_{\mu\nu}(q)$. Hence, 
\be
    q^\nu\left[ (a_3)_{\mu\nu}(q) + (a_4)_{\mu\nu}(q) \right] = 0\,.
\label{qa3a4}
\ee

%%%%%%%%%%%%%%%%%%%%%%%%%%%%%%%%%%%%%%%%%%%%%%%%%%%%%%%%
\section{Schwinger mechanism in Yang-Mills theories}\label{sec:smg}
%%%%%%%%%%%%%%%%%%%%%%%%%%%%%%%%%%%%%%%%%%%%%%%%%%%%%%%%

The BRST symmetry of the Yang-Mills Lagrangian given 
in \1eq{lagden} prohibits the inclusion of a mass term 
of the form $m^2 A^2_\mu$. Moreover, 
a symmetry-preserving regularization scheme, 
such as dimensional regularization, prevents 
the generation of a mass term at any finite 
order in perturbation theory. 
Nonetheless, as affirmed 
four decades ago~\cite{Cornwall:1979hz,Parisi:1980jy,Cornwall:1981zr,Bernard:1981pg,Bernard:1982my,Donoghue:1983fy}, 
the nonperturbative 
Yang-Mills dynamics endow the gluons with an 
effective mass, which  
sets the scale for all dimensionful quantities, 
and tames the instabilities originating from 
the infrared divergences of the perturbative expansion (\eg Landau pole). 
In addition, the presence of this 
mass causes the 
effective decoupling (screening) of the gluonic modes beyond a ``maximum gluon wavelength''~\cite{Brodsky:2008be},
and leads to the dynamical 
suppression of the Gribov copies, see, \eg\cite{Braun:2007bx,Binosi:2014aea,Gao:2017uox} and references therein.

The generation of a gluon mass proceeds  
through the nonperturbative realization  
of the Schwinger mechanism~\cite{Schwinger:1962tn,Schwinger:1962tp}. 
Even though the technical details associated with the implementation of this mechanism in a four-dimensional non-Abelian setting are particularly elaborate, 
the general underlying idea is relatively easy to convey. 

To that end, consider the dimensionless vacuum polarization
${\bf \Pi}(q)$, defined 
through ${\Pi}(q) = q^2 {\bf \Pi}(q)$, such that  
\be 
\Delta^{-1}({q})=q^2 [1 + i {\bf \Pi}(q)]\,.
\label{vacpol}
\ee
The Schwinger mechanism is based on the fundamental  
observation that, if ${\bf \Pi}(q)$ 
develops a pole at $q^2=0$ 
(to be referred to as ``{\it massless pole''}) then the 
vector meson (gluon) picks up a mass, regardless 
of any ``prohibition'' imposed by the gauge symmetry 
at the level of the original Lagrangian. 
Thus, in Euclidean space, the above sequence of ideas leads to  
\be
\lim_{q \to 0} {\bf \Pi}(q) = m^2/q^2 \,\,\Longrightarrow \,\,\lim_{q \to 0} \,\Delta^{-1}(q) = \lim_{q \to 0} \,(q^2 + m^2) \,\,\Longrightarrow \,\,\Delta^{-1}(0) = m^2\,,
\label{schmech}
\ee
and the gauge boson propagator saturates to a  non-zero value at the origin. This 
effect of infrared saturation of the propagator signifies the generation of a mass, which is  
identified with the positive residue  
of the pole.

At this descriptive level,  
Schwinger's argument 
is completely general, 
making no particular reference 
to the specific dynamics 
that would 
lead to the appearance of the 
required massless pole 
inside ${\bf \Pi}(q)$.
In fact, depending on the 
particular theory, the field-theoretic  
circumstances that trigger the crucial sequence captured by  \1eq{schmech} may be very distinct, see, \eg~\cite{Jackiw:1973tr,Jackiw:1973ha}. 
In the case of Yang-Mills theories, the 
origin of the massless poles is purely
nonperturbative~\cite{Eichten:1974et}: 
the strong dynamics produce scalar composite 
excitations, which carry color and have vanishing 
masses. These poles are carried by the 
fully-dressed vertices of the theory; and since  
these vertices enter 
in the gluon SDE shown in \fig{fig:SDEs} [upper (lower) panel for the 
QQ (QB) propagator], the 
massless poles find their way into the gluon self-energy 
(or, equivalently, the gluon vacuum polarization).
The detailed implementation of this idea 
has been presented in a series of works~\cite{Eichten:1974et,Smit:1974je,Cornwall:1981zr,Papavassiliou:1989zd,Aguilar:2008xm,Aguilar:2011xe,Aguilar:2011yb,Binosi:2012sj,Aguilar:2011xe,Aguilar:2011ux,Aguilar:2015bud,Aguilar:2016vin,Aguilar:2016ock,Papavassiliou:2022wrb}, 
and will be summarized in the rest of this section.

\begin{figure}[t]
  \centering
\includegraphics[width=0.8\textwidth]{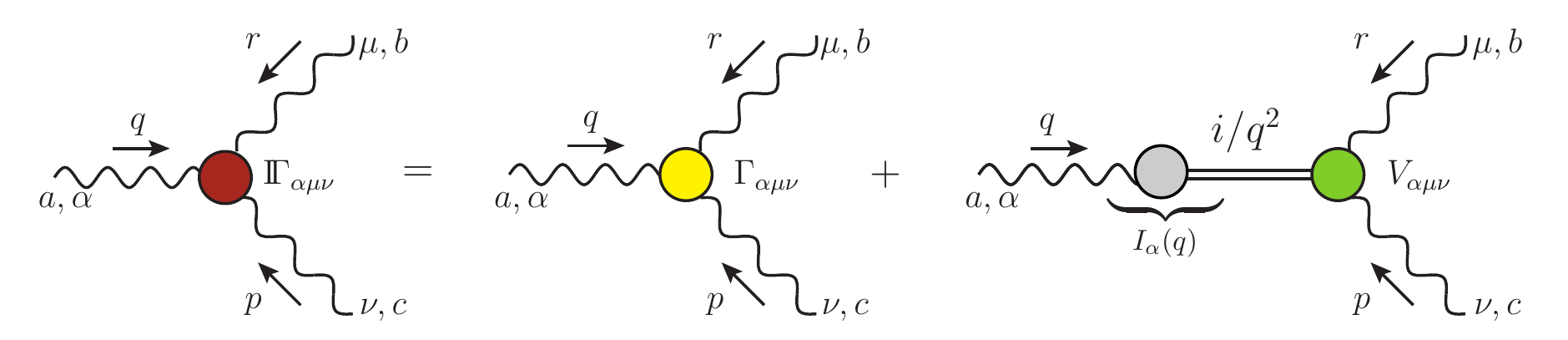} \\
\includegraphics[width=0.8\textwidth]{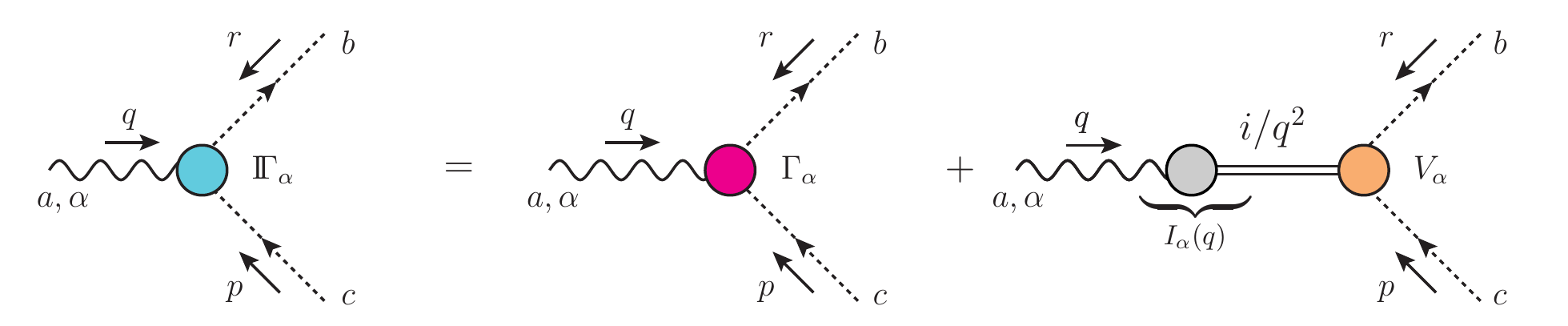}
\caption{The diagrammatic representation of the three-gluon and ghost-gluon vertices introduced in \1eq{fullgh}:
$\fatg_{\alpha\mu\nu}(q,r,p)$ (first row) and $\fatg_\alpha(r,p,q)$ (second row).
 The first term on the r.h.s. indicates the pole-free part, $\g_{\alpha\mu\nu}(q,r,p)$ or $\g_{\alpha}(r,p,q)$, 
while the second denotes the pole term  $V_{\alpha\mu\nu}(q,r,p)$ or $V_{\alpha}(r,p,q)$.}
\label{fig:poles}
\end{figure}

Let us focus for now on the conventional 
three-gluon and ghost-gluon vertices, $\fatg_{\alpha\mu\nu}(q,r,p)$ and $\fatg_\alpha(r,p,q)$,
respectively, introduced below \1eq{Zs_def}. When the formation of massless poles is triggered, these  
vertices assume the general form (see \fig{fig:poles})
\begin{eqnarray} 
\fatg_{\alpha\mu\nu}(q,r,p) &=& \g_{\alpha\mu\nu}(q,r,p) + V_{\alpha\mu\nu}(q,r,p) \,,\qquad
\nonumber\\
\fatg_\alpha(r,p,q) &=& \g_{\alpha}(r,p,q) + V_{\alpha}(r,p,q)\,,
\label{fullgh}
\end{eqnarray}
where $\g_{\alpha\mu\nu}(q,r,p)$ and $\g_\alpha(r,p,q)$ are their pole-free components, while 
$V_{\alpha\mu\nu}(q,r,p)$ and $V_{\alpha}(q,r,p)$ contain {\it longitudinally coupled} poles, whose 
special tensorial structure is given by 
\begin{eqnarray} 
V_{\alpha\mu\nu}(q,r,p) &=& \frac{q_\alpha}{q^2}  C_{\mu\nu}(q,r,p) + \frac{r_\mu}{r^2} A_{\alpha\nu}(q,r,p) + \frac{p_\nu}{p^2} B_{\alpha\mu}(q,r,p) \,,
\nonumber\\
V_\alpha(r,p,q) &=& \frac{q_\alpha}{q^2}\Cgh(r,p,q)\,,
\label{eq:Vgen}
\end{eqnarray}
such that 
\be
\label{eq:transvp}
{P}_{\alpha'}^{\alpha}(q){P}_{\mu'}^{\mu}(r){P}_{\nu'}^{\nu}(p) V_{\alpha\mu\nu}(q,r,p) = 0 \,,\qquad {P}_{\alpha'}^{\alpha}(q) V_\alpha(r,p,q) = 0\,.
\ee

%%%%%%%%%%%%%%%%%%%%%%%%%%%%%%%%%%
%Fig. 3 - Vertex splitting
%%%%%%%%%%%%%%%%%%%%%%%%%%%%%%%%%%
%%%%%%%%%%%%%%%%%%%%%%%%%%%%%%%%%%

We emphasize that 
the reason why 
$V_{\alpha\mu\nu}(q,r,p)$ and $V_{\alpha}(q,r,p)$ 
are longitudinally coupled may be directly inferred  
from their special decomposition, shown in \fig{fig:poles}. 
In particular, let us denote by 
$I_\alpha(q)$ 
the transition amplitude 
that connects a gluon with a massless composite scalar, 
depicted as a gray circle in \fig{fig:poles}. 
Since $I_\alpha(q)$ 
depends solely on the momentum $q$, and carries a single Lorentz index, $\alpha$, its general form is given by 
$I_\alpha(q) = q_\alpha I (q)$, where $I (q)$ is a scalar form factor~\cite{Aguilar:2011xe,Ibanez:2012zk}.
This observation accounts directly for the 
form of $V_{\alpha}(q,r,p)$ given in \1eq{eq:Vgen}; 
to deduce the form of  
$V_{\alpha\mu\nu}(q,r,p)$, one must, in addition,  
appeal to Bose symmetry, which   
imposes the structures 
$r_\mu/r^2$ and $p_\nu/p^2$ 
in the remaining two channels.

Returning to the SDE of \1eq{fig:SDEs}, 
the component $V_{\alpha\mu\nu}(q,r,p)$
will enter in it through graphs 
($d_1$) and $(d_4)$, while 
the component $V_{\alpha}(q,r,p)$ 
through graph ($d_3$).
Since $V_{\alpha\mu\nu}(q,r,p)$ has poles 
for each one of its three momenta, 
let us point out that 
only the pole associated with the $q$-channel, \ie the channel that carries
the momentum entering in the gluon propagator, is relevant for the Schwinger mechanism
that will generate mass for $\Delta(q)$.
In fact, in the Landau gauge that we employ, 
the gluon propagators inside the diagrams 
($d_1$) and $(d_4)$
are transverse, leading to a considerable 
reduction in the number of the 
form factors of $V_{\alpha\mu\nu}(q,r,p)$ that 
participate actively, since 
\be
{P}_{\mu'}^{\mu}(r){P}_{\nu'}^{\nu}(p) V_{\alpha\mu\nu}(q,r,p) = \frac{q_\alpha}{q^2} {P}_{\mu'}^{\mu}(r){P}_{\nu'}^{\nu}(p) C_{\mu\nu}(q,r,p)\,.
\label{eq:PPG}
\ee
Consequently, for the ensuing 
analysis, one requires only 
the tensorial decomposition of the component  $C_{\mu\nu}(q,r,p)$ in \1eq{eq:Vgen}, which is given by 
\be 
C_{\mu\nu}(q,r,p) = C_1 \, g_{\mu\nu}  + C_2\, r_\mu r_\nu  + C_3 \, p_\mu p_\nu  +  C_4 \, r_\mu p_\nu  + C_5 \,  p_\mu r_\nu  \,,
\label{eq:Cdec}
\ee
where \mbox{$C_j := C_j(q,r,p)$}. 
Then, the substitution of \1eq{eq:Cdec} into 
\1eq{eq:PPG}, and use of the relation $q+p+r =0$,  
reveals that only two form factors survive inside 
($d_1$) and $(d_4)$, namely 
\be
 {P}_{\mu'}^{\mu}(r){P}_{\nu'}^{\nu}(p) V_{\alpha\mu\nu}(q,r,p) = \frac{q_\alpha}{q^2} {P}_{\mu'}^{\mu}(r){P}_{\nu'}^{\nu}(p)
 \left[ C_1 \, g_{\mu\nu} + C_5 q_\mu q_\nu\right]\,.
\label{eq:PPG2}
\ee

Since the main function of the Schwinger mechanism is to make the gluon propagator saturate at the origin, 
it is important to explore the properties 
of the structures appearing in \1eq{eq:PPG2} near $q=0$. To that end, we expand the r.h.s. of 
\1eq{eq:PPG2}, keeping terms at most linear in $q$. 
After noticing that the term proportional to $C_5$ in
\1eq{eq:PPG2} is of order ${\cal O}(q^2)$, 
we end up with a single relevant form factor
associated with $V_{\alpha\mu\nu}(q,r,p)$,  
namely $C_1(q,r,p)$, which survives the 
$q \to 0$ limit of graphs ($d_1$) and $(d_4)$.
As for $V_\alpha(r,p,q)$, 
its unique component, $\Cgh(q,r,p)$, 
enters directly in ($d_3$).

The continuation of this analysis   
entails the Taylor expansion of $C_1(q,r,p)$ and $\Cgh(r,p,q)$ around $q=0$. In carrying out this 
expansion, one 
employs the following two key relations, 
\be
C_1(0,r,-r) = 0 \,,\qquad\qquad \Cgh(r,-r,0) = 0 \,.
\label{C1_0}
\ee
The first  one follows directly from the Bose symmetry of the three-gluon vertex, which implies that 
$C_1(q,r,p)=-C_1(q,p,r)$; as we will see in Sec.~\ref{sec:widis3g}, it may also be derived in a completely independent way from the fundamental STIs satisfied by the three-gluon vertex.
The justification of the second relation in \1eq{C1_0} is less straightforward;  
its derivation, presented in Appendix~\ref{app:poleBQI}, relies on the BQI~\cite{Binosi:2002ez,Binosi:2009qm} linking the conventional ghost-gluon vertex, $\fatg_\alpha(r,p,q)$, with its background counterpart, $\fatgt_\alpha(r,p,q)$.

Thus, after taking 
\1eq{C1_0} into account, the Taylor expansion of ${C}_1(q,r,p)$ and ${C}(r,p,q)$ around $q=0$ yields
\be
\label{eq:taylor_C}
\lim_{q \to 0} {C}_1(q,r,p) =  2 (q\cdot r) \Cfat(r) \, + \cdots\,,  \qquad
\lim_{q \to 0} {C}(r,p,q) =  2 (q\cdot r)\C(r) \, +   \cdots\,,
\ee
with
\be
\Cfat(r) := \left[ \frac{\partial {C}_1(q,r,p)}{\partial p^2} \right]_{q = 0}\,,\qquad\qquad
\C(r) :=   \left[ \frac{\partial {C}(r,p,q)}{\partial p^2} \right]_{q = 0}\,.
\label{eq:theCs}
\ee
The functions $\Cfat(r)$ and $\C(r)$ are of central 
importance 
for the rest of this review. 
In particular, there are three key points related to them that will be elucidated in detail 
in what follows: 
\begin{enumerate}
\item $\Cfat(r)$ and $\C(r)$ are  
the {\it BS amplitudes} describing the formation of  gluon-gluon
and ghost-antighost {\it colored} composite bound states, respectively, see  Sec.~\ref{sec:CBSE}.

\item The gluon mass is determined by  
certain integrals that involve 
$\Cfat(r)$ and $\C(r)$, given 
explicitly in Sec.~\ref{sec:massgen}.

\item 
$\Cfat(r)$ and $\C(r)$ lead to 
smoking-gun 
displacements of the WIs. 
In fact, the displacement induced by $\Cfat(r)$, 
has been confirmed by lattice QCD, 
by combining judiciously the results of several 
lattice simulations, see subsection~\ref{subsec:widis}. 

\end{enumerate}

We end this section by emphasizing that 
the BFM vertices develop poles in exactly the same way as their conventional counterparts. In particular, the main relations Eqs.~\eqref{fullgh}, \eqref{eq:Vgen}, \eqref{C1_0} and \eqref{eq:theCs} remain valid, with the only modification 
that all quantities carry hats or tildes; these BFM vertices   
will be used extensively in Sec.~\ref{sec:massgen}. Note that the conventional and background vertices, including their pole content, are related through appropriate BQIs, see \eg \2eqs{Cgh_0}{BQI_Cgl_0}.

%%%%%%%%%%%%%%%%%%%%%%%%%%%%%%%%%%%%%%%%%%%%%%%%%%%%%%%%
\section{Dynamical formation of massless poles}\label{sec:CBSE}
%%%%%%%%%%%%%%%%%%%%%%%%%%%%%%%%%%%%%%%%%%%%%%%%%%%%%%%%

One crucial aspect of the implementation 
of the Schwinger mechanism in a Yang-Mills context 
is that the poles that  
comprise the components $V_{\alpha\mu\nu}(q,r,p)$ and $V_{\alpha}(q,r,p)$ in \1eq{eq:Vgen} are {\it not} 
introduced by hand; rather, they are generated 
{\it dynamically}, as massless composite excitations that carry color. 
In fact, this subtle process is controlled by a system of coupled 
linear BSEs for the functions   
$\Cfat(r)$ and $\C(r)$, which play 
the role of the BS amplitudes for generating 
composite massless scalars out of two gluons and a ghost-antighost pair, 
respectively.

%%%%%%%%%%%%%%%%%%%%%%%%%%%%%%%%%%
%Fig. 4 - Coupled vertex SDEs
%%%%%%%%%%%%%%%%%%%%%%%%%%%%%%%%%%
\begin{figure}[t]
\centering
\includegraphics[width=0.95\textwidth]{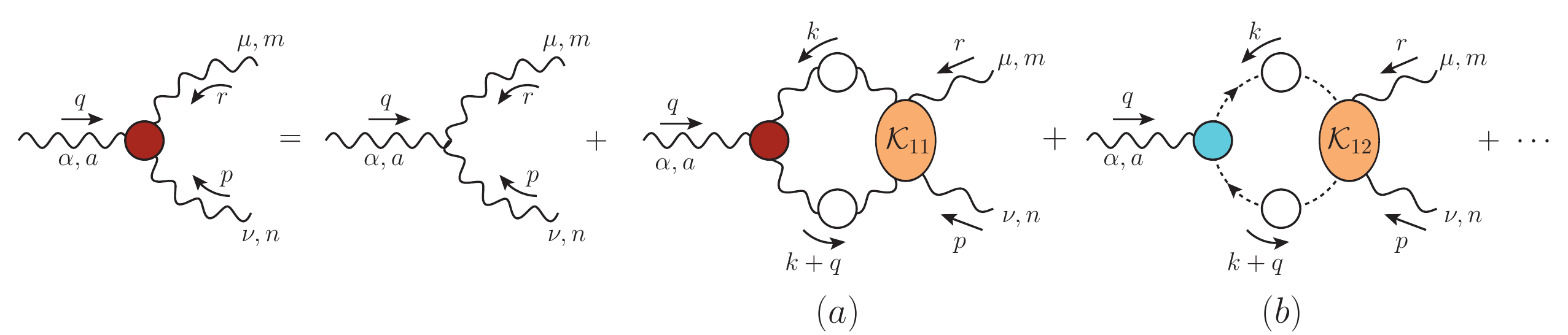}\\
\includegraphics[width=0.95\textwidth]{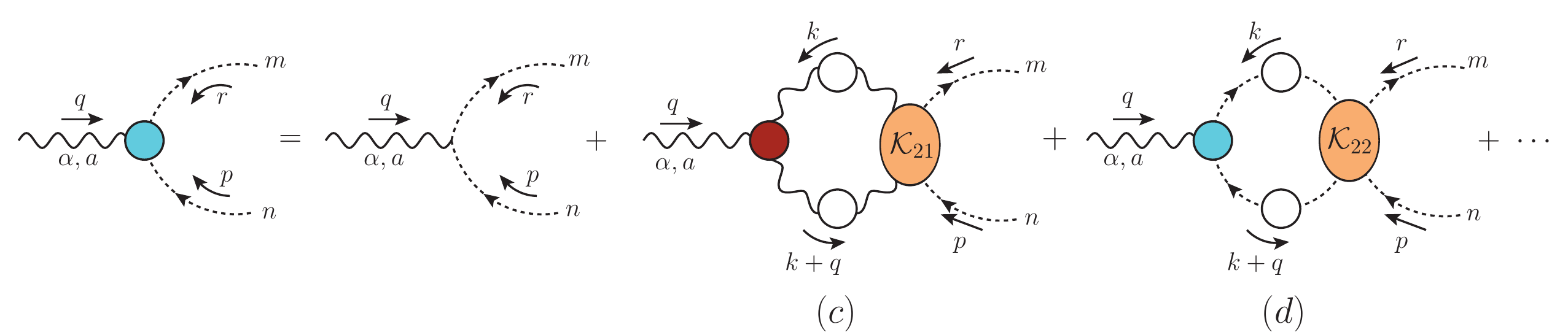}  
\caption{The coupled system of Schwinger-Dyson equations (SDEs) for the three-gluon and ghost-gluon vertices, $\fatg_{\alpha\mu\nu}(q,r,p)$ and $\fatg_\alpha(r,p,q)$, respectively. The orange ellipses represent four-point scattering kernels, denoted by ${\cal K}_{ij}$. We omit diagrams containing five-point scattering kernels.}
\label{fig:vert_SDEs}
\end{figure}
%%%%%%%%%%%%%%%%%%%%%%%%%%%%%%%%%%

The starting point for the 
derivations of the aforementioned BSEs 
are the SDEs 
for $\fatg_{\alpha\mu\nu}(q,r,p)$ and $\fatg_\alpha(r,p,q)$, 
shown diagrammatically in 
\fig{fig:vert_SDEs}, and given by~\cite{Aguilar:2021uwa}
\begin{eqnarray}
\fatg^{\alpha\mu\nu} &=& \gz^{\alpha\mu\nu} - \lambda \int_k \fatg^{\alpha\beta\gamma}\Delta_{\beta\rho} \Delta_{\gamma\sigma} {\cal K}_{11}^{\mu\nu\sigma\rho}
+ 2 \lambda  \int_k \fatg^{\alpha} D D {\cal K}_{12}^{\mu\nu} \,, \nonumber\\
\fatg^{\alpha} &=& \gz^\alpha - \lambda \int_k \fatg^{\alpha\beta\gamma}\Delta_{\beta\rho} \Delta_{\gamma\sigma}{\cal K}_{21}^{\sigma\rho} 
- \lambda \int_k \fatg^{\alpha} D D {\cal K}_{22} \,,
\label{BSE_inhom}
\end{eqnarray}
where 
\be  
\lambda := i g^2 C_{\rm A}/2 \,, \label{lambda_def}
\ee 
and the tree-level expressions for the vertices $\fatg^{\alpha\mu\nu}$ 
and $\fatg^{\alpha}$ are given by 
\be 
\gz^{\alpha\mu\nu}(q,r,p) = (q - r)^\nu g^{\alpha\mu} + (r - p)^\alpha g^{\mu\nu} + (p - q)^\mu g^{\nu\alpha} \,, \qquad\qquad \gz^\alpha(r,p,q) = r^\alpha\,.
\label{bare3g}
\ee
Note that, for compactness, all momentum arguments 
have been suppressed; they may be easily restored 
by appealing to \fig{fig:vert_SDEs}. 

The following steps are subsequently implemented: 

1. Substitute into both sides of \1eq{BSE_inhom} the expressions for the fully-dressed vertices given in \1eq{fullgh}.

2. In order to exploit \1eq{eq:PPG2}, multiply 
the first equation by the factor ${P}_{\mu'\mu}(r){P}^{\mu'}_{\nu}(p)$.

3. Take the limit of the system as $q\to 0$:  this  
activates \1eq{eq:taylor_C} and 
introduces 
the functions ${\mathbb C}(r)$ and ${\mathbb C}(r)$. 

4. Isolate 
the tensorial structures proportional to $q^{\alpha}$,
and match the terms on both sides.

\begin{figure}[t!]
\centering
\includegraphics[width=0.475\textwidth]{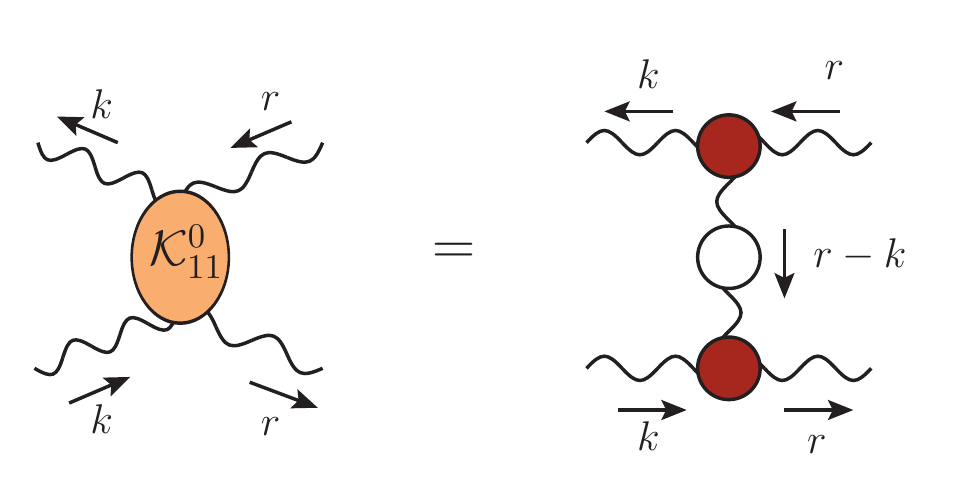} \hfil \includegraphics[width=0.475\textwidth]{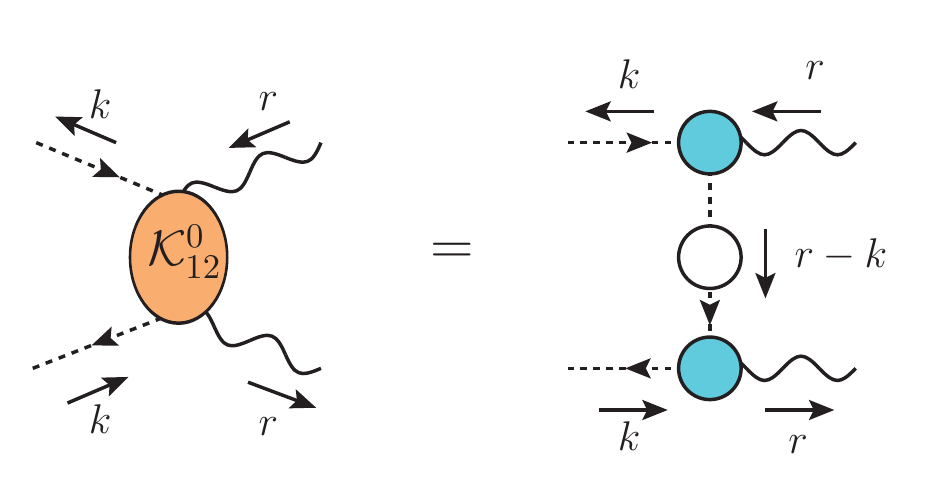} \\
\includegraphics[width=0.475\textwidth]{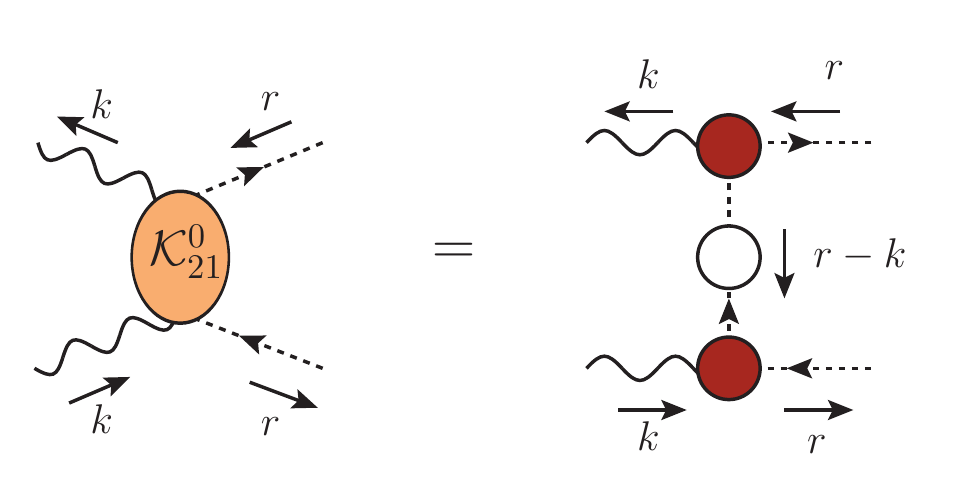} \hfil \includegraphics[width=0.475\textwidth]{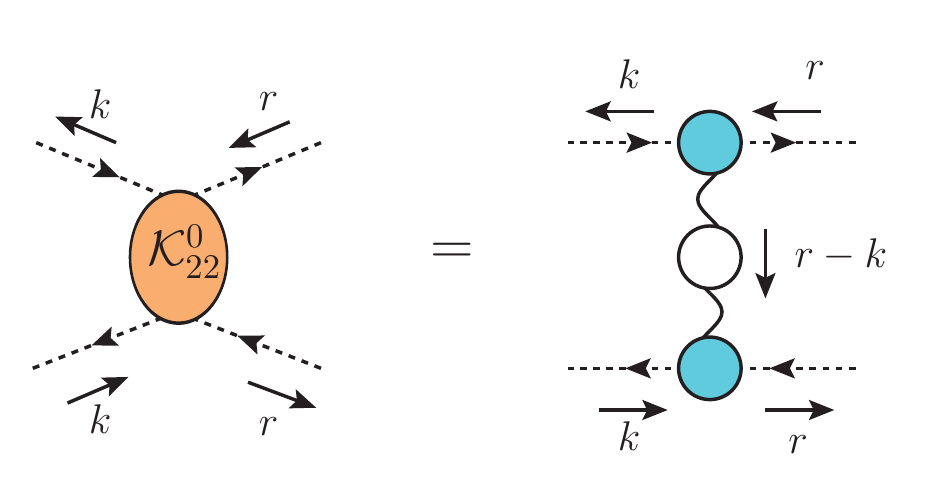}
\caption{The one-particle exchange approximations, ${\cal K}^0_{ij}$, of the kernels ${\cal K}_{ij}$ appearing in \fig{fig:vert_SDEs}.}
\label{fig:kernels}
\end{figure}

5. Employ the ``one-particle exchange'' approximation for the kernels  ${\cal K}_{ij}$, to be denoted by 
 ${\cal K}^0_{ij}$, shown in \fig{fig:kernels}. 

\begin{figure}[t]
\centering
\includegraphics[width=0.95\textwidth]{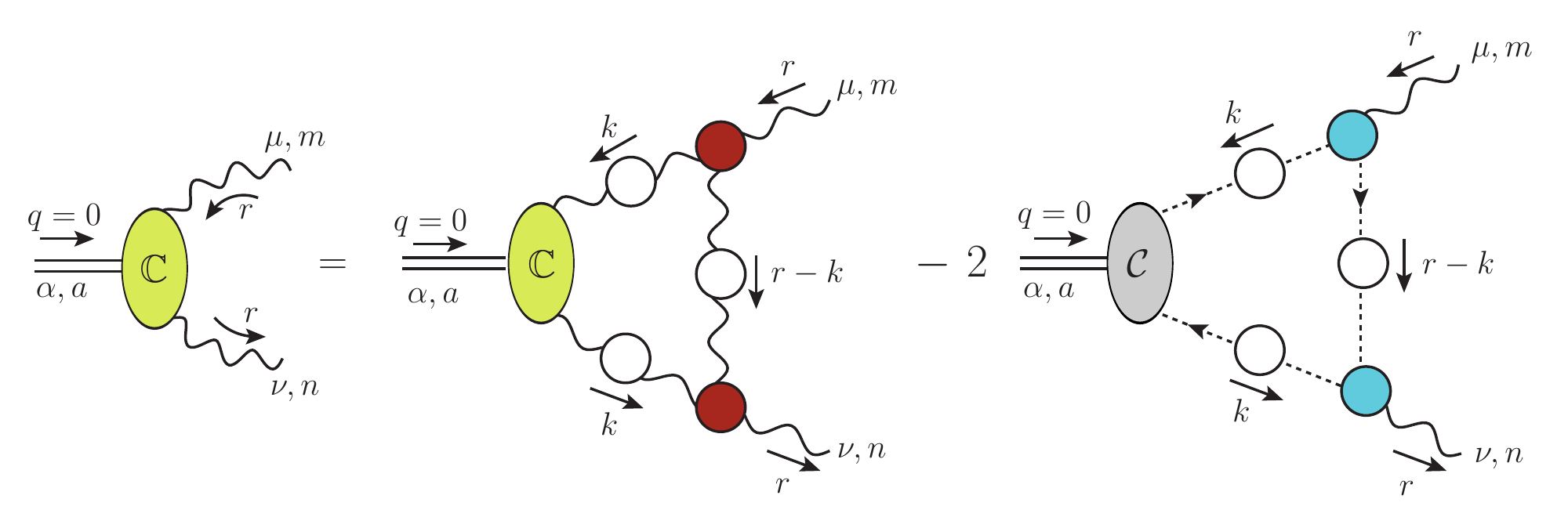}\\
\includegraphics[width=0.95\textwidth]{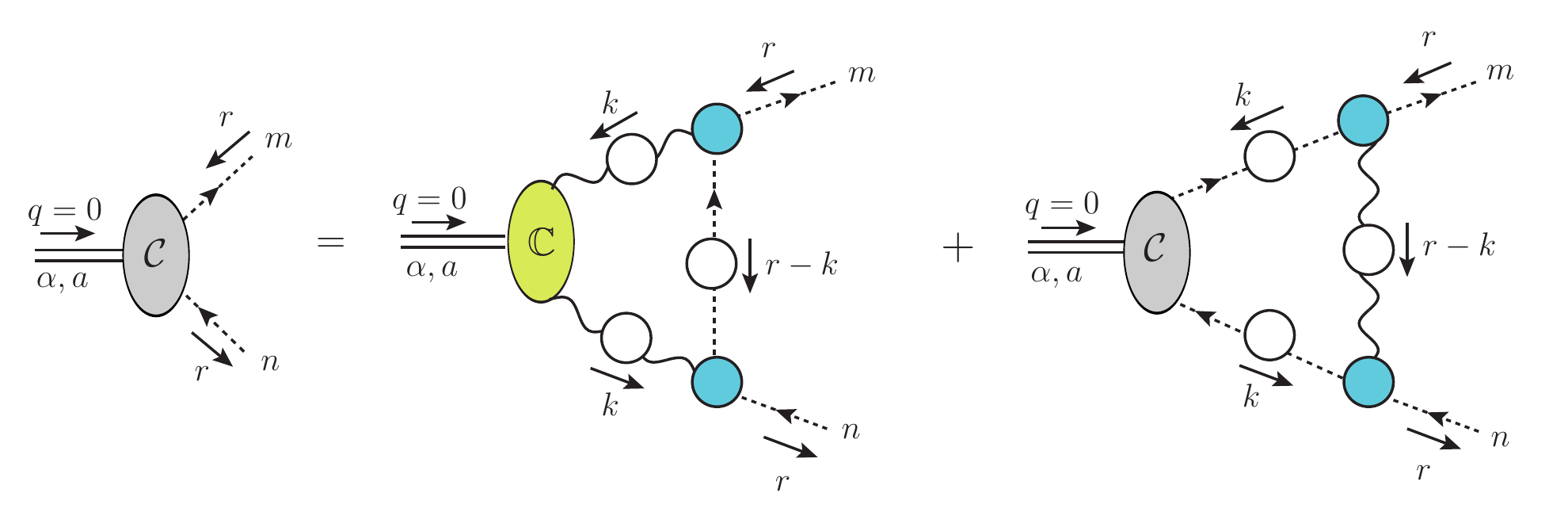}  
\caption{The diagrammatic representation of the coupled system of Bethe-Salpeter equations (BSEs) that governs the evolution of the 
functions ${\mathbb C}(r^2)$ and ${\cal C}(r^2)$. }
\label{fig:BSEs}
\end{figure}

Thus, we arrive at a system of homogeneous equations 
involving $\Cfat(r)$ and $\C(r)$, 
\begin{eqnarray}
\Cfat(r) &=& - \frac{\lambda}{3}\int_k \Cfat(k) \Delta^2(k)P_{\rho\sigma}(k)P_{\mu\nu}(r) \widetilde{\cal K}_{11}^{\mu\nu\sigma\rho} 
+ \frac{2 \lambda}{3} \int_k \C(k) D^2(k)P_{\mu\nu}(r) \widetilde{\cal K}_{12}^{\mu\nu}  \,, \nonumber\\
\C(r) &=& - \lambda \int_k \Cfat(k)\Delta^2(k)P_{\sigma\rho}(k) \widetilde{\cal K}_{21}^{\sigma\rho} 
- \lambda \int_k \C(k)D^2(k) \widetilde{\cal K}_{22} \,,
\label{BSE_hom}
\end{eqnarray}
where $\widetilde{\cal K}_{ij} := (r\cdot k /r^2)\, {\cal K}^0_{ij}(r,-r,k,-k)$;  
the system is diagrammatically depicted in 
\fig{fig:BSEs}.

Before turning to the numerical analysis, the 
BSE system must
be passed to the Euclidean space, following
standard conversion rules. 
In doing so we note 
that the integral measure is modified according to $d^4k \to i d^4k_{\s {\rm E}}$; this extra factor of $i$ combines with the $\lambda$ defined in 
\1eq{lambda_def} to give real expressions.

As announced, 
the system of coupled equations given in \1eq{BSE_hom} 
represents the BSEs that govern the formation of massless colored bound states out of two gluons and a ghost-antighost pair. The
functions $\Cfat(r)$ and $\C(r)$ are the corresponding BS amplitudes; finding nontrivial solutions 
for them, \ie something other than 
$\Cfat(r) = \C(r) =0$ identically,  
is crucial for the implementation of the 
Schwinger mechanism.

%%%%%%%%%%%%%%%%%%%%%%%%%%%%%%%%%%
%Fig. 5 - One particle exchange
%%%%%%%%%%%%%%%%%%%%%%%%%%%%%%%%%%
%%%%%%%%%%%%%%%%%%%%%%%%%%%%%%%%%%

%%%%%%%%%%%%%%%%%%%%%%%%%%%%%%%%%%
%Fig. 6 - Coupled BSEs
%%%%%%%%%%%%%%%%%%%%%%%%%%%%%%%%%%
%%%%%%%%%%%%%%%%%%%%%%%%%%%%%%%%%%

The equations 
in \1eq{BSE_hom} are linear and homogeneous in the unknown functions.
There are two main consequences arising from this fact.
First, the numerical solution of the system will be reduced  
to an eigenvalue problem.  
Second, the overall scale of the solutions is undetermined, since  
the multiplication of a given solution by an arbitrary real constant produces another solution
\footnote{The ambiguity originates from considering only leading terms in the expansion around $q = 0$, and 
may be resolved if further orders in $q$ are kept, see, \eg\mbox{\cite{Nakanishi:1969ph,Maris:1997tm,Blank:2010bp}}.}.

\begin{figure}[t!]
\centering
\includegraphics[width=0.475\textwidth]{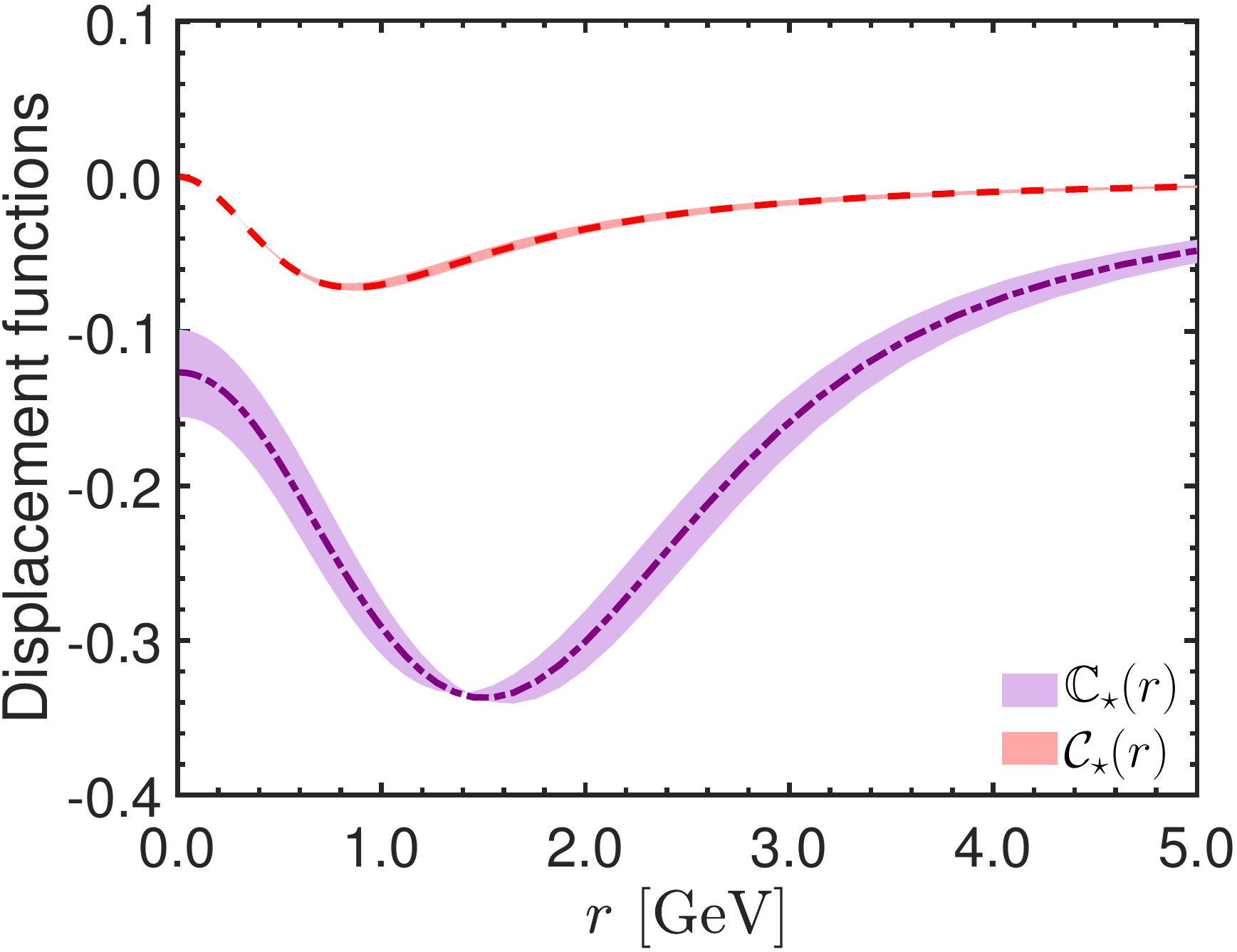}
\caption{
The solutions for $\CB(r)$ (purple dot-dashed) and $\Cc(r)$ (red dashed) obtained from the coupled BSE system of \1eq{BSE_hom}.
}
\label{fig:C_gl_Cgh}
\end{figure}

It turns out that the condition 
for obtaining nontrivial solutions, 
when expressed in terms 
of the strong coupling, 
$\alpha_s:=g^2/4 \pi$, states that 
they exist for 
$\alpha_s=0.63$, when the renormalization point $\mu =4.3$ GeV. 
The solutions obtained when 
$\alpha_s$ acquires this special value 
are shown in  \fig{fig:C_gl_Cgh}; they have undergone 
scale fixing\footnote{
The scale was fixed by requiring the best possible matching with the 
result obtained for $\Cfat(r)$ from the WI displacement, see  Sec.~\ref{sec:wilat}.}, and are denoted by 
$\CB(r)$ and $\Cc(r)$. 
Observe that $\CB(r)$ is significantly larger in magnitude than $\Cc(r)$, 
implying that the
three-gluon vertex accounts for the bulk of the gluon mass, as originally claimed in~\cite{Aguilar:2017dco}.

It is important to compare 
the value of 
$\alpha_s=0.63$, imposed by the 
BSE eigenvalue, 
with the expected value for $\alpha_s$ for the renormalization scheme employed: within the {\it asymmetric} momentum subtraction (MOM) scheme (see Sec.~\ref{sec:ghost_dyn}), we have that 
$\alpha_s=0.27$~\cite{Boucaud:2017obn}. This numerical discrepancy in the values of $\alpha_s$
is clearly an 
artifact of the truncation employed, and 
concretely of the 
approximation of the kernels ${\cal K}_{ij}$ by their one-particle exchange diagrams, ${\cal K}^0_{ij}$.
A preliminary analysis reveals that mild modifications
of the kernels ${\cal K}_{ij}$ lead to 
considerable variations in the 
value of $\alpha_s$, but leave 
the form of the solutions for $\CB(r)$ and $\Cc(r)$ practically unaltered.
This observation suggests that, while a more complete  
knowledge of the BSE kernels is required in order 
to bring $\alpha_s$ closer to its MOM value,
the solutions obtained with the present 
approximations should be considered 
as particularly stable.

%%%%%%%%%%%%%%%%%%%%%%%%%%%%%%%%%%
% Figure 7  - Bound state amplitudes obtained from BSE 
%%%%%%%%%%%%%%%%%%%%%%%%%%%%%%%%%%
%%%%%%%%%%%%%%%%%%%%%%%%%%%%%%%%%%

%%%%%%%%%%%%%%%%%%%%%%%%%%%%%%%%%%%%%%%%%%%%%%%%%%%%%%%%
\section{Generation of the gluon mass}\label{sec:massgen}
%%%%%%%%%%%%%%%%%%%%%%%%%%%%%%%%%%%%%%%%%%%%%%%%%%%%%%%%

%%%%%%%%%%%%

We next demonstrate in detail how the 
presence of the massless poles 
in the vertices that enter 
in the SDE of the gluon propagator 
generate a gluon mass. 

We start by pointing out that, 
since the fundamental STIs of the theory 
remain intact under the action 
of the Schwinger mechanism, 
\2eqs{pitr}{pitr2} remain valid, 
and the mass term 
$m^2 = \Delta^{-1}(0)$ 
will appear in the transverse combination 
$\Delta^{-1}(0)P_{\mu\nu}(q)$.  
However, the determination of 
the mass proportional to $g_{\mu\nu}$ 
exposes an entirely different array of principles compared to the 
corresponding 
computation for the 
$q_\mu q_\nu/q^2$ component. 

The calculation with respect to the 
$q_\mu q_\nu/q^2$ component 
is rather direct; since the massless poles in the vertices are themselves longitudinally coupled, their contribution to the $q_\mu q_\nu/q^2$ component of $\Pi_{\mu\nu}(q)$ is easily worked out, as will be illustrated in Subsec.~\ref{subsec:longcomp}. In contrast, the emergence of a mass 
proportional to $g_{\mu\nu}$ 
is intimately connected with 
a powerful relation, known as \emph{seagull identity}~\cite{Aguilar:2009ke,Aguilar:2016vin}, which in the absence of the Schwinger mechanism would enforce the masslessness of the propagator, as will be discussed in Subsec.~\ref{subsec:widis}. 
In fact, one main conceptual difference 
between the two approaches is that 
in the $g_{\mu\nu}$ case, 
the use of the PT-BFM-based version of the 
SDE given in \1eq{sdebq}  
is crucial for the emergence of the correct result. 

In order to simplify the 
technical aspects of the calculation without compromising its conceptual content, we will 
determine the 
contribution to the gluon mass 
due the pole 
in the ghost-gluon vertex, namely 
$V_\alpha(r,p,q)$ in the case of $\fatg_\alpha(r,p,q)$, 
and  $\widetilde{V}_\alpha(r,p,q)$  in the case of  $\widetilde{\fatg}_\alpha(r,p,q)$.
To that end, we will 
focus on the subset of self-energy graphs 
containing only ghost loops, \ie graph $(d_3)$ 
in the case of $\Pi_{\mu\nu}(q)$,  
and graphs $(a_3)$ and $(a_4)$ 
in the case of $\widetilde {\Pi}_{\mu\nu}(q)$, shown in the upper and lower 
row of \fig{fig:SDEs}, respectively.

%%%%%%%%%%%%%%%%%%%%%%%%%%%%%%%%%%%%%%%%%%%%%%%%%%%%%%%%
\subsection{Gluon mass from the \texorpdfstring{$q_\mu q_\nu$}{qmuqnu} component}\label{subsec:longcomp}
%%%%%%%%%%%%%%%%%%%%%%%%%%%%%%%%%%%%%%%%%%%%%%%%%%%%%%%%

Let us calculate the contribution to the gluon mass stemming from the ghost loop, \ie the diagram $(d_3)$ of \fig{fig:SDEs}, which, for general values of $q$, reads
\be 
(d_3)_{\mu\nu}(q) = g^2C_{\rm A}  \int_k (k+q)_\mu D(k+q)D(k)\fatg_\nu(-k,k+q,-q) \,. \label{d3_gen}
\ee 

To isolate the  $q_\mu q_\nu/q^2$ component of \1eq{d3_gen} at the origin, we first decompose the full vertex $\fatg_\nu(-k,k+q,-q)$ as in \2eqs{fullgh}{eq:Vgen}, and drop directly the pole-free part, since it does not contribute at $q = 0$. Then, denoting by $(d_3^V)_{\mu\nu}(q)$ the contribution of $V_\nu(-k,k+q,-q)$ to $(d_3)_{\mu\nu}(q)$, we obtain
\be 
(d_3^V)_{\mu\nu}(q) = - g^2C_{\rm A}  \frac{q_\nu}{q^2}\int_k (k+q)_\mu D(k+q)D(k) C(-k,k+q,-q)  \,. 
\ee 

Next, a Taylor expansion around $q = 0$, using \2eqs{C1_0}{eq:taylor_C}, yields
\be 
(d_3^V)_{\mu\nu}(q) = - 2g^2C_{\rm A}\frac{q_\nu q^\rho}{q^2}\int_k k_\mu k_\rho D^2(k)\C(k) \,. \label{d3long_step2}
\ee 
Evidently, the integral above can only be proportional to $g_{\mu\rho}$, such that
\be 
(d_3^V)_{\mu\nu}(q) = - \frac{2g^2C_{\rm A}}{d}\left(\frac{q_\mu q_\nu }{q^2}\right)\int_k k^2 D^2(k)\C(k) \,, \label{d3long_step3}
\ee 
where the tensor structure $q_\mu q_\nu/q^2$ is already isolated.

Then, let us denote by $\Delta^{-1}_{\rm gh}(0)$ the contribution to the mass originating in the $q_\mu q_\nu/q^2$ of the ghost loop. Noting that the contribution of $(d_3^V)_{\mu\nu}(q)$ to the propagator is $i$ times the negative of its $q_\mu q_\nu/q^2$ form factor, we obtain that
\be
\Delta^{-1}_{\rm gh}(0) = \frac{4\lambda}{d}  \int_k k^2  D^2(k)\C(k) \,.
\label{mcalt2}
\ee
At this point, we set $d=4$ and renormalize \1eq{mcalt2}. This leads to the appearance of the \emph{finite} renormalization constant of the ghost-gluon vertex, 
$Z_1$.

Next, we express the result in terms of the ghost dressing function $F$, pass to Euclidean space, and employ hyperspherical coordinates, to obtain the final expression
\be
\Delta^{-1}_{\rm gh}(0) = 
\hat\lambda \,Z_1\int_0^\infty\!\! dy \, F^2(y) \,{\cal C}(y) \,,
\label{eq:msm}
\ee
where $\hat\lambda := C_\mathrm{A}\alpha_s/8\pi$.

The derivation 
of the contributions from the diagrams $(d_1)$ and $(d_4)$ 
proceeds in a completely analogous way, but is algebraically 
more involved, see~\cite{Aguilar:2016vin} for details.

It is instructive to consider how the result of 
\1eq{eq:msm} emerges in the context of \1eq{sdebq}.  
To this end, we consider the ghost block $\pt^{(2)}_{\mu\nu}(q)$ of \fig{fig:SDEs}, whose diagrams have the expressions given in \1eq{a3a4};  clearly, only diagram $(a_3)_{\mu\nu}(q)$ can contribute to the $q_\mu q_\nu$ component of $\pt^{(2)}_{\mu\nu}(q)$. 

Then, we decompose $\widetilde\fatg_\alpha(r,p,q)$
in complete analogy with \2eqs{fullgh}{eq:Vgen}, \ie
\be
\fatgt_\alpha(r,p,q) = \widetilde\Gamma_{\alpha}(r,p,q) + \frac{q_\alpha}{q^2}{\widetilde C}(r,p,q) \,,
\label{ghsm}
\ee
and expand the $(a_3)_{\mu\nu}(q)$ of \1eq{a3a4} around $q = 0$, isolating its $q_\mu q_\nu/q^2$ component. These steps eventually lead to 
\be
{\widetilde\Delta}^{-1}_{\rm gh}(0) = \frac{4\lambda}{d} \int_k k^2  D^2(k)\Ctilde(k)\,,
\label{mcalt}
\ee
where $\Ctilde(q)$ is defined in the exact same way as  $\C(q)$, namely  through \1eq{eq:theCs} but with tildes over all relevant quantities.
It is now easy to establish that \1eq{mcalt} is completely equivalent to  \1eq{mcalt2}, simply 
by multiplying both of its sides by $Z_1 F(0)$,  
and then using \1eq{BQI_Cgh}  on the r.h.s. and 
\2eqs{sdebq}{F0_G0} on the l.h.s.

Hence, when the mass is computed through the $q_\mu q_\nu/q^2$ component of the self-energy, the contributions originating from the ghost diagrams of either the BQ or the QQ propagator furnish the same result. The same is not true 
for the calculation through the $g_{\mu\nu}$ component, since the ghost diagram $(d_3)_{\mu\nu}$ of the QQ propagator is not by itself transverse, and a meaningful analysis is preferably  carried out within the BFM.

%%%%%%%%%%%%%%%%%%%%%%%%%%%%%%%%%%%%%%%%%%%%%%%%%%%%%%%%
\subsection{Gluon mass from the \texorpdfstring{$g_{\mu\nu}$}{gmunu} component:
seagull identity and 
Ward identity displacement}\label{subsec:widis}
%%%%%%%%%%%%%%%%%%%%%%%%%%%%%%%%%%%%%%%%%%%%%%%%%%%%%%%%

The fact that the activation of the Schwinger mechanism 
is crucial for the self-consistent generation of a 
gluon mass may be best appreciated 
in conjunction with the so-called 
{\it seagull identity}~\cite{Aguilar:2009ke,Aguilar:2016vin}. The content of this identity is that 
\be 
\int_k k^2\frac{\partial f(k)}{\partial k^2} + \frac{d}{2}\int_k \,f(k) = 0 \,,
\label{sea}
\ee
for functions $f(k)$ that satisfy Wilson's criterion~\cite{Wilson:1972cf}; the cases of physical interest are \mbox{$f(k) = \Delta(k), D(k)$}. 
The general demonstration of the validity of 
\1eq{sea} has been given in~\cite{Aguilar:2016vin}; 
for a detailed discussion of how \1eq{sea} 
prevents the photon from acquiring a mass 
in scalar electrodynamics, see~\cite{Aguilar:2015bud}.

What is so special about 
\1eq{sea} is that, 
within 
the PT-BFM formalism, the l.h.s. of 
\1eq{sea} coincides with the contributions of loop 
diagrams to the $g_{\mu\nu}$ component of the gluon mass. 
Therefore, \1eq{sea} enforces the nonperturbative
masslessness of the gluon 
in the absence of the Schwinger mechanism: even if a massive gluon propagator 
(made ``massive'' through a procedure 
other than the Schwinger mechanism) 
were to be substituted 
inside \1eq{sea}, one would obtain  
zero as contribution to the gluon mass! 
For example, the simple choice $f= (k^2-m^2)^{-1}$,   
reduces the l.h.s of \1eq{sea} to (dimensionally regularized) text-book integrals, which add up to give 
precisely zero~\cite{Aguilar:2015bud}.

In order to appreciate in some detail how the seagull identity prevents the 
$g_{\mu\nu}$ component of the propagator from acquiring a mass  
in the absence of the Schwinger mechanism, 
let us consider once again the ghost block $\pt^{(2)}_{\mu\nu}(q)$ of \fig{fig:SDEs}; 
now both graphs, ($a_3$) and ($a_4$), contribute to the $g_{\mu\nu}$ component. 

Let us assume that the Schwinger mechanism is turned off; 
at the level of the Bcc vertex this means that 
${\widehat V}_\alpha(r,p,q)$ vanishes identically, and  
\mbox{$\fatgt_\alpha(r,p,q) = \gt_\alpha(r,p,q)$}.  
Consequently, $\gt_\alpha(r,p,q)$ saturates 
the STI of \1eq{st2}, 
\be 
q^\alpha \gt_\alpha(r,p,q) = {D}^{-1}(p) - {D}^{-1}(r)\,.
\label{STI1}
\ee
Since the form-factors of the vertex $\gt_\alpha(r,p,q)$ 
do not contain any poles, the derivation from \1eq{STI1}
of the corresponding WI  proceeds in the standard text-book way: 
both sides of \1eq{STI1} undergo a Taylor expansion around $q = 0$, and 
terms at most linear in $q$ are retained. 
Thus, one arrives at the simple QED-like WI 
\be
{\widetilde \Gamma}_\alpha(r,-r,0) = \, \frac{\partial {D}^{-1}(r)}{\partial r^\alpha} \quad \Longrightarrow \quad D^2(r) {\widetilde \Gamma}_\alpha(r,-r,0) = - 2 r_\alpha \frac{\partial {D}(r)}{\partial r^2} \,.
\label{WInopole}
\ee

We now compute the $g_{\mu\nu}$ component of $\pt^{(2)}_{\mu\nu}(q)$ at $q=0$, 
or, equivalently,  ${\widetilde\Delta}^{-1}_{\rm gh}(0)$. 
From \1eq{a3a4}, we see that $(a_4)_{\mu\nu}$ is proportional to $g_{\mu\nu}$ in its entirety. 
On the other hand, $(a_3)_{\mu\nu}(q)$ contains both
$g_{\mu\nu}$ and $q_{\mu}q_{\nu}$ components; however, the latter vanishes in the limit $q\to 0$ if the vertex is pole-free. Then, it is straightforward to show that, as $q\to0$,  
\be
{\widetilde\Delta}^{-1}_{\rm gh}(0)  = \frac{2\lambda}{d} \left[ \int_k   k_\mu D^2(k)\widetilde{\Gamma}^\mu(-k,k,0) + d \int_k D(k) \right] \,.
\label{Pi2}
\ee

At this point, employing the WI of \1eq{WInopole} (with $r \to -k$),  we get
\be
{\widetilde\Delta}^{-1}_{\rm gh}(0) = \frac{4\lambda }{d}\underbrace{\left[ \int_k k^2 \frac{\partial {D}^{-1}(k)}{\partial k^2} + \frac{d}{2}\int_k  D(k) \right]}_{\rm seagull\,\, identity} = 0 \,.
\label{Pi2_0}
\ee
Hence, the WI satisfied by the vertex in the absence of the Schwinger mechanism triggers the seagull identity, which, in turn, 
enforces the masslessness of the propagator.

When the Schwinger mechanism gets activated, 
the STIs satisfied by the 
vertices of the theory retain their original form, but are 
resolved through the nontrivial participation of the terms containing the massless poles~\mbox{\cite{Eichten:1974et,Poggio:1974qs,Smit:1974je,Cornwall:1981zr,Papavassiliou:1989zd,Aguilar:2008xm,Binosi:2012sj,Aguilar:2016vin}}.
In particular, the full vertex $\fatgt_\alpha(r,p,q)$ satisfies precisely \1eq{st2}, namely
\begin{eqnarray}
q^\alpha \fatgt_\alpha(r,p,q) &=& q^\alpha \widetilde\Gamma_{\alpha}(r,p,q) + \widetilde{C}(r,p,q)
\nonumber\\
 &=& {D}^{-1}(p) - {D}^{-1}(r)\,.
\label{STI1sm}
\end{eqnarray}
Notice in particular that the contraction of $\fatgt_\alpha(r,p,q)$ by $q^\alpha$
cancels the massless pole in $q^2$, 
leading to a completely pole-free result.
Therefore, the WI obeyed by $\widetilde\Gamma_{\alpha}(r,p,q)$ may be derived as before, through a standard  Taylor expansion, leading to  
\be
q^\alpha \widetilde\Gamma_{\alpha}(r,-r,0) = - \widetilde{C}(r,-r,0) + q^\alpha \left\{ \frac{\partial {D}^{-1}(r)}{\partial r^\alpha}
- \left[\frac{\partial \widetilde{C}(r,p,q)}{\partial q^\alpha}\right]_{q=0}\right\}  \,.
\label{lhssm}
\ee
Evidently, the unique zeroth-order contribution appearing in \1eq{lhssm},
namely $\widetilde{C}(r,-r,0)$, must vanish, 
\be
\widetilde{C}(r,-r,0) = 0 \,.
\label{Cant}
\ee
Note that this particular 
property may be independently derived 
from the antisymmetry of  $\widetilde{C}(r,p,q)$ under $r \leftrightarrow p$, 
$\widetilde{C}(r,p,q) = - \widetilde{C}(p,r,q)$, which is a consequence imposed by the ghost-antighost symmetry of the
$B(q){\bar c}(r) c(p)$ vertex. The above result, together with \1eq{Cgh_0}, is used to prove \1eq{C1_0} in App.~\ref{app:poleBQI}.

Thus, \1eq{lhssm} becomes
\be
q^\alpha \widetilde\Gamma_{\alpha}(r,-r,0) = q^\alpha \left\{ \frac{\partial {D}^{-1}(r)}{\partial r^\alpha}
- 2 r_\alpha \Ctilde(r) \right\}  \,, \qquad \Ctilde(r) :=   \left[ \frac{\partial {\widetilde C}(r,p,q)}{\partial p^2} \right]_{q = 0}\,,
\label{lhssm2}
\ee
and the matching of the terms linear in $q$ yields the WI 
\be
\widetilde\Gamma_{\alpha}(r,-r,0)  = \frac{\partial {D}^{-1}(r)}{\partial r^\alpha} - \underbrace{2 r_\alpha\,\Ctilde(r)}_{\rm WI\, displacement}\,.
\label{WIdis}
\ee
Comparing \2eqs{WInopole}{WIdis}, it becomes clear that 
the Schwinger mechanism induces a characteristic displacement to the WIs 
that are 
satisfied by the pole-free parts of the vertices~\cite{Aguilar:2016vin}.

%%%%%%%%%%%%%%%%%%%%%%%%%%%%%%%%%%%%%%%%%%%

Returning to \1eq{Pi2}, but now substituting in it the displaced version of \1eq{WInopole}, namely 
\be 
D^2(k) \widetilde{\Gamma}^\mu(-k,k,0) = 2 k^\mu \left[\frac{\partial D(k)}{\partial k^2} + D^2(k) \widetilde{\cal C}(k)\right]\,.
\label{derdis}
\ee
When \1eq{derdis} is substituted into  \1eq{Pi2}, the first term of its r.h.s. triggers the seagull identity and vanishes, exactly as before; 
however, the second term survives, furnishing precisely the result given in 
\1eq{mcalt}. 

Completely analogous procedures may be applied to 
the remaining two blocks, $\pt^{(1)}_{\mu\nu}(q)$ and $\pt^{(3)}_{\mu\nu}(q)$,
by exploiting the Abelian STIs of \2eqs{st1}{st3}, respectively~\cite{Binosi:2012sj}.

%%%%%%%%%%%%%%%%%%%%%%%%%%%%%%%%%%%%%%%%%%%%%%%%%%%%%%%%
\section{Renormalization group invariant interaction strength }\label{sec:dhat}
%%%%%%%%%%%%%%%%%%%%%%%%%%%%%%%%%%%%%%%%%%%%%%%%%%%%%%%%

%%%%%%%%%%%%%%%%%%%%%%%%%%%%%%%%%%
The PT-BFM formalism  
provides the natural framework 
for the construction of 
the RGI version of the naive 
one-gluon exchange interaction. 

To fix the ideas, 
recall that 
in QED, the one-photon exchange interaction, 
defined as $\alpha \Delta_{\rm A}(q)$, 
where $\alpha :=e^2/4\pi$ is the 
hyper-fine structure constant and 
$\Delta_{\rm A}(q)$ the photon propagator, 
is an RGI combination, by virtue of 
the relation $Z_e = Z_A^{-1/2}$;  
see comments following \1eq{Zg_ZA}.
Moreover, this particular combination 
is universal (process-independent) because 
it may be identified within any two-to-two 
scattering process, regardless of the nature 
of the initial and final states 
(electrons, muons, taus, etc).
Instead, in QCD, the corresponding combination 
$\alpha_s \Delta(q)$ is (trivially) universal but not RGI. 
When the vertices that connect the gluon to 
the external particles are ``dressed'' ($\Gamma_0\to\Gamma$), 
the combination $\Gamma \, \alpha_s \Delta \, \Gamma$ \, 
becomes RGI;  however, it is no longer 
process-independent, because the vertices $\Gamma$ 
contain information on the characteristics of the 
external particles, \eg the $\Gamma$ is not 
the same if the external particles are quarks 
or gluons. This apparent conundrum 
may be resolved by resorting to the PT, 
which reconciles harmoniously 
the notions of RGI and process-independence.

\begin{figure}[h]
\centering
\includegraphics[scale = 0.6] {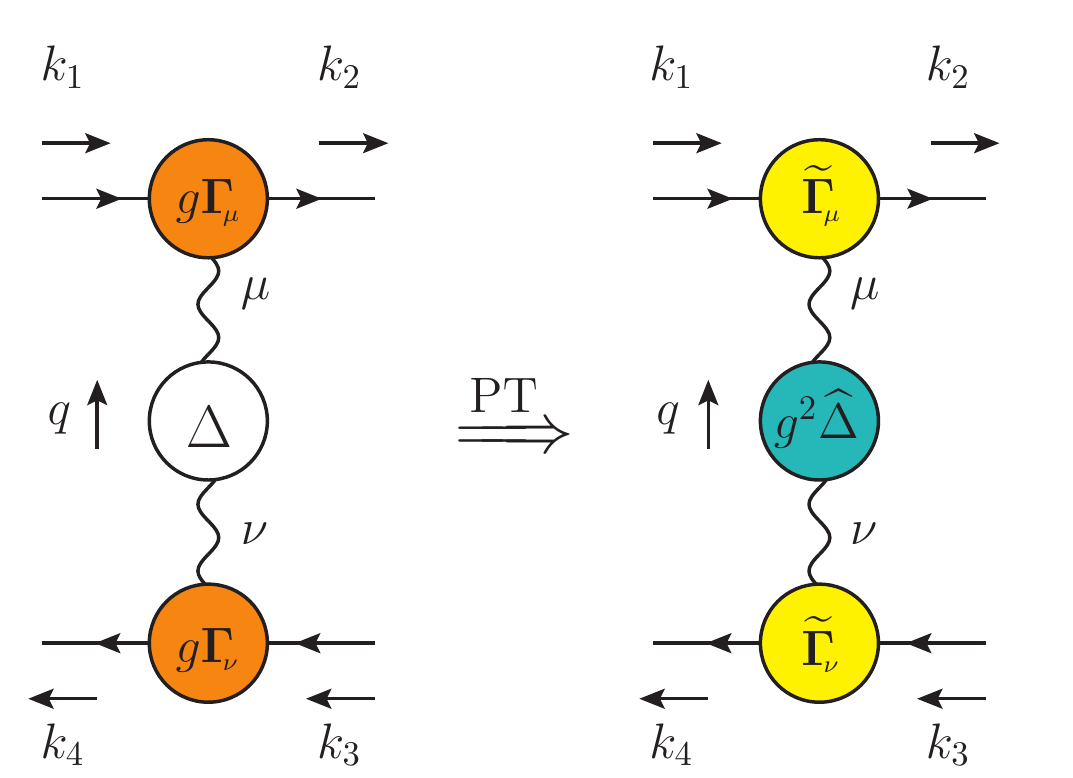} \hspace{0.4cm} \includegraphics[scale = 0.6 ]{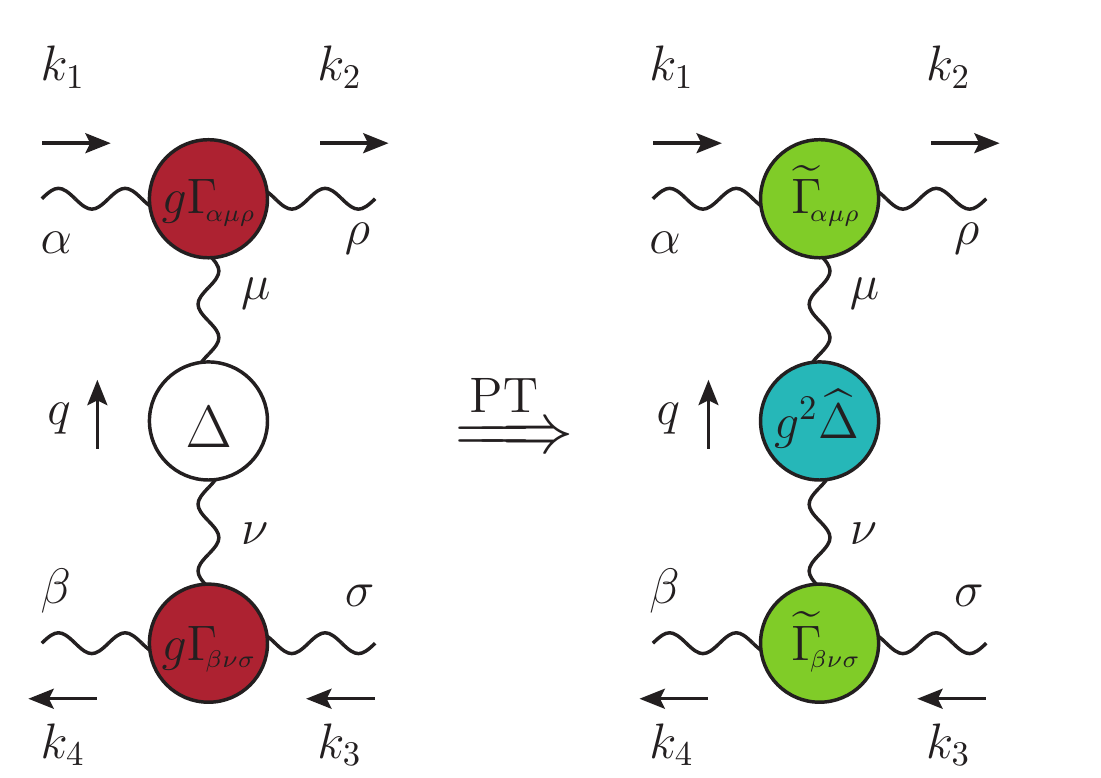}
\caption{ Diagrammatic representation of 
the basic PT rearrangement in the case of 
quark-antiquark scattering, 
corresponding to the S-matrix element ${\cal T}_{q{\bar q}\to q{\bar q}}$ of \1eq{qq_scat} \, 
(left),
and gluon-gluon scattering, corresponding to ${\cal T}_{gg\to gg}$ of \1eq{gg_scat} \, (right).
}
\label{fig:one_gluon_PT}
\end{figure}

Within the PT framework, 
the starting point of the 
construction are ``on-shell'' processes~\cite{Cornwall:1981zr,Cornwall:1989gv,Pilaftsis:1996fh,Binosi:2002ft,Binosi:2009qm},
such as those depicted in \fig{fig:one_gluon_PT}.
The fundamental observation 
is that the dressed vertices appearing there 
contain propagator-like
contributions, which may be unambiguously identified  
by means of a well-defined diagrammatic procedure.
After discarding terms that vanish on shell, 
the contributions extracted from a vertex 
have a two-fold effect: ({\it i}) the genuine vertex contributions left behind form a new vertex, $\widetilde \Gamma$, which satisfies Abelian STIs,  and 
({\it ii}) when the propagator-like 
pieces from both vertices are allotted 
to the conventional propagator, $\Delta_{\mu\nu}(q)$,  the 
resulting effective propagator,  $\widehat\Delta_{\mu\nu}(q)$, captures
all RG logarithms associated with the running of the coupling; for example, 
at one loop and for large $q^2$, one has 
\be 
\widehat\Delta^{-1}(q) \approx q^2\left[ 1 + b g^2 \ln(q^2/\mu^2) \right] \,,
\label{deltarc}
\ee 
where $b= 11 C_A/48 \pi^2 $ is the 
first coefficient of the Yang-Mills 
$\beta$ function. We emphasize that the PT construction goes 
through to all orders in perturbation theory,
as well as nonperturbatively, and all key 
properties of the PT Green's function persist 
unaltered~\cite{Binosi:2002ft,Binosi:2003rr}.  

The correspondence between the PT and the BFM may be summarized by  
stating that the PT rearrangement outlined above 
amounts effectively to replacing the Q-type gluon that 
is being exchanged (carrying momentum $q$) 
by a B-type gluon~\cite{Denner:1994nn,Hashimoto:1994ct,Papavassiliou:1994yi,Pilaftsis:1996fh}; external (on-shell) fields are always of the Q-type.
Thus, the notation used above for the PT effective 
Green's functions (``tildes'' and ``hats'') corresponds precisely to 
the BFM notation introduced in Section~\ref{sec:prel}. 
Note that the formal expression of all PT rearrangements implemented diagrammatically are 
the BQIs that relate conventional Green's functions to their BFM counterparts~\cite{Binosi:2009qm}. 
For example, 
in the case of the quark-gluon vertex, 
we have that  
the vertices ${\mathbf \Gamma}_\mu(q,k_1,-k_2)$ [with external fields $Q^a_\mu(q)q^b(k_1){\bar q}^c(-k_2)$] and ${\widetilde {\mathbf \Gamma}}_\mu(q,k_1,-k_2)$ [$B^a_\mu(q)q^b(k_1){\bar q}^c(-k_2)$] are related by the BQI~\cite{Aguilar:2014lha}
\be
{\widetilde {\mathbf \Gamma}}_\mu(q,k_1,-k_2) = [ 1 + G(q) ]  {\mathbf \Gamma}_\mu(q,k_1,-k_2) 
+ \cdots \,, 
\label{quark_BQI}
\ee
where the ellipsis denotes terms that vanish on shell. Similarly, the BQI of \1eq{BQI_BQQ}, when evaluated on-shell, yields a completely analogous result, to wit, 

\be 
\fatgt_{\mu\alpha\rho}(q,k_1,-k_2) = 
[ 1 + G(q)]  \fatg_{\mu\alpha\rho}(q,k_1,-k_2) 
+ \cdots \,.
\label{gluon_BQI}
\ee
It is now clear how the PT 
gives rise to 
a process-independent 
propagator-like component: regardless of 
the process (\ie the type of vertex 
connecting the internal gluon to the 
external states), each vertex contributes 
to the conventional $\Delta(q)$
a factor of $[ 1 + G(q)]^{-1}$, finally 
leading to the BQI of \1eq{propBQI}~\cite{Binosi:2014aea}. 

The culmination of the above sequence of ideas 
is reached by noting that, by virtue of \1eq{Zg_ZA},  the combination 
\be 
{\widehat d}(q) := \alpha_s{\widehat \Delta(q)} = \frac{\alpha_s\Delta(q)}{[1 + G(q)]^2} \,, \label{dhat}
\ee 
is RGI: 
it retains exactly the same form 
before and after renormalization, and, consequently, does 
not depend on the renormalization point $\mu$~\cite{Cornwall:1981zr}. 
The quantity ${\widehat d}(q)$
has mass dimension of $-2$, 
and is 
known in the literature as the ``RGI running interaction strength''~\cite{Binosi:2014aea}.

%%%%%%%%%%%%%%%%%%%%%%%%%%%%%%%%%%
% Figure 8  - One gluon exchange PT
%%%%%%%%%%%%%%%%%%%%%%%%%%%%%%%%%%
%%%%%%%%%%%%%%%%%%%%%%%%%%%%%%%%%%

The steps leading to the natural appearance of 
${\widehat d}(q)$ within any given process 
may be summarized 
in the case of quark-antiquark, or gluon-gluon 
scattering. 

Consider the S-matrix elements ${\cal T}_{q{\bar q}\to q{\bar q}}$, for the scattering of a quark and an antiquark, and ${\cal T}_{gg\to gg}$, for the scattering of two gluons. The quark-antiquark scattering is depicted in the left panel of \fig{fig:one_gluon_PT}.  Using the BQI of \1eq{propBQI} we obtain
\begin{align} 
{\cal T}_{q{\bar q}\to q{\bar q}} =& \,\, \left[ g {\mathbf\Gamma}_\mu(q,k_1,-k_2) \right] \Delta(q) P^{\mu\nu}(q) \left[ g {\mathbf \Gamma}_\nu(-q,k_3,-k_4) \right] \nonumber\\
\overset{{\rm PT}}
{=}& \,\, \left\lbrace g [1 + G(q)]^{-1} {\widetilde {\mathbf\Gamma}}_\mu(q,k_1,-k_2) \right\rbrace \Delta(q) P^{\mu\nu}(q) \left\lbrace g [1 + G(q)]^{-1}{\widetilde {\mathbf \Gamma}}_\nu(-q,k_3,-k_4) \right\rbrace \nonumber\\
\overset{{\rm PT}}
{=}& \,\,  {\widetilde {\mathbf\Gamma}}_\mu(q,k_1,-k_2)  \left\lbrace g^2 [1 + G(q)]^{-2}\Delta(q) \right\rbrace P^{\mu\nu}(q) {\widetilde {\mathbf \Gamma}}_\nu(-q,k_3,-k_4) \nonumber\\
\overset{{\rm PT}}
{=} & \,\, {\widetilde {\mathbf \Gamma}}_\mu(q,k_1,-k_2) \underbrace{\left[ g^2 {\widehat \Delta}(q) \right]}_{4\pi{\widehat d}(q)} P^{\mu\nu}(q) {\widetilde {\mathbf \Gamma}}_\nu(-q,k_3,-k_4) \,, \label{qq_scat}
\end{align}
where we omit color structures.

Similarly, the scattering of two gluons, depicted in the right panel of \fig{fig:one_gluon_PT}, yields
\begin{align}
{\cal T}_{gg\to gg} =& \,\, \left[ g \g_{\alpha\mu\rho}(k_1,q,-k_2) \right] \Delta(q) P^{\mu\nu}(q) \left[ g \g_{\beta\nu\sigma}(k_3,-q,-k_4)\right] \nonumber\\
\overset{{\rm PT}}{=} & \,\,  \left\lbrace g [ 1 + G(q)]^{-1}{\widetilde \g}_{\alpha\mu\rho}(k_1,q,-k_2) \right\rbrace \Delta(q) P^{\mu\nu}(q) \left\lbrace g [ 1 + G(q)]^{-1}{\widetilde \g}_{\beta\nu\sigma}(k_3,-q,-k_4)\right\rbrace \nonumber\\
\overset{{\rm PT}}{=} & \,\, {\widetilde \g}_{\alpha\mu\rho}(k_1,q,-k_2) \left\lbrace g^2 [ 1 + G(q) ]^{-2}\Delta(q) \right\rbrace P^{\mu\nu}(q) {\widetilde \g}_{\beta\nu\sigma}(k_3,-q,-k_4)\nonumber\\
\overset{{\rm PT}}{=} & \,\, {\widetilde \g}_{\alpha\mu\rho}(k_1,q,-k_2) \underbrace{\left[ g^2 {\widehat \Delta}(q) \right]}_{4\pi{\widehat d}(k)} P^{\mu\nu}(q) {\widetilde \g}_{\beta\nu\sigma}(k_3,-q,-k_4) \,. \label{gg_scat}
\end{align}
Evidently, the same ${\widehat d}(q)$, defined in \1eq{dhat}, appears naturally 
in both \2eqs{qq_scat}{gg_scat}: it is, in that sense, a process-independent RGI 
interaction capturing faithfully the 
one-gluon exchange dynamics~\cite{Cornwall:1981zr,Watson:1996fg,Binosi:2002vk,Aguilar:2009nf,Binosi:2014aea,Binosi:2016nme,Cui:2019dwv,Roberts:2020hiw}. 

\begin{figure}[h]
\centering
\includegraphics[width=0.475\textwidth]{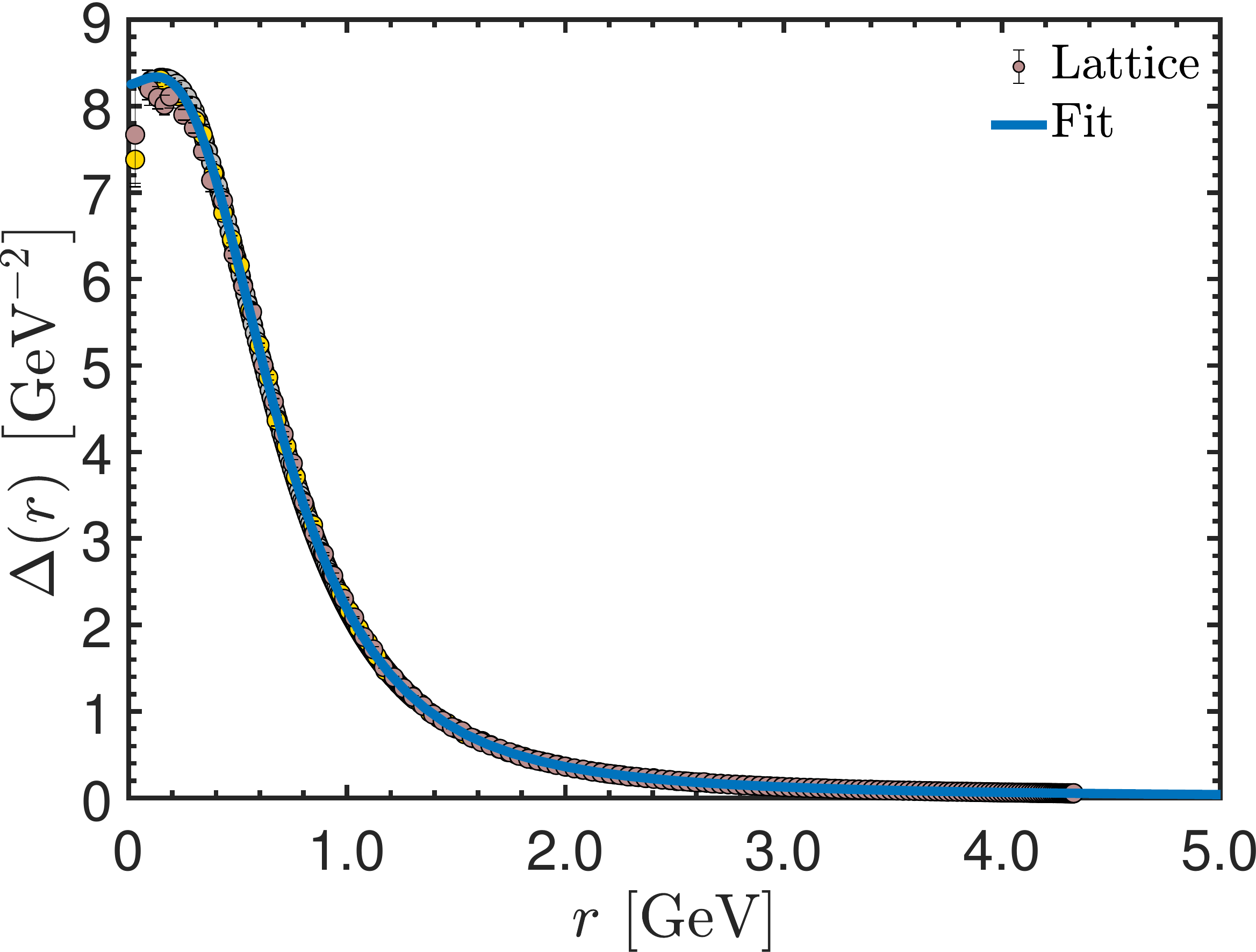} \hfil \includegraphics[width=0.475\textwidth]{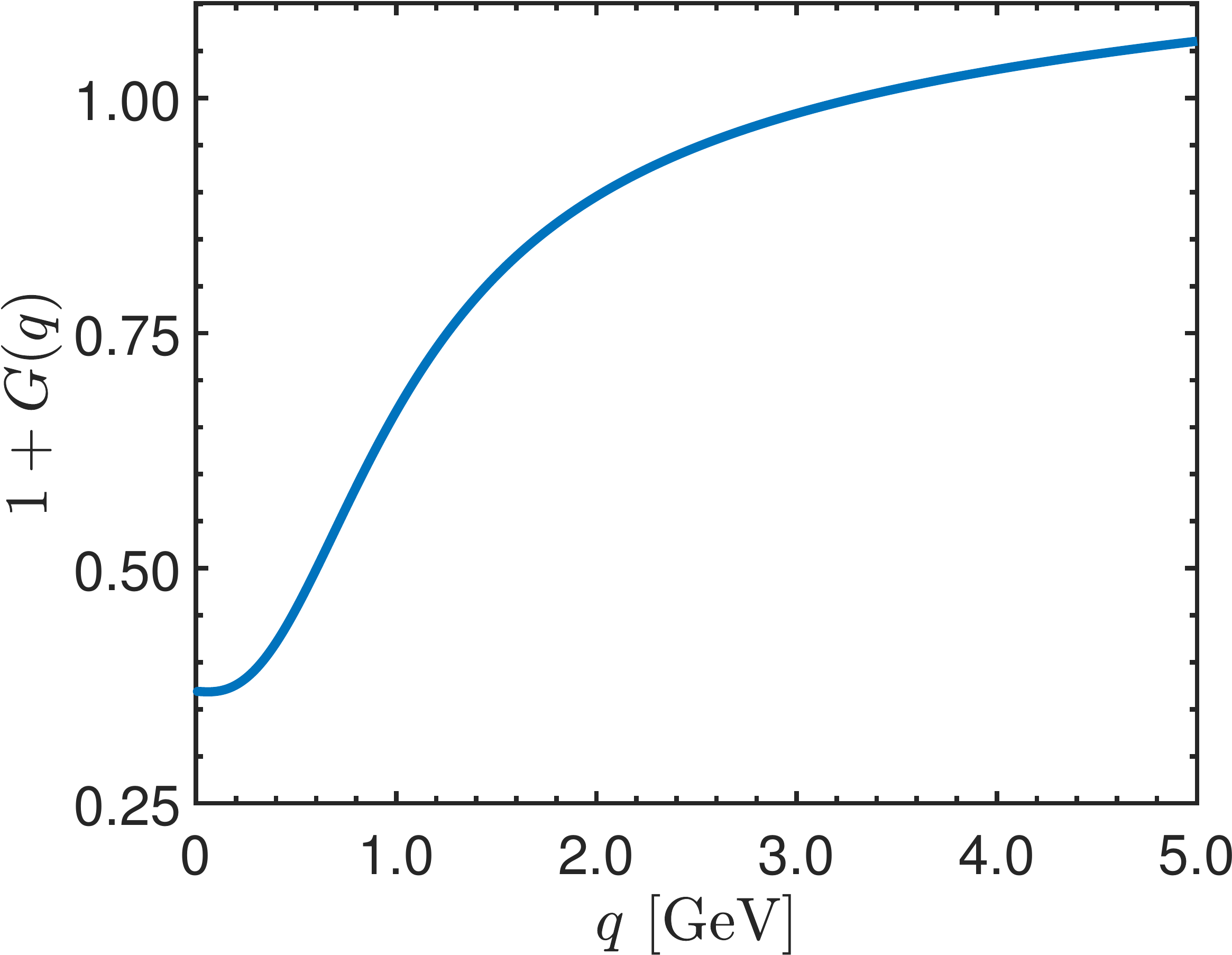} \\
\includegraphics[width=0.475\textwidth]{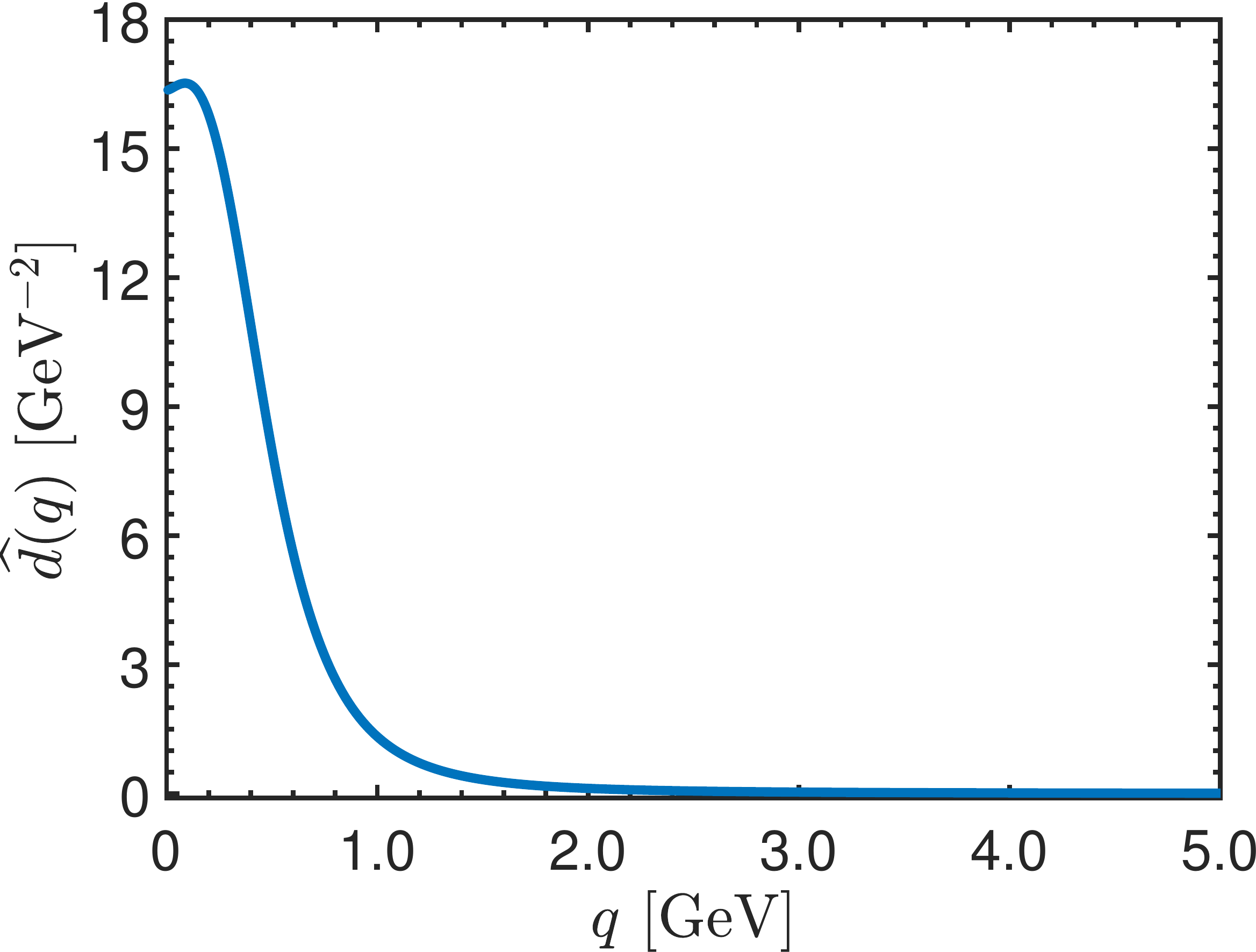} \hfil \includegraphics[width=0.475\textwidth]{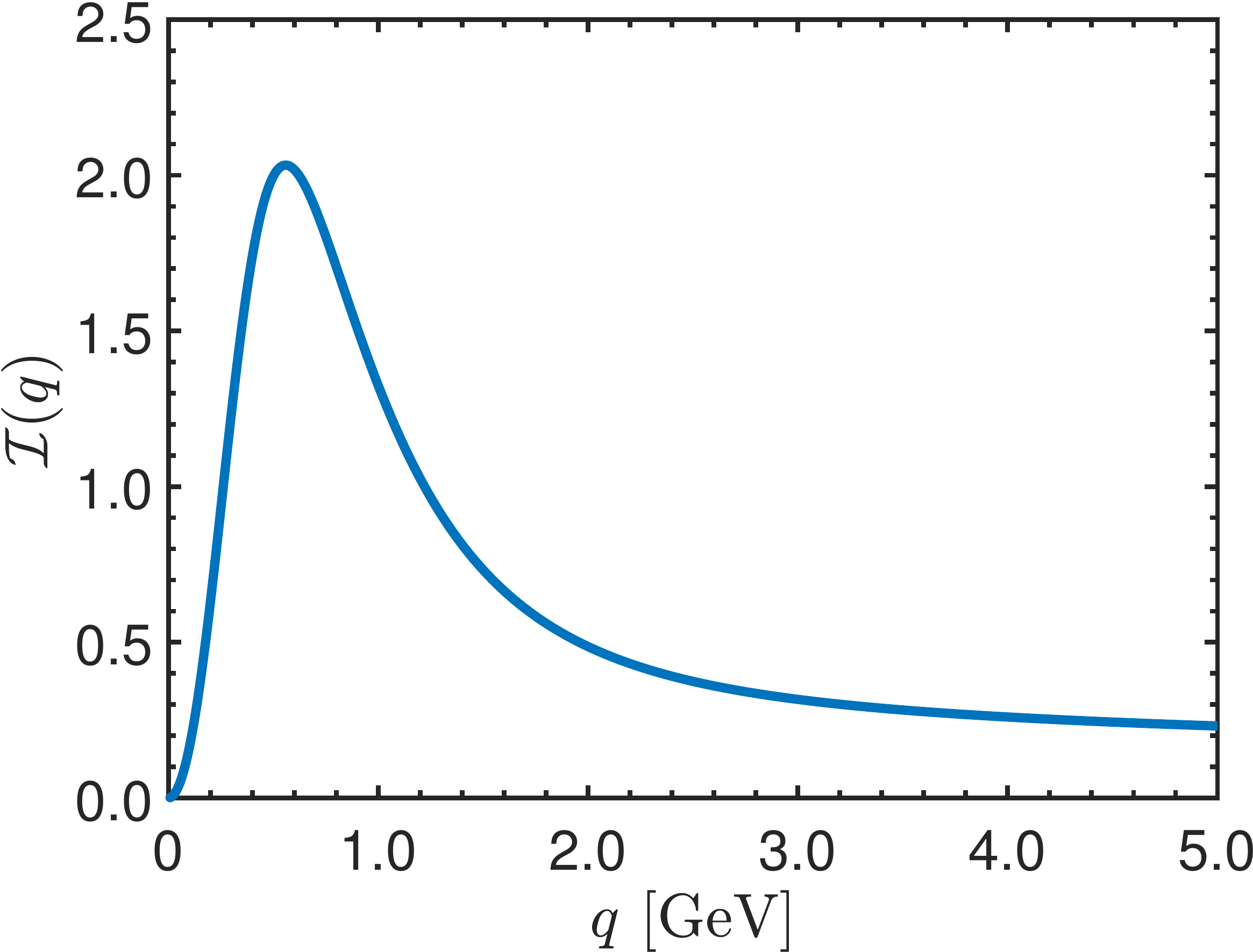}
\caption{ Top left: Gluon propagator, $\Delta(q)$, from lattice simulations of Ref.~\cite{Aguilar:2021okw} (points) and a fit given by Eq.~(C11) of~\cite{Aguilar:2021uwa} (blue continuous). Top right: The auxiliary function $1 + G(q)$, defined in \1eq{Lambda_GL}. Bottom left: The renormalization group invariant (RGI) running interaction strength ${\widehat d}(q)$ defined in \1eq{dhat}, computed using the $\Delta(q)$ and $1 + G(q)$ shown in the top panels, with $\alpha_s=0.27$~\cite{Boucaud:2017obn} and $Z_1 = 0.9333$ [see Sec.~\ref{sec:ghost_dyn}]. Bottom right: The corresponding dimensionless RGI interaction ${\cal I}(q)$, defined in \1eq{Ical}.
}
\label{fig:gluon_prop}
\end{figure}

The actual determination of 
${\widehat d}(q)$ proceeds by means 
of the second equality in \1eq{dhat}, \ie  
by combining the standard gluon 
propagator, $\Delta(q)$, together with the 
function $1 + G(q)$.
In the top left panel of \fig{fig:gluon_prop} we show lattice data for the conventional gluon propagator from~\cite{Aguilar:2021okw} (points) and a physically motivated fit (blue continuous), given by Eq.~(C11) of~\cite{Aguilar:2021uwa}. In the top right panel of the same figure we show the $1+G(q)$ auxiliary function, which can be computed by contracting \1eq{Lambda_GL} with $P^{\mu\nu}(q)/3$ (see, \eg~\cite{Aguilar:2009nf}), using the results of~\cite{Aguilar:2018csq} for the ghost-gluon kernel, $H_{\nu\mu}(r,p,q)$. 
Then, in the bottom left panel of \fig{fig:gluon_prop} we show the ${\widehat d}(q)$ that results from combining the fit for $\Delta(q)$ and the $1 + G(q)$ shown in the top panels of the same figure and using $\alpha_s=0.27$~\cite{Boucaud:2017obn} and $Z_1 = 0.9333$ [see Sec.~\ref{sec:ghost_dyn}].

%%%%%%%%%%%%%%%%%%%%%%%%%%%%%%%%%%
% Figure 9  - Gluon prop, derivative, 1 + G, and dhat
%%%%%%%%%%%%%%%%%%%%%%%%%%%%%%%%%%
%%%%%%%%%%%%%%%%%%%%%%%%%%%%%%%%%%

From the ${\widehat d}(q)$ of \1eq{dhat} one may define the dimensionless RGI interaction~\cite{Binosi:2014aea}, ${\cal I}(q)$, 
\be 
{\cal I}(q) := q^2 {\widehat d}(q) \,. 
\label{Ical}
\ee
As explained in~\cite{Binosi:2014aea}, this quantity provides the strength required in order to describe ground-state hadron observables using SDEs in the matter sector of the theory. 
In that sense, ${\cal I}(q)$  
bridges a longstanding gap that has existed between nonperturbative continuum QCD and  
ab initio predictions of basic hadron properties.

%%%%%%%%%%%%%%%%%%%%%%%%%%%%%%%%%%%%%%%%%%%%%%%%%%%%%%%%
\section{Three-gluon vertex and its planar degeneracy}\label{sec:planar_deg}
%%%%%%%%%%%%%%%%%%%%%%%%%%%%%%%%%%%%%%%%%%%%%%%%%%%%%%%%

The three-gluon vertex, $\fatg_{\alpha\mu\nu}(q,r,p)$, plays a pivotal role in the dynamics of QCD~\cite{Papavassiliou:2022umz}, manifesting its non-Abelian nature through the gluon self-interaction. In fact, the most celebrated perturbative feature of 
QCD, namely  
asymptotic freedom, hinges on the properties of this particular interaction vertex. 
Its importance in the nonperturbative domain 
has led to an intense effort for unveiling its elaborate  features~\cite{Alkofer:2004it,Pelaez:2013cpa, Aguilar:2013vaa,Blum:2014gna,Eichmann:2014xya,Williams:2015cvx,Blum:2015lsa,Cyrol:2016tym,Corell:2018yil,Boucaud:2017obn,Huber:2018ned, Aguilar:2019jsj,Aguilar:2019uob,Aguilar:2019kxz,Parrinello:1994wd,Alles:1996ka,Parrinello:1997wm,Boucaud:1998bq,Cucchieri:2006tf,Cucchieri:2008qm,Athenodorou:2016oyh,Duarte:2016ieu,Boucaud:2017obn,Vujinovic:2018nqc,Aguilar:2019jsj,Barrios:2022hzr,Pinto-Gomez:2022brg,Pinto-Gomez:2022qjv}. 
Indeed, as we have seen in Secs.~\ref{sec:smg} and \ref{sec:CBSE}, the pole structure of the three-gluon vertex is crucial for the onset of the Schwinger mechanism 
and the dynamical generation of a gluon mass. Moreover, its pole-free part provides highly nontrivial contributions to the SDEs of several Green's functions, most notably the gluon propagator (cf.~\fig{fig:SDEs}), as well as in the Bethe-Salpeter and Faddeev equations 
that determine the properties of glueballs~\cite{Meyers:2012ka,Sanchis-Alepuz:2015hma,Souza:2019ylx,Huber:2020ngt,Huber:2021yfy} and hybrid mesons~\cite{Xu:2018cor}, respectively.

For general momenta, $\fatg_{\alpha\mu\nu}(q,r,p)$ is a particularly complicated function, comprised by 14 tensor structures
and their associated form factors~\cite{Ball:1980ax}.  
Fortunately, 
in the Landau gauge, considerable simplifications take place, making the treatment of the three-gluon vertex 
less cumbersome.
Indeed, in the latter gauge, quantities of interest require only the knowledge of the {\it transversely projected} three-gluon vertex~\cite{Eichmann:2014xya,Blum:2014gna,Huber:2016tvc,Aguilar:2022thg}, $\gbar_{\alpha\mu\nu}(q,r,p)$, defined as
\begin{align}
\label{eq:barGamma}
\gbar_{\alpha \mu \nu}(q,r,p) &=  \fatg^{\alpha' \mu' \nu'}(q,r,p) P_{\alpha' \alpha}(q) P_{\mu' \mu}(r) P_{\nu' \nu}(p) \nonumber\\ 
&= \Gamma^{\alpha' \mu' \nu'}(q,r,p) P_{\alpha' \alpha}(q) P_{\mu' \mu}(r) P_{\nu' \nu}(p) \,.  
\end{align}
Note that $\gbar_{\alpha \mu \nu}(q,r,p)$ does not contain massless poles, by virtue of \1eq{eq:transvp}.
Furthermore, $\gbar_{\alpha \mu \nu}(q,r,p)$ can be parametrized in terms of only $4$ independent tensor structures, \ie
\be
\gbar^{\alpha \mu \nu}(q,r,p) = \sum_{i=1}^4 \widetilde{\g}_i(q^2,r^2,p^2) \,
\tlambda_i^{\alpha\mu\nu}(q,r,p) \,. 
\label{eq:expl}
\ee
Due to the Bose symmetry of $\gbar_{\alpha\mu\nu}(q,r,p)$, the $\tlambda_i^{\alpha\mu\nu}(q,r,p)$ can be chosen to be individually Bose symmetric, such that its form factors $\widetilde{\Gamma}_i(q^2,r^2,p^2)$ are symmetric under the exchange of any two arguments~\cite{Pinto-Gomez:2022brg}. In fact, they can only depend on three totally symmetric combinations of momenta. 

Quite remarkably, lattice~\cite{Pinto-Gomez:2022brg,Pinto-Gomez:2022qjv,Pinto-Gomez:2022woq} and continuum~\cite{Eichmann:2014xya,Blum:2014gna,Huber:2016tvc} studies alike, have demonstrated that, to a very good level of accuracy, the $\widetilde{\g}_i$ depend exclusively on a single judiciously chosen variable. Specifically, the $\widetilde{\g}_i$ computed on the lattice in~\cite{Pinto-Gomez:2022brg,Pinto-Gomez:2022qjv,Pinto-Gomez:2022woq} can be parametrized in terms of the special Bose symmetric combination
\be
s^2 = \frac{1}{2}\left( q^2 + r^2 + p^2 \right) \,.  \label{eq:s2}
\ee
Thus, the $\widetilde{\g}_i$ are the same for any combination of $q^2$, $r^2$, and $p^2$ that fulfils \1eq{eq:s2} for a given value of $s^2$. This property has been denominated \emph{planar degeneracy}, because \1eq{eq:s2} with fixed $s$ defines a plane, normal to the vector $(1,1,1)$, in the first octant of the coordinate system $(q^2,r^2, p^2)$.

In particular, the form factor $\widetilde{\Gamma}_1(q^2,r^2,p^2)$ of the classical tensor structure is rather accurately approximated by
\begin{equation}
\widetilde{\Gamma}_1(q^2,r^2,p^2) \approx \widetilde{\Gamma}_1(s^2,s^2,0) \approx \Ls(s) 
\,. 
\label{eq:Gamma1allkin}    
\end{equation}
In the above equation, $\Ls$ is the single transverse form factor of the three-gluon vertex in the soft gluon limit~\cite{Aguilar:2021uwa}, and is obtained in lattice simulations as the $q=0$ limit of the following totally transverse projection~\cite{Aguilar:2021lke}
\be 
\Ls(r) = \left. \frac{\g_0^{\alpha\mu\nu}(q,r,p)P_{\alpha\alpha'}(q)P_{\mu\mu'}(r)P_{\nu\nu'}(p)\fatg^{\alpha'\mu'\nu'}(q,r,p)}{\g_0^{\alpha\mu\nu}(q,r,p)P_{\alpha\alpha'}(q)P_{\mu\mu'}(r)P_{\nu\nu'}(p)\g_0^{\alpha'\mu'\nu'}(q,r,p)} \right\vert_{q\to 0} \,. \label{Lsg_def}
\ee 

\begin{figure}[h]
\centering
\includegraphics[width=0.475\textwidth]{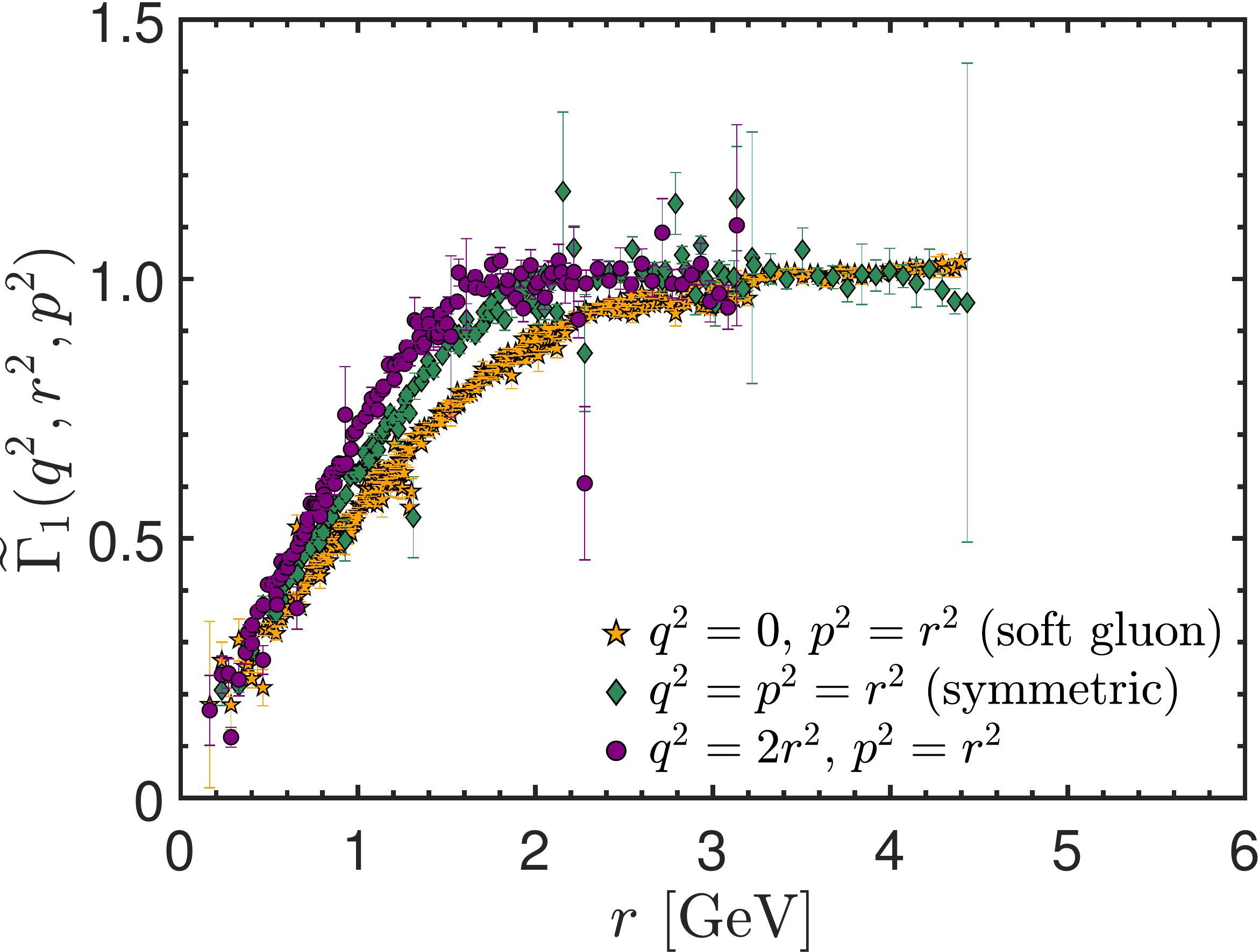} \hfil \includegraphics[width=0.475\textwidth]{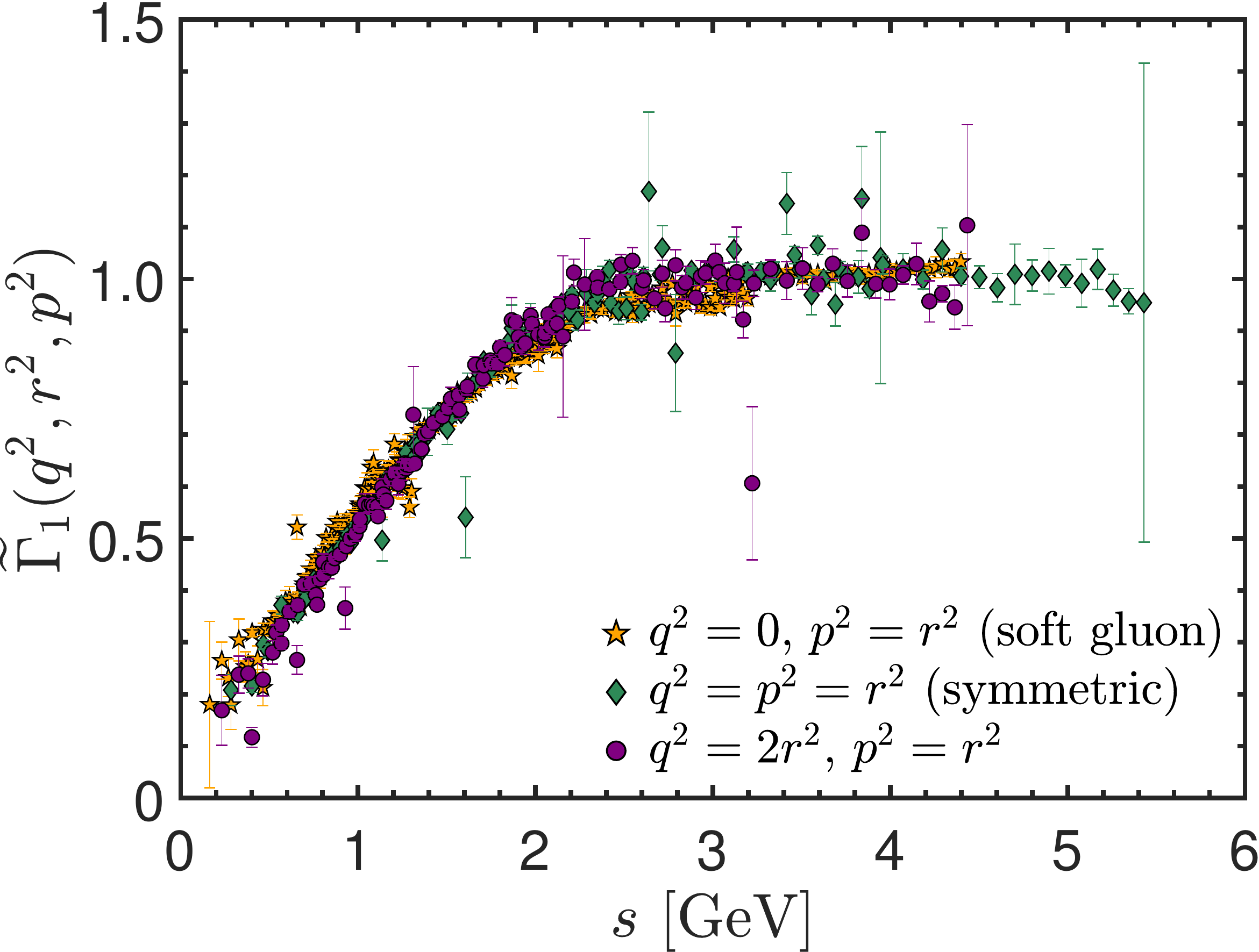}
\caption{ Lattice data from Ref.~\cite{Pinto-Gomez:2022brg} for the classical form factor, $\widetilde{\g}_1(q^2,r^2,p^2)$, of the transversely projected three-gluon vertex in three different kinematic configurations: the soft-gluon ($q = 0$, $p^2 = r^2$, orange stars), the symmetric limit ($q^2 = p^2 = r^2$, green diamonds), and the case $p^2 = r^2$ with $q^2 = 2r^2$ (purple circles). In the left panel $\widetilde{\g}_1(q^2,r^2,p^2)$ is plotted as a function of $r$, while in the right it is plotted as a function of the Bose symmetric variable $s$ defined in \1eq{eq:s2}.
}
\label{fig:planar}
\end{figure}

A particular realization of the planar degeneracy property is shown in \fig{fig:planar}, where we show the classical form factor $\widetilde{\g}_1(q^2,r^2,p^2)$, obtained from the  
lattice simulation of~\cite{Pinto-Gomez:2022brg}; we   
consider three different kinematic configurations, characterized by a single momentum. Specifically, the orange stars correspond to the soft-gluon limit, $q = 0$, which implies $p^2 = r^2$; the green diamonds denote the symmetric limit, where all of the momenta have the same magnitude, $q^2 = p^2 = r^2$; and the purple circles represent points with $p^2 = r^2$ and $q^2 = 2 r^2$. When plotted against the momentum $r$, the three configurations of $\widetilde{\g}_1(q^2,r^2,p^2)$ produce three clearly distinct curves; however, 
when plotted in terms of the Bose symmetric variable $s$ of \1eq{eq:s2}, they become statistically indistinguishable, 
manifesting the validity of \1eq{eq:Gamma1allkin}.

%%%%%%%%%%%%%%%%%%%%%%%%%%%%%%%%%%
% Figure 10  - Planar degeneracy
%%%%%%%%%%%%%%%%%%%%%%%%%%%%%%%%%%
%%%%%%%%%%%%%%%%%%%%%%%%%%%%%%%%%%

In addition to the planar degeneracy property, lattice~\cite{Aguilar:2021lke,Pinto-Gomez:2022brg,Pinto-Gomez:2022qjv,Pinto-Gomez:2022woq} and continuum~\cite{Eichmann:2014xya,Blum:2014gna,Huber:2016tvc,Aguilar:2019jsj} results show a clear dominance of the classical form factor $\widetilde{\Gamma}_1$ over the remaining ones. Based on these considerations, the special approximation 
\begin{equation}
\label{eq:compact}
\gbar^{\alpha\mu\nu}(q,r,p) \approx L_{\textrm{sg}}\left(s \right) {\overline \Gamma}_{\!0}^{\alpha\mu\nu}(q,r,p) \,,
\end{equation} 
has been put forth, 
where ${\overline \Gamma}_{\!0}^{\alpha\mu\nu}(q,r,p)$ is the tree-level value of $\gbar^{\alpha\mu\nu}(q,r,p)$, 
\ie \1eq{eq:barGamma} with $\Gamma^{\alpha' \mu' \nu'}(q,r,p) 
\to {\Gamma}^{\alpha' \mu' \nu'}_{\!0}(q,r,p)$.  
\1eq{eq:compact}  
provides an accurate and  exceptionally compact approximation for $\gbar^{\alpha\mu\nu}(q,r,p)$ in general kinematics. 

\begin{figure}[h]
\centering
\includegraphics[width=0.475\textwidth]{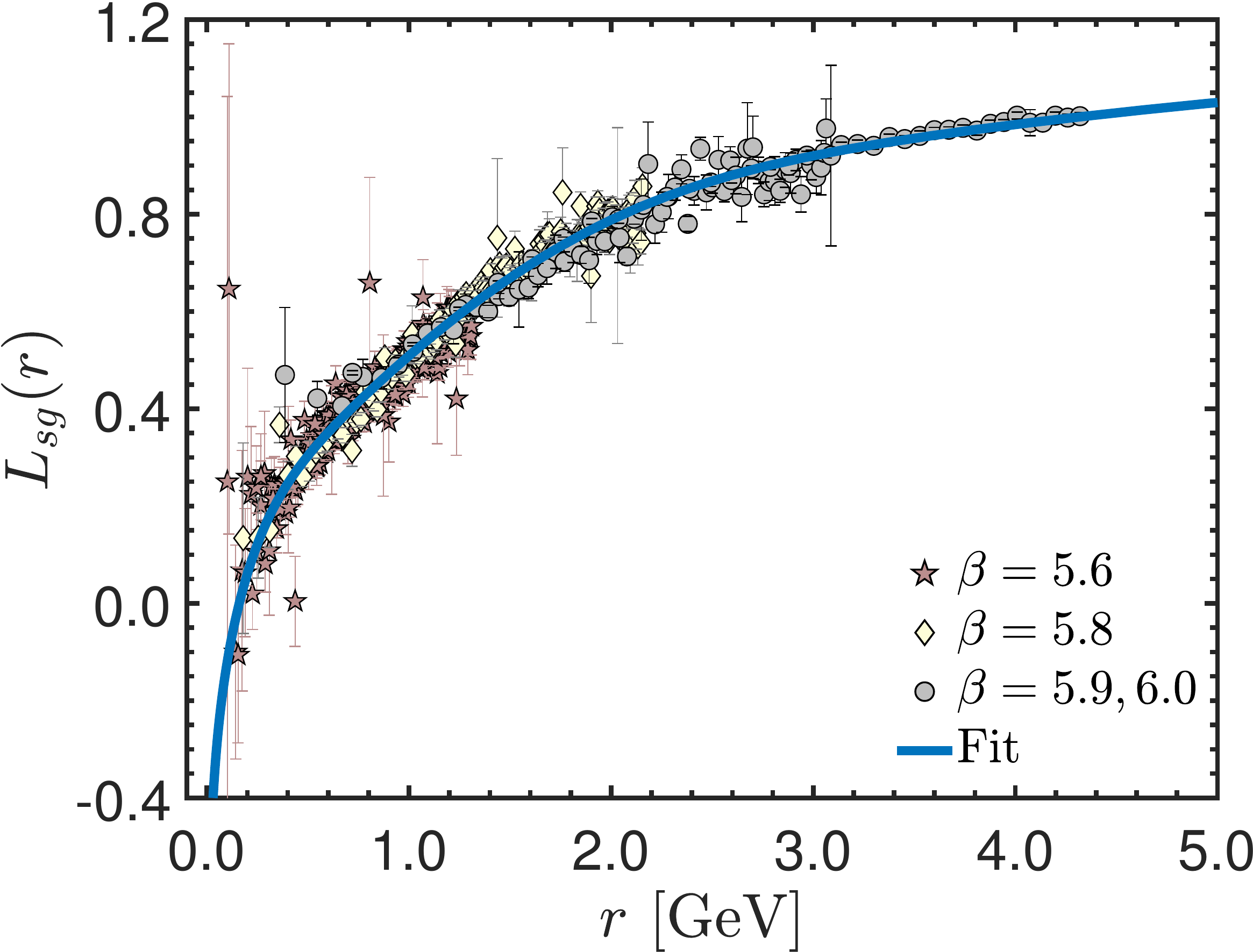}
\caption{ Lattice data from Ref.~\cite{Aguilar:2021lke} for $\Ls(q)$, compared to the fit for it given by Eq.~(C12) of \cite{Aguilar:2021uwa} (blue continuous curve).
}
\label{fig:Lsg}
\end{figure}

We emphasize that the shape of $\Ls(r)$ has been very precisely determined through dedicated lattice studies with large-volume simulations~\cite{Athenodorou:2016oyh,Boucaud:2017obn,Aguilar:2021lke,Aguilar:2021okw}. The outcome of this exploration is shown in \fig{fig:Lsg},  where we plot the lattice data of~\cite{Aguilar:2021lke} for $\Ls(r)$, together with a physically motivated fit given by Eq.~(C12) of \cite{Aguilar:2021uwa} (blue continuous curve). 

The approximation given by \1eq{eq:compact}, with the fit for $\Ls$ shown in \fig{fig:Lsg}, will be used explicitly in Secs.~\ref{sec:ghost_dyn} and \ref{sec:WSDE}, where the $\gbar^{\alpha\mu\nu}(q,r,p)$ 
in general kinematics 
will be needed as input for the determination of other physically important quantities.

%%%%%%%%%%%%%%%%%%%%%%%%%%%%%%%%%%
% Figure 11  - Lsg
%%%%%%%%%%%%%%%%%%%%%%%%%%%%%%%%%%
%%%%%%%%%%%%%%%%%%%%%%%%%%%%%%%%%%

%%%%%%%%%%%%%%%%%%%%%%%%%%%%%%%%%%%%%%%%%%%%%%%%%%%%%%%%
\section{Ghost dynamics from Schwinger-Dyson equations}\label{sec:ghost_dyn}
%%%%%%%%%%%%%%%%%%%%%%%%%%%%%%%%%%%%%%%%%%%%%%%%%%%%%%%%

\begin{figure}[ht!]
  \centering
  \includegraphics[width=1\textwidth]{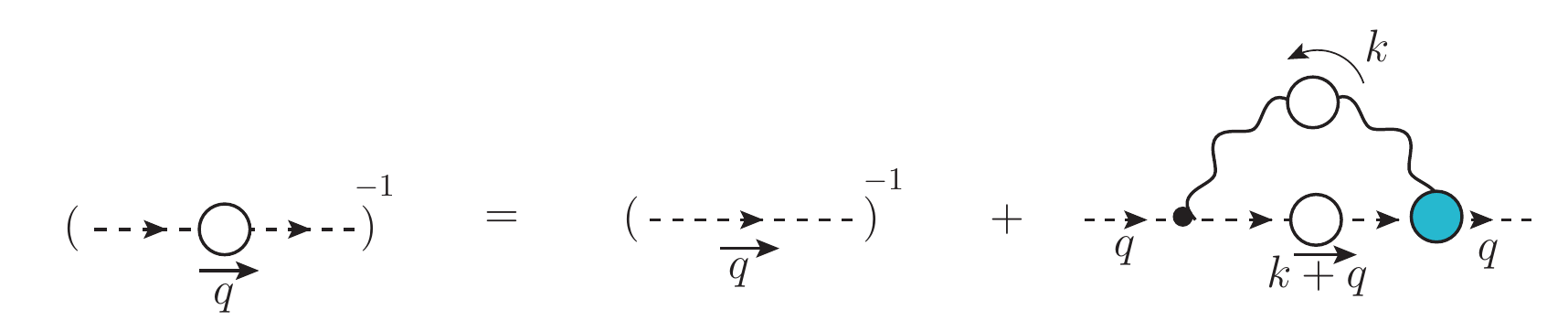}
  \\
  \includegraphics[width=\textwidth]{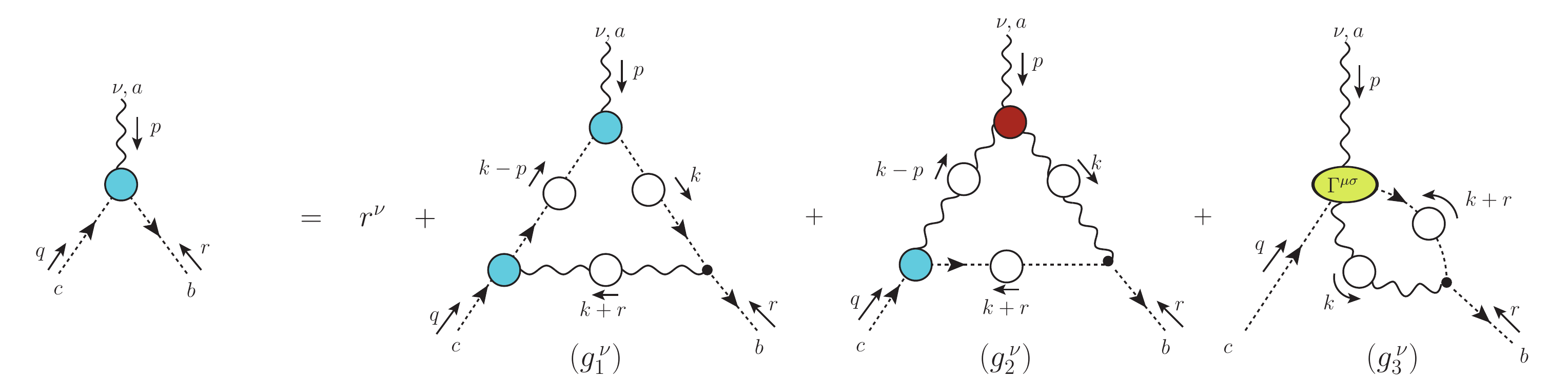}	
  \caption{ Top: SDE governing the momentum evolution of the ghost propagator. Bottom: SDE for the ghost-gluon vertex, $\fatg_{\nu}(r,q,p)$. }
\label{fig:ghost_SDE}
\end{figure}

We next turn our attention to the ghost sector 
of the theory, whose scrutiny 
is important for several reasons. First, it has been connected to particular scenarios of color confinement~\cite{Kugo:1979gm,Nakanishi:1990qm}. Second, the Green's functions 
associated with the ghost sector 
appear as ingredients in the SDEs of several key functions, such as the gluon propagator and the three-gluon vertex~\cite{Cucchieri:2006tf,Cucchieri:2008qm,Alkofer:2008jy,Aguilar:2013vaa,Pelaez:2013cpa,Blum:2014gna,Eichmann:2014xya,Williams:2015cvx,Blum:2015lsa,Cyrol:2016tym,Duarte:2016ieu,Athenodorou:2016oyh,Boucaud:2017obn,Aguilar:2019jsj,Aguilar:2019uob,Aguilar:2019kxz}, affecting their nonperturbative behavior in nontrivial ways, as will be discussed in Sec.~\ref{sec:ghost_loops}. Third, the SDEs governing the ghost sector 
are simpler than their gluonic counterparts, 
because they are comprised by fewer diagrams; in fact, the SDE of the ghost propagator 
contains a single diagram, see \fig{fig:ghost_SDE}. Fourth, 
in the Landau gauge, the validity of Taylor's theorem~\cite{Taylor:1971ff} 
facilitates considerably the task of renormalization. 

Consequently, the SDEs of the ghost sector are an excellent testing ground 
for $(a)$ probing the impact of the gluonic 
Green's functions that contribute to them~\cite{Aguilar:2021okw}; $(b)$ assessing the reliability of truncation schemes~\cite{Huber:2017txg,Aguilar:2022wsh}; and $(c)$ testing the agreement between lattice and continuum approaches.

One of the 
central results of numerous studies in the continuum ~\cite{Aguilar:2008xm,Dudal:2008sp,Boucaud:2008ky,Boucaud:2008ji,Kondo:2009gc,Boucaud:2011ug,Pennington:2011xs,Dudal:2012zx,Aguilar:2013xqa,Cyrol:2016tym,Huber:2018ned,Aguilar:2018csq,Aguilar:2021okw} as well as a variety of lattice simulations~\cite{Ilgenfritz:2006he,Cucchieri:2007md,Bogolubsky:2007ud,Cucchieri:2008fc,Bogolubsky:2009dc,Ayala:2012pb,Boucaud:2018xup,Cui:2019dwv}
may be summarized by stating that
the ghost propagator, $D(q)$, remains massless, while 
the corresponding dressing function,  $F(q)$, saturates at the origin.
As we will discuss in Sec.~\ref{sec:ghost_loops}, 
the nonperturbative masslessness of the ghost has important implications for the infrared behavior of the gluon propagator and the three-gluon vertex.

In what follows 
we provide a concrete example of the state-of-the-art SDE analysis of the ghost
sector, by solving the coupled system of equations that governs the ghost-dressing function
and the ghost-gluon vertex. In order to obtain a closed system of equations, we use lattice
results for the gluon propagator, the three-gluon vertex, and the value of the coupling
constant in the particular renormalization scheme employed.

The main points of this analysis may be summarized as follows.

({\it i}) We begin by considering  
the coupled system of SDEs given by Fig.~\ref{fig:ghost_SDE},  which determines the ghost propagator and ghost-gluon 
vertex. The treatment will be simplified  
by neglecting diagram $(d_3^{\nu})$ of \fig{fig:ghost_SDE}, thus eliminating the dependence on the ghost-ghost-gluon-gluon vertex, $\Gamma^{\mu\sigma}$.
This is a particularly robust truncation,  
because the impact of the neglected 
diagram on the ghost-gluon vertex has been shown to be less than $2\%$~\cite{Huber:2017txg}.

({\it ii}) 
Note that due to the fully transverse nature of the gluon propagators in the Landau gauge, 
in conjunction with the fact that 
various projections need to be implemented 
during the treatment of this system, the pole parts $V$ of all fully dressed 
vertices appearing in \fig{fig:ghost_SDE} will be annihilated; thus, we will have 
throughout the replacement $\fatg\to\g$.

({\it iii})
We proceed by decomposing the pole-free part, $\g_\nu(r,q,p)$, of the ghost-gluon vertex into its most general Lorentz structure, namely
\be 
\g_\nu(r,q,p) = r_\nu B_1(r,q,p) + p_\nu B_2(r,q,p) \,, \label{Bi_def}
\ee 
whose scalar form factors reduce to $B_1^0 = 1$ and $B_2^0 = 0$ at tree level.
Evidently, due to the transversality of the gluon propagator, only the classical tensor $r_\nu$, accompanied by the form factor
$B_1$, will survive in all SDE diagrams of \fig{fig:ghost_SDE}.

%%%%%%%%%%%%%%%%%%%%%%%%%%%%%%%%%%
% Figure 12  - Coupled ghost SDEs
%%%%%%%%%%%%%%%%%%%%%%%%%%%%%%%%%%
%%%%%%%%%%%%%%%%%%%%%%%%%%%%%%%%%%

({\it iv}) The SDE of \fig{fig:ghost_SDE} is given by
\be  
F^{-1}(r) = 1 + 2\lambda \int_k f(k,r) B_1(-r,k+r, -k)\Delta(k)D(k+r) \,, \label{ghost_SDE}
\ee 
where $\lambda$ is given by \1eq{lambda_def}, and we define
\be 
f(k,r) := 1 - \frac{ (r \cdot k)^2 }{ r^2 k^2 } \,. \label{fqk_def}
\ee 

({\it v})
Next, we note that the form factor $B_1(r,q,p)$ can be extracted from $\g_\nu(r,q,p)$ through the projection
\be 
B_1(r,q,p) = \varepsilon^\nu \g_\nu(r,q,p)\,, \qquad \varepsilon^\nu := \frac{p^2 r^\nu - (r\cdot p)p^\nu }{r^2 p^2 - (r\cdot p)^2} \,. \label{B1_proj}
\ee 
Hence, acting with $\varepsilon^\nu$ on the diagrams in the second line of \fig{fig:ghost_SDE}, we obtain
\be 
B_1(r,q,p) = 1 - \lambda\left[ a(r,q,p) - b(r,q,p) \right] \,, \label{B1_SDE}
\ee 
where
\begin{align}
a(r,q,p) &= q^\alpha r^\mu\varepsilon^\nu \!\!\int_k D(k)D(k-p)\Delta(k+r)B_1(p-k,q,k+r)B_1(-k,k-p,p)P_{\alpha\mu}(k+r)k_\nu \,, \nonumber\\
b(r,q,p) &= q^\alpha r^\mu\varepsilon^\nu \!\!\int_k \Delta(k)\Delta(k-p)D(k+r)B_1(k+r,q,p-k) \gbar_{\nu\mu\alpha}(p,-k,k-p) \,. \label{ab_gen}
\end{align} 

({\it vi})
At this point, we invoke the property of the planar degeneracy of $\gbar_{\alpha\mu\nu}(q,r,p)$, discussed in Sec.~\ref{sec:planar_deg}. Employing \1eq{eq:compact} into the SDE for $B_1$, the term $b(r,q,p)$ of \1eq{ab_gen} becomes
\be 
b(r,q,p) = q^\alpha r^\mu\varepsilon^\nu \!\!\int_k \Delta(k)\Delta(k-p)D(k+r)B_1(k+r,q,p-k) \gbar^0_{\nu\mu\alpha}(p,-k,k-p)\Ls({\bar s}) \,, \label{b_compact}
\ee
with ${\bar s}^2 = p^2 + k^2 - 2 (k\cdot p)$.

We emphasize that although \1eq{eq:compact} constitutes in general an approximation, there is one particular kinematic limit in which the expression for $b(r,q,p)$ given in \1eq{b_compact} becomes exact. Specifically, in the soft gluon limit ($p=0$), it can be shown \emph{exactly} that \cite{Aguilar:2021okw}
\be 
P_{\mu}^{\mu'}(k)P_\nu^{\nu'}(k)\g_{\alpha\mu'\nu'}(0,k,-k) = 2\Ls(k)k_\alpha P_{\mu\nu}(k) \,. \label{PPGamma}
\ee 
Then, starting from either the general expression for $b(r,q,p)$ of \1eq{ab_gen} and using \1eq{PPGamma}, or from the approximate version given by \1eq{b_compact}, it can easily be shown that the $p = 0$ limit is the same. As such, the use of \1eq{eq:compact} yields not only an excellent approximation in general kinematics, but also the exact soft gluon limit for the contribution of the three-gluon vertex to the form factor $B_1$.

({\it vii})
Now we consider the renormalization of the coupled system of equations. Since the ghost-gluon vertex is finite in the Landau gauge~\cite{Taylor:1971ff}, most SDE treatments~\cite{Schleifenbaum:2004id,Boucaud:2008ky,Huber:2012kd,Aguilar:2013xqa,Aguilar:2018csq,Aguilar:2021okw} of the ghost sector employ the so-called Taylor renormalization scheme, defined in such a way that the finite renormalization constant of the ghost-gluon vertex has the exact value $Z_1 = 1$~\cite{Taylor:1971ff,Boucaud:2008gn,Blossier:2010ky,Zafeiropoulos:2019flq,Aguilar:2021okw}.

However, in order to employ \1eq{eq:compact} most expeditiously, it is more convenient to renormalize in the so-called \emph{asymmetric} MOM scheme, because this is precisely the scheme employed in the lattice calculations of $\Ls$~\cite{Athenodorou:2016oyh,Boucaud:2017obn,Aguilar:2021lke,Aguilar:2021okw}. Specifically, this scheme is defined by imposing the normalization  conditions~\cite{Aguilar:2021lke,Aguilar:2021okw}
\be 
\Delta^{-1}_{{\rm \s{R}}}(\mu) = \mu^2 \,, \qquad F_{{\rm \s{R}}}(\mu) = 1\,, \qquad \Ls^{{\rm \s{R}}}(\mu) = 1 \,. \label{asym_def}
\ee 
Past this point, we denote by ${\widetilde Z}_1$ the \emph{finite} value of the ghost-gluon renormalization constant in the asymmetric MOM scheme.
Evidently, \2eqs{Zs_def}{Bi_def} imply that
$B_1^{{\rm \s{R}}} = {\widetilde Z}_1 B_1$.

The renormalization of \2eqs{ghost_SDE}{B1_SDE} proceeds by substitution of the unrenormalized quantities by their renormalized counterparts, following \1eq{Zs_def}, and imposing \1eq{asym_def} for $F(\mu^2)$. 

Note that, in principle, 
${\widetilde Z}_1$, may be determined from the 
relation ${\widetilde Z}_1 = Z_3 Z_c Z_A^{-1}$, imposed by the 
corresponding STI~\cite{Celmaster:1979km}; 
however, these renormalization constants are not available to us, given that the associated  
Green's functions have been 
obtained from the lattice. Therefore,  
${\widetilde Z}_1$ is treated as an adjustable parameter, whose value 
is determined
by requiring that the solution of the SDE for $F(q)$ reproduces the corresponding lattice data of~\cite{Boucaud:2018xup,Aguilar:2021okw} as well as possible.

({\it viii})
Finally, we transform \2eqs{ghost_SDE}{B1_SDE} from Minkowski to Euclidean space, using standard conversion rules. Note that, once in Euclidean space, we will express the functional dependence of $B_1(r,q,p)$ in terms of the squared momenta of the antighost and gluon legs, $r^2$ and $p^2$, and the angle, $\theta$, between them, \ie $B_1(r,q,p) \equiv B_1(r^2,p^2,\theta)$.

The result of these manipulations is that \2eqs{ghost_SDE}{B1_SDE} become
\begin{align}  
F^{-1}(r) &= 1 - \frac{\alpha_s C_{\rm A}{\widetilde Z}_1 }{2\pi^2} \int_0^\infty dk^2 k^2 \Delta(k) \int_0^\pi d \phi \, s_\phi^4 \nonumber\\
& \times \left[ B_1(r^2,k^2,\phi)\frac{F(\sqrt{z})}{z} - B_1(\mu^2,k^2,\phi)\frac{F(\sqrt{u})}{u} \right] \,, \label{ghost_SDE_euc}
\end{align}
and
\be 
B_1(r^2,p^2,\theta) = {\widetilde Z}_1 - \frac{\alpha_s C_{\rm A}{\widetilde Z}_1}{8\pi^2}\left[ {\overline a} + 2 {\overline b} \right] \,, \label{B1_SDE_euc}
\ee 
respectively, with
\begin{align}
{\overline a} &= \frac{ 1 }{s_\theta} \int_0^\infty dk^2 k^2 F(k) \int_0^\pi d\phi s_\phi^3 \frac{\Delta(\sqrt{z})}{z} \int_0^\pi d\omega s_\omega \frac{F(\sqrt{v})}{v}B_1(k^2,p^2,\alpha)B_1(v,z,\beta) {\cal K}_a \,, \label{ab_euc} \\
{\overline b} &= \frac{ 1 }{s_\theta} \int_0^\infty dk^2 k^2 \Delta(k) \int_0^\pi d\phi s_\phi^3 \frac{F(\sqrt{z})}{z} \int_0^\pi d\omega s_\omega \frac{\Delta(\sqrt{v})}{v}B_1(z,v,\beta) \Ls({\overline s}) {\cal K}_b\,.\nonumber 
\end{align} 
In the above equations we 
employ the 
notation $c_x := \cos x$ and $s_x := \sin x$, 
and define the following variables
\begin{align*}
&r\cdot k := r k c_\phi \,,  &  & p\cdot k := p k ( c_\theta c_\phi + s_\theta s_\phi c_\omega) \,, \nonumber \\
&z := r^2 + k^2 + 2 r k c_\phi \,, & & u := \mu^2 + k^2 + 2 \mu k c_\phi \,, \nonumber\\
&{\overline s}^2 := ( p^2 + k^2 + v)/2 \,, & & v := p^2 + k^2 - 2 p k ( c_\theta c_\phi + s_\theta s_\phi c_\omega) \,, \nonumber\\
&\alpha := \pi - \cos^{-1}\left[ c_\theta c_\phi + s_\theta s_\phi c_\omega \right] \,, & & \beta := \cos^{-1}\left[ \frac{ k (p c_\theta c_\phi + p s_\theta s_\phi c_\omega - r c_\phi ) + p r c_\theta - k^2 }{ \sqrt{v z} }\right]\,. \label{angs}
\end{align*}
Finally, the kernels ${\cal K}_a$ and ${\cal K}_b$ are given by
\begin{align*}
{\cal K}_a =& (c_\theta c_\omega s_\phi - c_\phi s_\theta )\left[ ks_\phi(p c_\theta + r ) - p c_\theta c_\omega(k c_\phi + r ) \right] \,, \nonumber \\
{\cal K}_b =& c_\omega \left\{ k^2 p c_\phi \left[ c_\theta p \left( s_\theta^2 ( s_\phi^2 s_\omega^2 - 4 s_\phi^2 + 1 ) + s_\phi^2 \right) + r \left( s_\phi^2 - s_\theta^2 ( 2 s_\phi^2 + 1 ) \right) \right] \right. \nonumber\\
&  - k^3 \left[ s_\phi^2 \left( r c_\theta - 2 p s_\theta^2 + p \right) + p s_\theta^2 \right] + k p^2 \left[ s_\phi^2 \left( 2 s_\theta^2 ( p - r c_\theta ) - r c_\theta - p \right) + s_\theta^2 ( r c_\theta - p ) \right] \nonumber\\
& \left. \left. - c_\phi p^3 r s_\theta^2\right\} + s_\theta s_\phi \left\{ c_\theta p \left[ r \left( p^2-k^2 ( s_\omega^2 + s_\phi^2 s_\omega^2 - 2 s_\phi^2 ) \right) - c_\phi k ( s_\omega^2 - 2 ) (k^2 + p^2) \right] \right.\right. \nonumber\\
& + k \left[ c_\phi k^2 r - c_\phi p^2 r \left( s_\theta^2 ( s_\omega^2 - 2 ) + s_\omega^2 \right) + k p^2 \left(  3 s_\theta^2 s_\phi^2 s_\omega^2 - 2 s_\theta^2 s_\omega^2 - 4 s_\theta^2 s_\phi^2 + 3 s_\theta^2 \right.\right. \nonumber\\
& \left. \left.\left. + (3-2 s_\omega^2 ) s_\phi^2 - 2 \right) \right] \right\} \,.
\end{align*}

We are now in position to solve \2eqs{ghost_SDE_euc}{B1_SDE_euc} numerically. We choose the renormalization point at $\mu = 4.3$ GeV, and employ for $\Delta(q)$ and $\Ls(q)$ the fits to lattice data shown in Figs.~\ref{fig:gluon_prop} and \ref{fig:Lsg}, respectively. Note that 
for large momenta 
these fits recover 
the behavior dictated by the corresponding anomalous 
dimensions~\cite{Aguilar:2021uwa}. For the strong coupling, we use the value $\alpha_s(4.3 \text{ GeV}) = 0.27$, determined from the lattice simulations of~\cite{Boucaud:2017obn}.

Below we discuss the main results of this analysis:

The value of ${\widetilde Z}_1$ was obtained by solving the SDE system for various values of this constant until the $\chi^2$ of the comparison between the solution for $F(q)$ and the lattice data of~\cite{Boucaud:2018xup,Aguilar:2021okw} was minimized. This procedure yields ${\widetilde Z}_1 = 0.9333\pm 0.0075$.

\begin{figure}[ht!]
  \centering
  \includegraphics[width=0.475\textwidth]{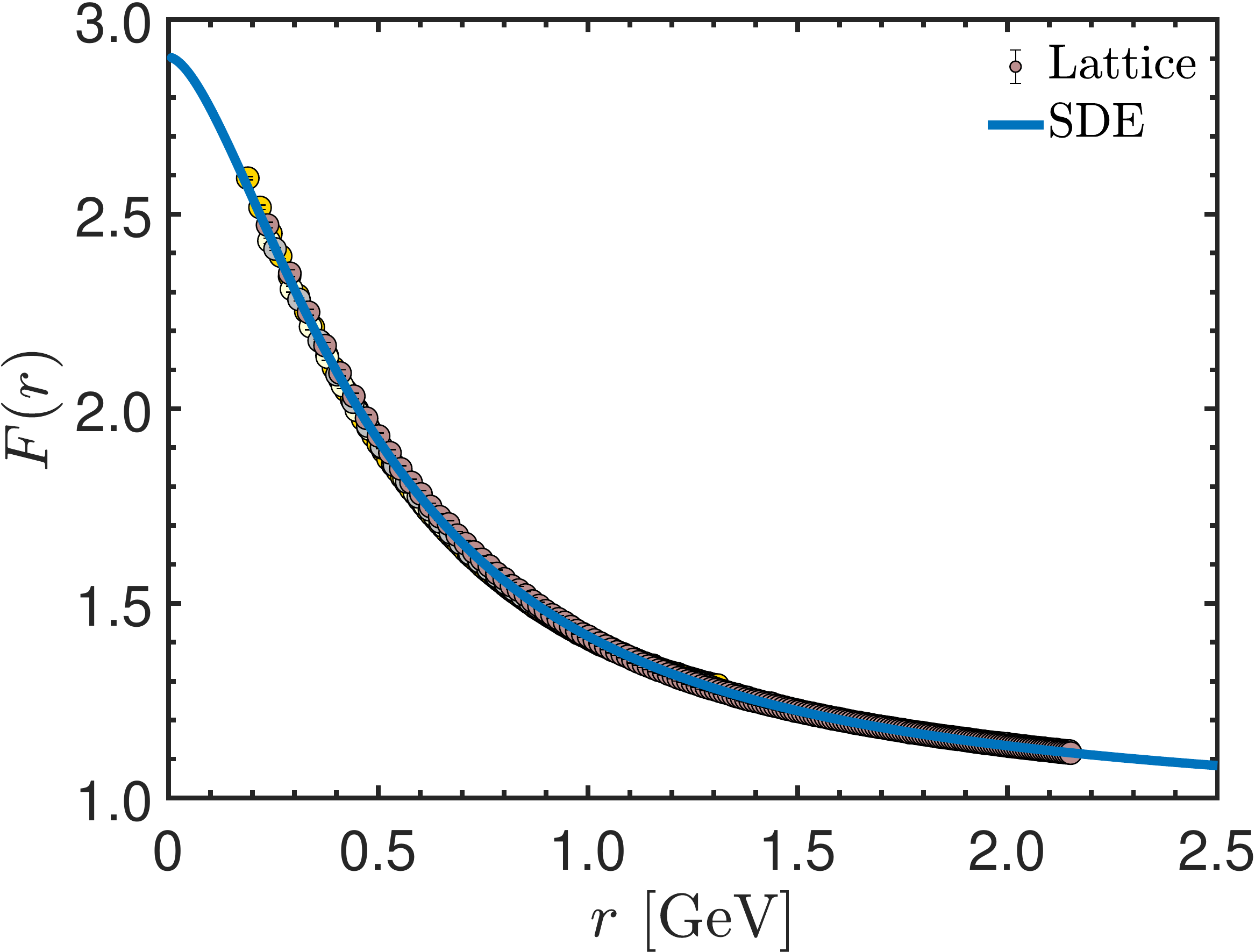} \hfil \includegraphics[width=0.475\textwidth]{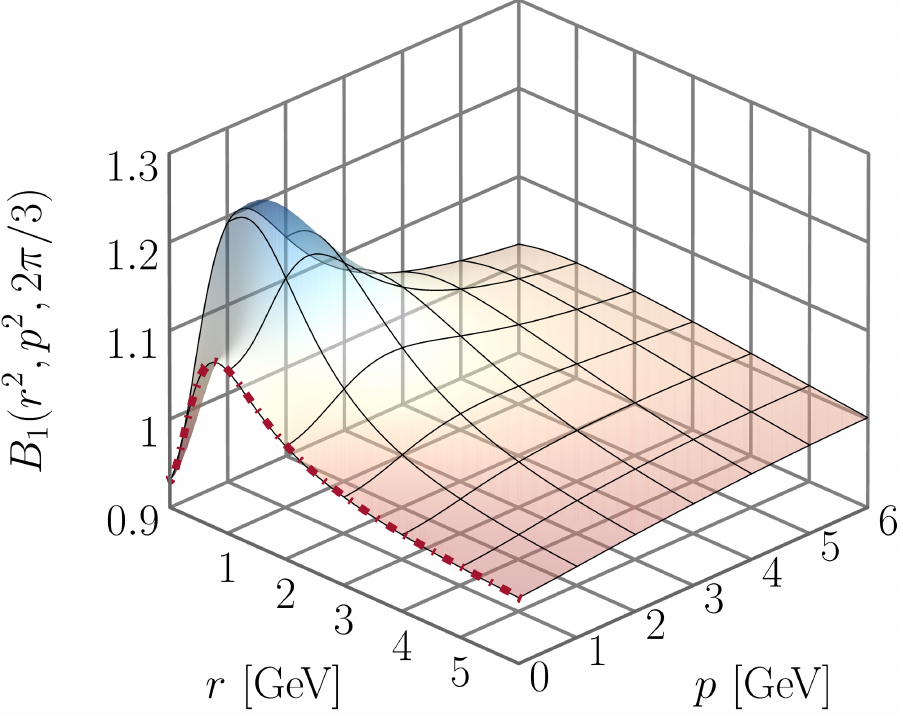}
  \caption{ Left: ghost dressing function $F(q)$ obtained from the coupled system of SDEs of \2eqs{ghost_SDE}{B1_SDE} (blue continuous line) compared to the lattice data of Ref.~\cite{Boucaud:2018xup,Aguilar:2021okw}. Right: The corresponding result for $B_1(r^2,p^2,\theta)$ for arbitrary magnitudes of the antighost and gluon momenta, $r$ and $p$, respectively, and a representative value of $\theta = 2\pi/3$ for the angle between them. The red dot-dashed curve highlights the soft gluon limit ($p=0$).}
\label{fig:B1_genkin}
\end{figure}

In the left panel of \fig{fig:B1_genkin} we show 
as a blue continuous line the SDE result for $F(q)$, with the above value of ${\widetilde Z}_1$. The result is compared to the lattice data of~\cite{Boucaud:2018xup,Aguilar:2021okw}, which have been cured from discretization artifacts. As it turns out, the SDE and lattice results for $F$ agree within $1\%$.

We next consider the form factor $B_1$. In the right panel of \fig{fig:B1_genkin} we show $B_1(r^2,p^2,\theta)$ as a surface, for arbitrary values of the magnitudes of the momenta $r$ and $p$, and for 
the angle $\theta$ formed between them at $\theta = 2\pi/3$. In the same panel, we highlight as a red dot-dashed curve the soft gluon limit\footnote{The soft gluon limit is approached by taking $p\to0$ in $B_1(r^2,p^2,\theta)$;  in the nonperturbative case, this limit is independent of the value of $\theta$. } $B_1(r^2,0,2\pi/3)$ of the general kinematics $B_1(r^2,p^2,2\pi/3)$.

%%%%%%%%%%%%%%%%%%%%%%%%%%%%%%%%%%
% Figure 13  - F and B1 general, results
%%%%%%%%%%%%%%%%%%%%%%%%%%%%%%%%%%
%%%%%%%%%%%%%%%%%%%%%%%%%%%%%%%%%%

The only available SU(3) lattice data  
for $B_1$ have been obtained in the soft gluon limit~\cite{Ilgenfritz:2006he,Sternbeck:2006rd},
and have sizable error bars. Furthermore, they 
have been computed within the Taylor scheme, while in the present work we used the asymmetric MOM scheme. Nevertheless, we can meaningfully compare our SDE results with those of the lattice, and perform a statistical analysis to assess their agreement.

Specifically, 
denoting by $B_1^{\rm \s{T}}$ the Taylor scheme value of the form factor $B_1$, it can easily be shown that
\be 
B_1(r^2,p^2,\theta) = {\widetilde Z}_1 B_1^{\rm \s{T}}(r^2,p^2,\theta) \,, \label{B1_Taylor}
\ee 
which allows us to carry out the desired comparison.

\begin{figure}[ht!]
  \centering
  \includegraphics[width=0.475\textwidth]{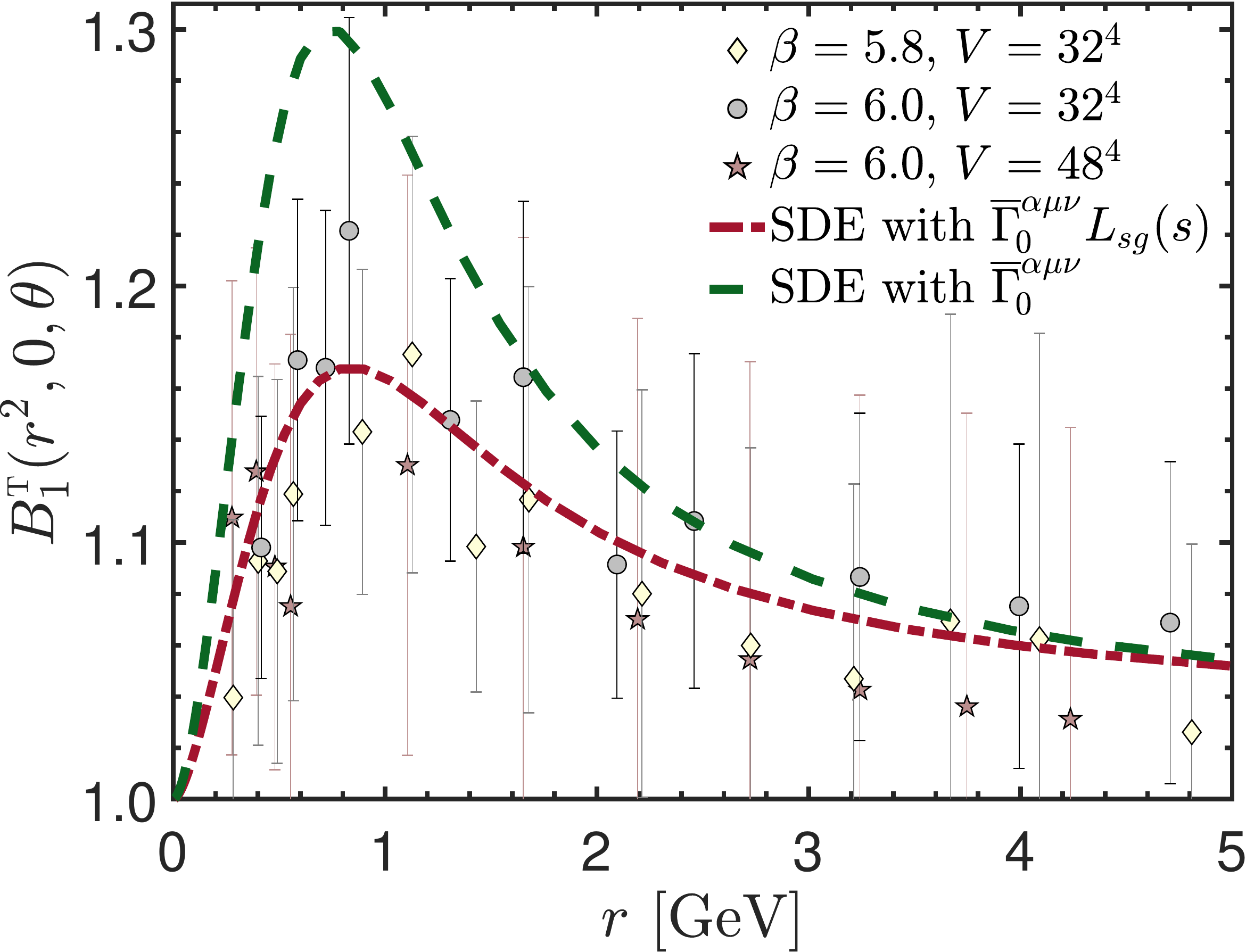}
  \caption{ Soft gluon limit, $B_1^{\rm \s{T}}(r^2,0,\theta)$, of the classical form factor of the ghost-gluon vertex in Taylor scheme. The points correspond to the lattice data of Ref.~\cite{Ilgenfritz:2006he,Sternbeck:2006rd}. The red dot-dashed line shows the SDE solution with the three-gluon vertex dressed according to \1eq{eq:compact}, while the green dashed represents the SDE solution with tree-level three-gluon vertex.}
\label{fig:B1_lattice}
\end{figure}

Then, we use \1eq{B1_Taylor} to compute $B_1^{\rm \s{T}}(r^2,0,\theta)$ from the $B_1(r^2,0,2\pi/3)$ slice (red dot-dashed curve) in the right panel of \fig{fig:B1_genkin}, and compare the result to the lattice data of~\cite{Ilgenfritz:2006he,Sternbeck:2006rd} (points) in \fig{fig:B1_lattice}. Evidently, the SDE determination agrees with the lattice results.

%%%%%%%%%%%%%%%%%%%%%%%%%%%%%%%%%%
% Figure 14  - B1 lattice
%%%%%%%%%%%%%%%%%%%%%%%%%%%%%%%%%%
%%%%%%%%%%%%%%%%%%%%%%%%%%%%%%%%%%

In order to quantify this agreement, we next conduct a $\chi^2$ analysis. To this end, we consider only the 22 lattice points $r_i$ in the interval \mbox{$r_i \in [0.3,2.5]$~GeV}, where the signal is most pronounced. Then, we compute the $\chi^2$ of the data through
\be 
\chi^2_j = \sum_i \frac{[B_1^{\rm lat}(r_i^2,0,\theta) - g_j(r_i)]^2}{\epsilon_{B_1}(r_i^2,0,\theta)} \label{chi2_B1_def}\,,
\ee 
where $B_1^{\rm lat}(r_i^2,0,\theta)$ are the lattice points shown in \fig{fig:B1_lattice}, $\epsilon_{B_1}(r_i^2,0,\theta)$ their respective errors, and $g_j(r_i)$ are three hypotheses which we will compare to the lattice data.
Specifically, for the $g_j$ we consider the three cases
\be 
g_j(r_i) = 
\begin{cases}
   1 & \text{if }j = 1\,, \\
   \text{SDE with }\gbar^{\alpha\mu\nu} = {\overline \Gamma}_{\!0}^{\alpha\mu\nu}\Ls(s) & \text{if }j = 2\,, \\
   \text{SDE with }\gbar^{\alpha\mu\nu} ={\overline \Gamma}_{\!0}^{\alpha\mu\nu} & \text{if }j = 3\,,
\end{cases}
\label{chi2_cases}
\ee
\ie $g_1$ is the tree-level value of $B_1$, $g_2$ the solution of the SDE using \1eq{eq:compact} for dressing the three-gluon vertex, corresponding to the red dot-dashed curve of \fig{fig:B1_lattice}, and $g_3$ is the solution of the SDE obtained setting the three-gluon vertex to tree-level, which amounts to the substitution $\Ls\to1$ in \1eq{B1_SDE_euc}, and is represented by a green dashed curve in \fig{fig:B1_lattice}.

Then, for each $\chi_j^2$ we compute the probability $P_j$ that normally distributed errors would yield a $\chi^2$ at least as large as $\chi_j^2$, through
\be 
P_{j} = \int_{\chi^2_j}^\infty \chi_{\rm \s{PDF}}^2(22,x) dx = \left.\frac{\Gamma(n_r/2,\chi^2/2)}{\Gamma(n_r/2)}\right\vert_{n_r = 22}^{\chi^2=\chi^2_j}\,. \label{P_B1}
\ee 
In the above equation, $\chi_{\rm \s{PDF}}^2(n,x)=x^{n/2-1}e^{-x/2}/[2^{n/2}\Gamma(n/2)]$ denotes the $\chi^2$ probability distribution function with $n$ degrees of freedom, while $\Gamma(z,x)$ is the incomplete $\Gamma$ function.

The results of the above analyses are collected in Table~\ref{B1_stats}. We note that the case $g_1$, \ie the tree-level value of $B_1$, is discarded at $5.1\sigma$ confidence level. As for case $g_3$, it is discarded at the $3.4\sigma$ level. On the other hand, the SDE result with dressed three-gluon vertex, $g_2$, is statistically indistinguishable from the lattice data.

%%%%%%%%%%%%%%%%%%%%%%%%%%%%%%%%%%%%%
%  Table II - Statistical results B1
%%%%%%%%%%%%%%%%%%%%%%%%%%%%%%%%%%%%%
\begin{table}[htb]
	\caption{Statistical results of the $\chi^2$ analysis for the three hypotheses given in \1eq{chi2_cases} for the form factor $B_1$. For each case (first column), we give the corresponding $\chi^2_j$ computed from \1eq{chi2_B1_def} (second column), probability $P_j$ computed from \1eq{P_B1} (third row), and the same $P_j$ expressed in terms of confidence levels $\sigma$ (fourth row).}
	\label{B1_stats}
	\begin{center} 
	\begin{tabular}{|>{\centering\arraybackslash}m{1.5cm}|>{\centering\arraybackslash}m{2cm}|>{\centering\arraybackslash}m{2.5cm}|>{\centering\arraybackslash}m{2.5cm}|}
		\hline 	
		\vspace{0.1cm}
		Case ($j$) & $\chi^2_j$ & $P_j$ & Confidence level in $\sigma$ \\
  	\hline 	
		\vspace{0.1cm}
		1 & 71.37 & $4.0\times10^{-7}$ & $5.1$\\
  	\hline 	
		\vspace{0.1cm}
		2 & 3.399 & 1 - $1.8\times10^{-6}$ & $2.2\times 10^{-6}$\\
  	\hline 	
		\vspace{0.1cm}
		3 & 50.03 & $5.8\times10^{-4}$ & $3.4$\\  
		\hline 	
	\end{tabular}
	\end{center}
\end{table}
%%%%%%%%%%%%%%%%%%%%%%%%%%%%%%%%%%%%%%

Lastly, we point out that for both $F$ and $B_1$ we find a good qualitative agreement with various related studies~\cite{Schleifenbaum:2004id,Huber:2012kd,Aguilar:2013xqa,Cyrol:2016tym,Mintz:2017qri,Aguilar:2018csq,Huber:2018ned,Aguilar:2019jsj,Huber:2020keu,Barrios:2020ubx}, including kinematics other than the soft gluon limit considered in \fig{fig:B1_lattice}.

%%%%%%%%%%%%%%%%%%%%%%%%%%%%%%%%%%%%%%%%%%%%%%%%%%%%%%%%
\section{Divergent ghost loops and their impact on the QCD Green's functions}\label{sec:ghost_loops}
%%%%%%%%%%%%%%%%%%%%%%%%%%%%%%%%%%%%%%%%%%%%%%%%%%%%%%%%

The masslessness of the ghost propagator, discussed in Sec.~\ref{sec:ghost_dyn}, has important implications for the infrared behavior of other Green's functions. Specifically, while the saturation of the gluon propagator renders gluon loops infrared finite, ghost loops furnish infrared divergent contributions~\cite{Aguilar:2013vaa}, akin to those encountered in perturbation theory. In this section, we highlight with two characteristic examples 
how the effects of ghost loops 
manifest themselves at the level of the two- and 
three-point functions. Specifically, 
the ghost loops induce the appearance 
of a moderate maximum in the gluon propagator  
and are responsible for 
the zero-crossing and the 
logarithmic divergence at the origin displayed  
by the dominant form factors of the 
three-gluon vertex.

%%%%%%%%%%%%%%%%%%%%%%%%%%%%%%%%%%
% Figure 15  - Gluon maximum
%%%%%%%%%%%%%%%%%%%%%%%%%%%%%%%%%%
\begin{figure}[ht!]
  \centering
  \includegraphics[width=0.475\textwidth]{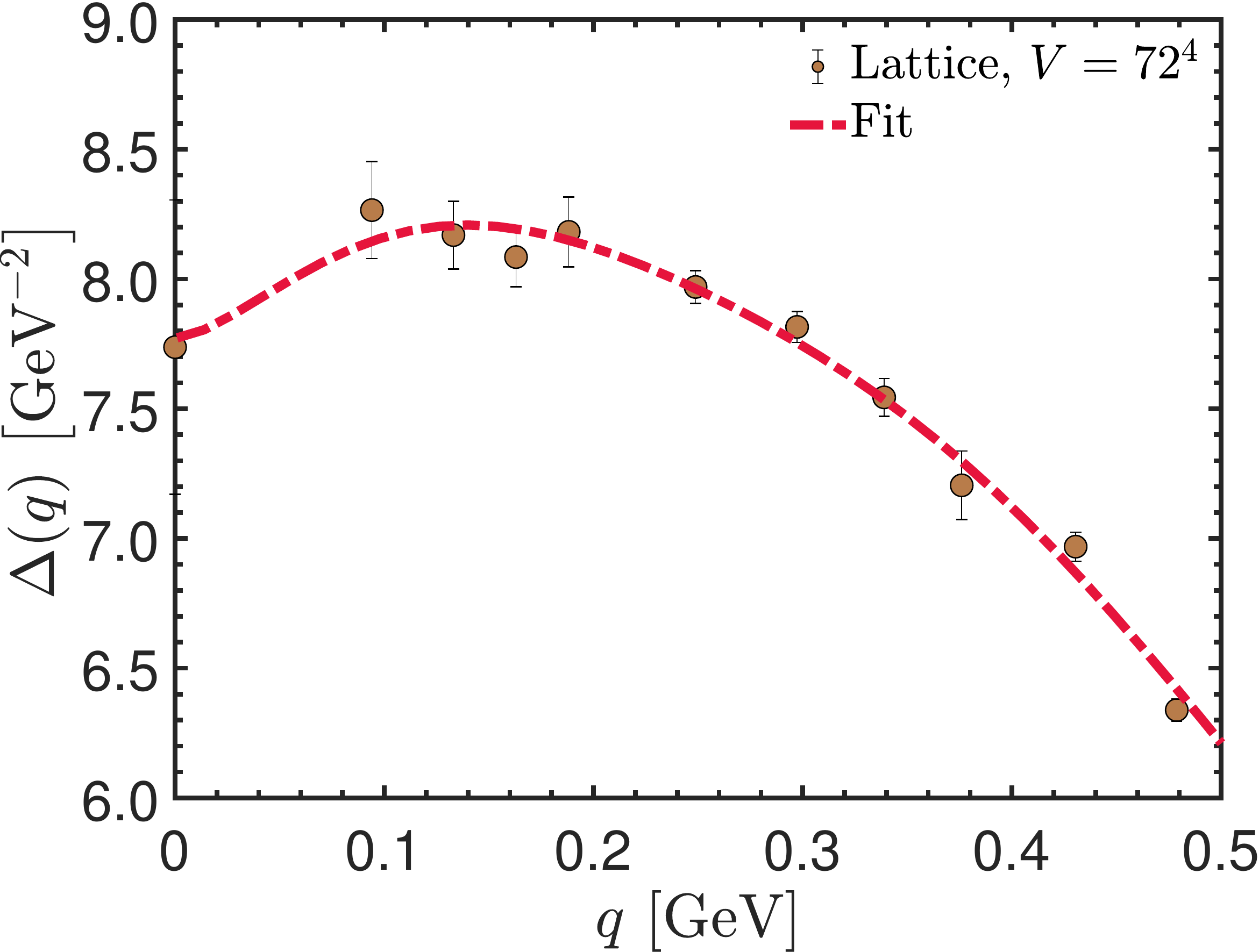} \hfil \includegraphics[width=0.475\textwidth]{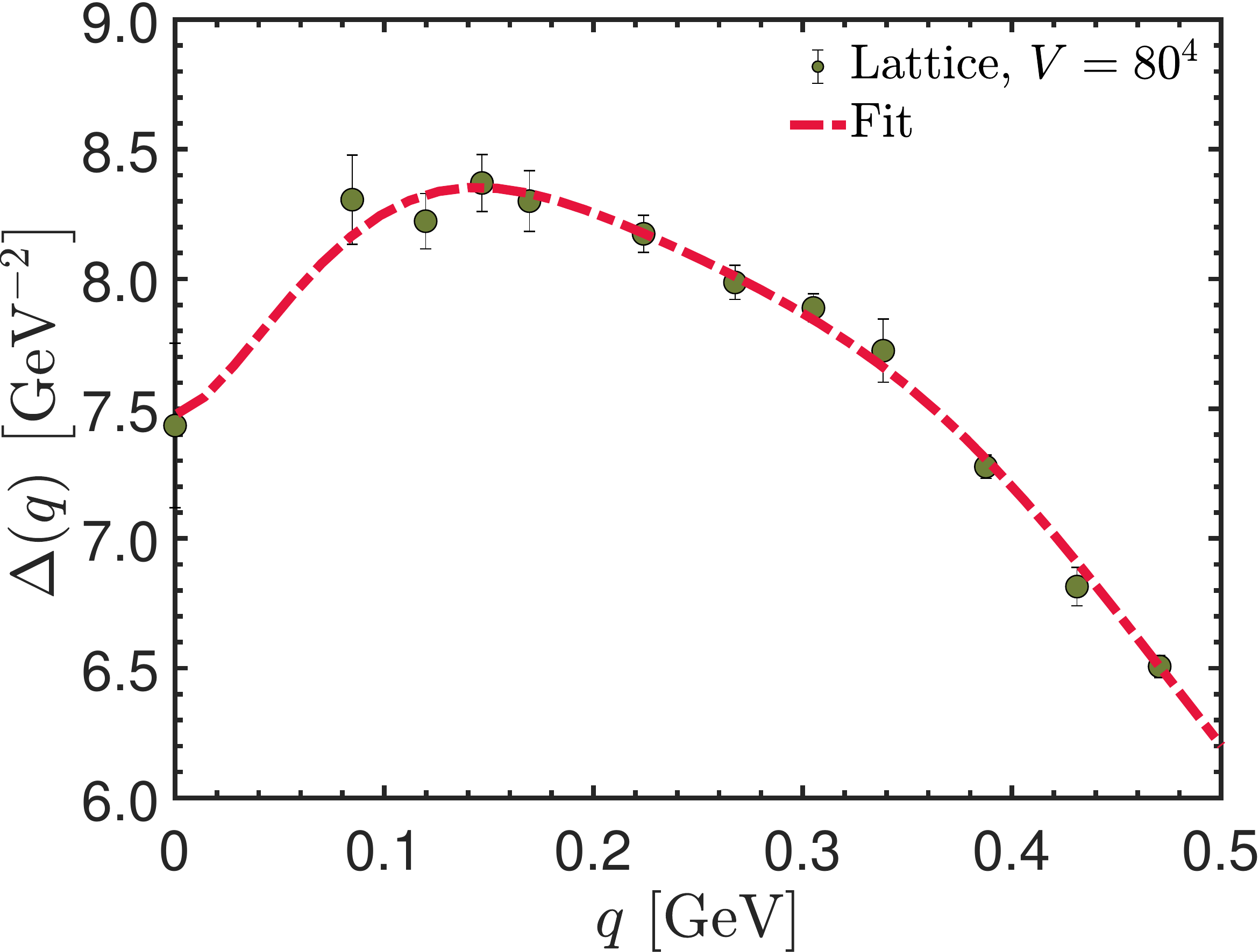}
  \caption{ Lattice data for the gluon propagator in the deep infrared. The data displayed correspond to the two lattice setups with the largest volumes of~\cite{Bogolubsky:2007ud}, namely, $V = 72^4$ (left) and $V = 80^4$ (right). The red dashed lines are smooth fits from which the position of the maximum can be estimated. }
\label{fig:gluon_zoom}
\end{figure}
%%%%%%%%%%%%%%%%%%%%%%%%%%%%%%%%%%

The basic observation at the level of the gluon 
SDE shown in 
\fig{fig:SDEs} is that, 
the ghost loop of $(d_3)$,  
due to the 
masslessness of its ingredients, furnishes  ``unprotected'' 
logarithms, \ie  terms of the type $\ln q^2$, which diverge as $q^2 \to 0$. 
Instead, gluonic loops contain infrared finite 
gluon propagators, and, therefore, 
give rise to contributions that remain
finite as $q^2 \to 0$, 
\ie they may be described in terms of 
``protected'' logarithms of the type 
$\ln (q^2+m^2)$. 

The circumstances described above may be modeled 
by 
\be 
\Delta^{-1}(q) = \underbrace{q^2 + m^2+ c_1 q^2 \ln\left(\frac{q^2 + \rho m^2}{\Lambda^2}\right)}_{f(q)} + c_2 q^2 \ln\left(\frac{q^2}{\Lambda^2}\right) \,, \label{Delta_IR}
\ee 
where $m$ is the gluon mass, 
$\Lambda$ the mass scale of QCD,
and $c_1$, $c_2$ and $\rho$ are constants; note that 
$\Delta^{-1}(0) = f(0) = m^2$ 

Differentiating \1eq{Delta_IR} with respect to $q^2$ we obtain
\be 
\frac{d\Delta^{-1}(q)}{dq^2} = \frac{df(q)}{dq^2} + c_2\left[ 1 + \ln\left(\frac{q^2}{\Lambda^2}\right) \right] \,. \label{gluon_der}
\ee 

The second term on the r.h.s. of \1eq{gluon_der} is infrared divergent, and necessarily dominates the behavior of the derivative of the propagator for sufficiently small $q$. Moreover, the value of the coefficient $c_2$ can be computed explicitly by expanding the ghost block $\pt^{(2)}_{\mu\nu}(q)$ of \fig{fig:SDEs} around $q = 0$ and using \1eq{sdebq}, which yields
\be 
c_2 = \frac{\alpha_s C_{\rm A} {\widetilde Z}_1^2 F^2(0) }{48\pi} \,. \label{c2}
\ee 
Therefore, $d\Delta^{-1}(q)/dq^2$ has the asymptotic behavior
\be 
\lim_{q \to 0} \frac{d\Delta^{-1}(q)}{dq^2} = \left[\frac{\alpha_s C_{\rm A} {\widetilde Z}_1^2 F^2(0) }{48\pi}\right]\ln\left(\frac{q^2}{\Lambda^2}\right) \,, \label{Delta_div}
\ee 
which diverges to $-\infty$ as $q\to 0$.
Now, since the gluon propagator is a decreasing function in the ultraviolet, we have that
$d\Delta^{-1}(q)/dq^2$ is positive for large momenta.
Therefore, 
there must exist a special momentum, 
denoted by $q_\star$, such that 
$[d\Delta(q)/dq^2]_{q=q_\star}=0$, which corresponds to a maximum\footnote{Note that $d\Delta^{-1}(q)/dq^2$ is an increasing function, since it is negative in the infrared and positive in the ultraviolet, \ie $d^2\Delta^{-1}(q)/d(q^2)^2> 0$. Therefore, assuming that $d\Delta^{-1}(q)/dq^2$ only crosses zero once, $q = q_\star$ must be a maximum of $\Delta(q)$.} of $\Delta(q)$.

The maximum of $\Delta(q)$, 
predicted by means of the simple arguments  
presented above, is 
observed in lattice simulations of the gluon 
propagator~\cite{Bogolubsky:2007ud,Bogolubsky:2009dc,Aguilar:2021okw}. In particular, it is clearly 
visible in \fig{fig:gluon_zoom}, where 
the data from 
the two largest volume lattice setups of~\cite{Bogolubsky:2007ud} are shown. 
The red dashed lines represent smooth functions,  fitted to each of the data sets, in the window $q\in [0,0.5]$ GeV. For each of the volumes considered,  $V = 72^4$ (left panel) and $V = 80^4$ (right panel), 
the estimate obtained for $q_{\star}$ 
is $q_{\star} = 140$ MeV.

It is interesting to observe in passing that 
the existence of a maximum of $\Delta(q)$ has an   
interesting implication on the form of the 
spectral function of the gluon propagator~\cite{Cyrol:2018xeq,Binosi:2019ecz,Kern:2019nzx,Horak:2021pfr,Horak:2021syv,Horak:2022myj}.
In particular, the standard K\"all\'en-Lehmann representation~\cite{Kallen:1952zz,Lehmann:1954xi} 
states that 
\be
\Delta (q) = \int_0^{\infty} \!\! d \lambda^2 \, \frac{\rho (\lambda^2)}{q^2 + \lambda^2}\,,
\label{KL}
\ee
where  $\rho (\lambda^2)$ is the gluon spectral function 
(with a factor $1/\pi$ absorbed in it). 
Thus, the differentiation of both sides 
of \1eq{KL} with respect to $q^2$ yields 
\be
\frac{ d\Delta(q) }{dq^2} =
- \int_0^{\infty} \!\! d \lambda^2 \, \frac{\rho (\lambda^2)}{(q^2+ \lambda^2)^2}\,.
\label{derspec}
\ee
Then, from \1eq{derspec} follows that 
the existence of a maximum for $\Delta (q)$ at $q=q_{\star}$
leads necessarily to the violation of reflection  positivity~\cite{Osterwalder:1973dx,Osterwalder:1974tc,Alkofer:2000wg,Cornwall:2013zra}, because the condition    
\be
\int_0^{\infty} \!\! d \lambda^2 \, \frac{\rho (\lambda^2)}{(q^2_{\star} + \lambda^2)^2} = 0\,,  
\label{specmax}
\ee
may be fulfilled only if $\rho(\lambda^2)$ 
reverses its sign. 
Note that an analogous argument based on the existence of an inflection point 
has been presented recently in~\cite{Ding:2022ows}.

\begin{figure}[h]
\centering
\includegraphics[width=0.8\textwidth]{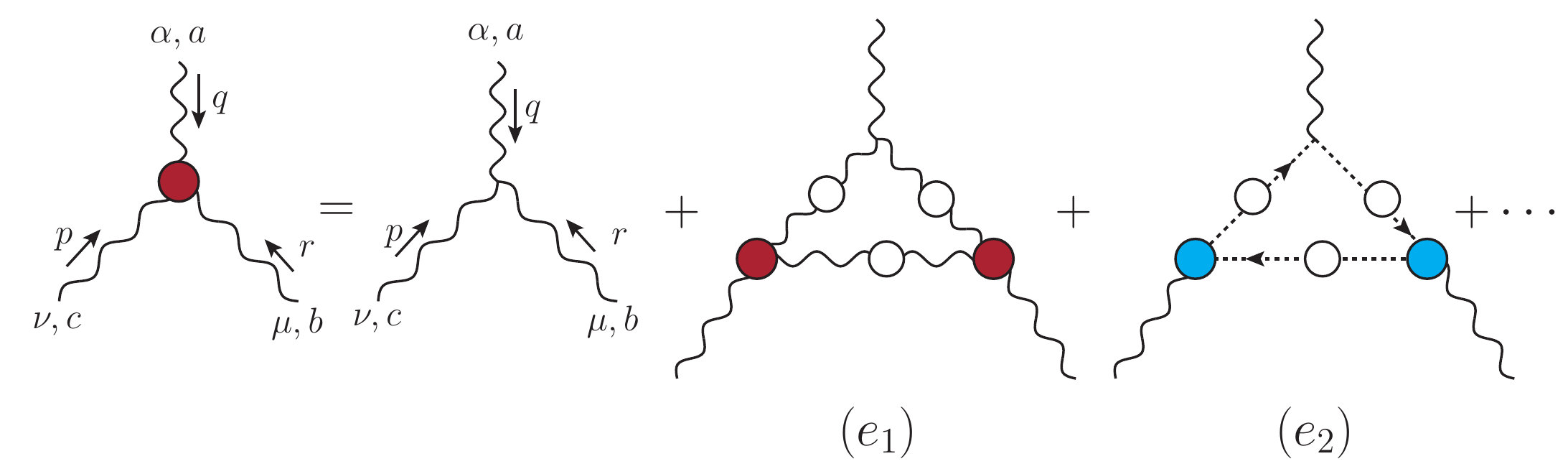}
\caption{ The SDE of the three-gluon vertex at the one-loop dressed level. The diagrams $(e_1)$ and $(e_2)$ are the 
gluon and the ghost triangle
contributions entering in the skeleton expansion of the three-gluon vertex. }
\label{fig:supression}
\end{figure}

\begin{figure}[ht!]
  \centering
  \includegraphics[width=0.475\textwidth]{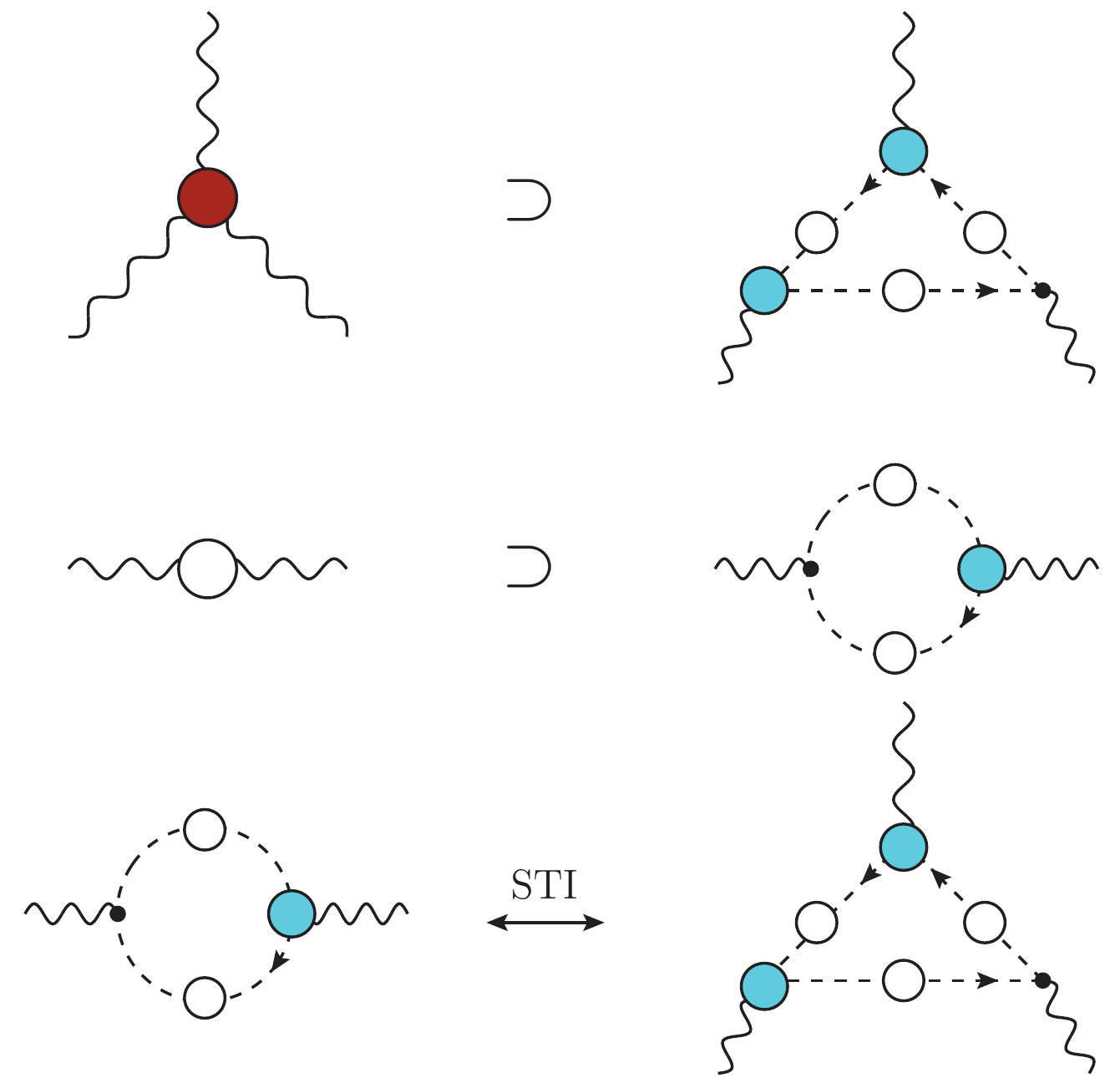}
  \caption{ The ghost triangle present in the three-gluon vertex SDE (top) and the ghost loop contributing to the gluon propagator in the corresponding equation (middle). The infrared divergences arising from these diagrams are connected through the Slavnov-Taylor identity (STI) of \1eq{st1_conv}, as shown schematically in the bottom panel.}
\label{fig:STI_to_SDE}
\end{figure}

Turning to the three-gluon vertex, it is well known 
that the corresponding ghost loops 
induce characteristic features to the form factors 
associated with its classical (tree-level) tensors. 
There are two 
complementary continuum 
descriptions of the dynamics that determine 
the behavior of these form  
factors: ${(\it i)}$  the SDE of the three-gluon vertex~\cite{Huber:2012kd,Blum:2014gna,Eichmann:2014xya,Williams:2015cvx}, 
depicted diagrammatically in \fig{fig:supression}, 
and ${(\it ii)}$ 
the STI of \1eq{st1_conv}~\cite{Aguilar:2013vaa}, which, in the limit 
of vanishing gluon momentum, and when the 
displacement function and the 
ghost sector are neglected, yields the approximate WI 
\be 
{\fatg}_{\alpha\mu\nu}(0,r,-r) \approx \frac{\partial \Delta^{-1}_{\mu\nu}(r)}{\partial r^\alpha} \,,
\label{stiappr}
\ee
which transmits the properties of the 
propagator 
derivative to the vertex form factors, as shown schematically in \fig{fig:STI_to_SDE}. 

In the simplified kinematic circumstances where only 
a single representative momentum is considered, to be denoted by $r$, the conclusions drawn by either method may be qualitatively described in 
terms of a simple model, namely  
\be
L(r) = b_0 + b_{\rm gl} \ln \left(\frac{r^2+m^2}{\Lambda^2}\right)
+ b_{\rm gh} \ln \left(\frac{r^2}{\Lambda^2}\right)\,,
\label{Lsupp}
\ee
where $L(r)$ denotes the particular combination of form factors, such that, at tree level,
$L_0(r)=1$, and $b_0$, $b_{\rm gl}$, and $b_{\rm gh}$ are positive constants.
The model in \1eq{Lsupp} encompasses two important cases studied on the lattice~\cite{Athenodorou:2016oyh,Duarte:2016ieu,Boucaud:2017obn,Aguilar:2019uob}, namely 
${(\it i)}$  \emph{the soft  gluon limit},  $L (r)\to \Ls(r)$, corresponding to the kinematic choice 
$q\to 0 \,,\quad p=-r\,, \quad \theta:=\widehat{pr}=\pi$, 
defined in \1eq{Lsg_def}, and 
${(\it ii)}$ 
  \emph{totally symmetric limit}, $L (r)\to L_{sym}(r)$, 
  corresponding to 
$q^2 = p^2= r^2 \,, \quad   \theta:= \widehat{q r} = \widehat{q p} = \widehat{r p} = 2\pi/3$. 

Upon inspection of \1eq{Lsupp} we note 
that, as \mbox{$r \to 0$}, the term with the unprotected logarithm will eventually dominate,
forcing $L(r)$ to reverse its sign (zero crossing), and finally 
display a logarithmic divergence, \mbox{$L(0)\to -\infty$}. Given that, 
in practice,  $b_{\rm gl}$ is considerably larger than $b_{\rm gh}$, 
the unprotected logarithm overtakes 
the protected one rather deep in the
infrared: the location of the zero-crossing is at about $160$ MeV~\cite{Boucaud:2017obn}. 
Consequently, in the intermediate region of momenta,
which is considered relevant for the onset of nonperturbative dynamics, we have \mbox{$L(r) < 1$};  
this effect is known in the literature as the infrared suppression of the three-gluon vertex.

%%%%%%%%%%%%%%%%%%%%%%%%%%%%%%%%%%
% Figure 16  - 3g SDE
%%%%%%%%%%%%%%%%%%%%%%%%%%%%%%%%%%
%%%%%%%%%%%%%%%%%%%%%%%%%%%%%%%%%%

\begin{figure}[h]
\centering
\includegraphics[width=0.475\textwidth]{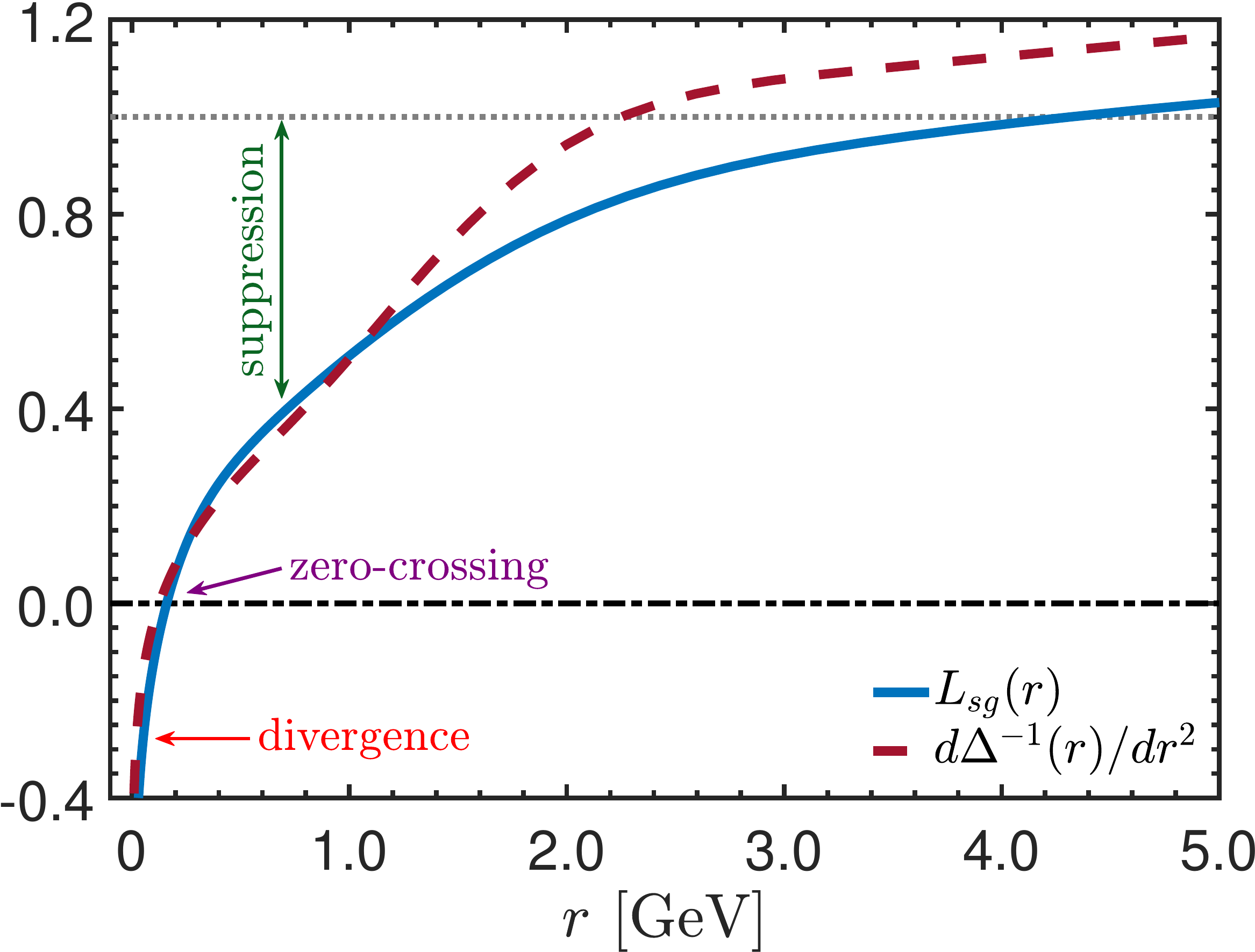}
\caption{ Comparison of $\Ls(r)$ (blue continuous) from \fig{fig:Lsg} and $d\Delta^{-1}(r)/dr^2$ (red dashed) resulting from the fit for $\Delta(r)$ of \fig{fig:gluon_prop}. Note that both display the characteristic features of infrared suppression with respect to their tree-level values (which is $1$ for both quantities), zero-crossing, and logarithmic divergence at the origin.}
\label{fig:gluon_der_and_Lsg}
\end{figure}

Most importantly, 
the special features of infrared suppression, zero-crossing, and 
logarithmic divergence at the origin 
have been corroborated through a variety of lattice results~\cite{Cucchieri:2008qm,Athenodorou:2016oyh,Duarte:2016ieu,Boucaud:2017obn,Sternbeck:2017ntv,Aguilar:2019uob,Aguilar:2021lke}, as shown, \eg in \fig{fig:Lsg}. The central curve of this figure is presented as 
the blue line 
in \fig{fig:gluon_der_and_Lsg}, where the aforementioned 
characteristics have been explicitly marked for the benefit of the reader. Note the close proximity of the blue curve 
to the 
$d\Delta^{-1}(r)/dr^2$  (red dashed line), especially below 
1 GeV.

%%%%%%%%%%%%%%%%%%%%%%%%%%%%%%%%%%
% Figure 17 - IR features - Lsg and dDelta^{-1}(r)dr^2
%%%%%%%%%%%%%%%%%%%%%%%%%%%%%%%%%%
%%%%%%%%%%%%%%%%%%%%%%%%%%%%%%%%%%

%%%%%%%%%%%%%%%%%%%%%%%%%%%%%%%%%%
% Figure 18  - Ghost loops
%%%%%%%%%%%%%%%%%%%%%%%%%%%%%%%%%%
%%%%%%%%%%%%%%%%%%%%%%%%%%%%%%%%%%

We end this section by pointing out that, 
in the case of Yang-Mills in $d = 3$~\cite{Gross:1980br,Jackiw:1980kv,Appelquist:1981vg,Deser:1982vy,Cornwall:1988ad,Cornwall:1992cu,Alexanian:1995rp,Cornwall:1995ac,Cornwall:1996jb,Buchmuller:1996pp,Jackiw:1997jga,Cornwall:1997dc,Karabali:1998yq,Eberlein:1998yk,Aguilar:2010zx,Aguilar:2013vaa,Huber:2016tvc,Corell:2018yil}, the situation is qualitatively similar to the one described above, but the divergences induced due to the masslessness of the ghost are stronger. Speciﬁcally, as may be already established at the level of a simple one-loop calculation ~\cite{Aguilar:2010zx}, the 
first derivative of the gluon propagator diverges at the origin as $1/q$ rather than a $\ln q^2$. 
Consequently, the corresponding effects are signiﬁcantly enhanced; in 
particular, the maximum of the gluon propagator is 
considerably more pronounced, becoming 
plainly visible on the lattice~\cite{Cucchieri:2009xxr}. 
Similarly, an abrupt negative divergence is observed in the corresponding vertex form factors~\cite{Cucchieri:2006tf,Maas:2020zjp}.

%%%%%%%%%%%%%%%%%%%%%%%%%%%%%%%%%%%%%%%%%%%%%%%%%%%%%%%%
\section{Ward identity displacement of the three-gluon vertex}\label{sec:widis3g}
%%%%%%%%%%%%%%%%%%%%%%%%%%%%%%%%%%%%%%%%%%%%%%%%%%%%%%%%

In complete analogy to the case of the ghost-gluon vertex discussed in subsection~\ref{subsec:widis}, the WI satisfied by the pole-free part of the three-gluon vertex is also displaced in the presence of longitudinally coupled massless poles. Quite importantly,  the associated  
displacement function, $\Cfat(r)$, {\it coincides}  with the 
BS amplitude that controls the formation of a (colored) 
scalar bound state with vanishing mass 
out of a gluon pair. The displacement of the 
WI circumvents the seagull cancellation involving the 
gluon propagator [\ie $f=\Delta$ in \1eq{sea}], furnishing 
to the $g_{\mu\nu}$ component the mass originating from graphs $(d_1)$ and $(d_4)$ in \fig{fig:SDEs}. 
In addition, it permits the indirect determination 
of the displacement function $\Cfat(r)$ from the lattice; 
this is particularly important, given that, 
by virtue of \1eq{eq:transvp}, the lattice ``observables'' do not perceive 
directly the presence of the massless poles.

The starting point of the analysis is the STI satisfied by the three-gluon vertex, $\fatg_{\alpha\mu \nu}(q,r,p)$, given by \1eq{st1_conv}. In order to eliminate the poles in $r$ and $p$, thus isolating the displacement of the WI originating from the pole in the channel $q$, we contract that equation with $P_{\mu'}^\mu(r)P_{\nu'}^\nu(p)$. Note that this procedure also eliminates any longitudinal pole terms in the $H_{\sigma\mu}(p,q,r)$ and $H_{\sigma\nu}(r,q,p)$.

Then, we decompose $\fatg_{\alpha\mu\nu}(q,r,p)$ into pole-free and longitudinally coupled massless pole parts, as in \1eq{fullgh}, and use \1eq{eq:PPG2}, to obtain
\be 
P_{\mu'}^\mu(r)P_{\nu'}^\nu(p)\left[ q^\alpha\g_{\alpha\mu\nu}(q,r,p) +  g_{\mu\nu}C_1(q,r,p) + q_\mu q_\nu C_5(q,r,p) \right] = P_{\mu'}^\mu(r)P_{\nu'}^\nu(p) R_{\nu\mu}(p,q,r) \,, \label{PPSTI}
\ee 
where
\be 
R_{\nu\mu}(p,q,r) := F(q)\left[ \Delta^{-1}(p) P_\nu^\sigma(p) H_{\sigma\mu}(p,q,r) - \Delta^{-1}(r) P_\mu^\sigma(r) H_{\sigma\nu}(r,q,p) \right] \,.
\ee 

At this point, we expand \1eq{PPSTI} around $q = 0$ and match coefficients of equal orders. At zeroth order in this expansion we obtain immediately that 
\be
C_1(0,r,-r) = 0 \,,
\label{C1_0b}
\ee
in exact analogy to \1eq{Cant}. Note that we have arrived once again at the result of \1eq{C1_0}, but through an entirely different path:
while \1eq{C1_0} is enforced by the Bose symmetry of the three-gluon vertex, \1eq{C1_0b} is a direct consequence of the STI that this vertex satisfies.

We next gather the terms 
in the expansion of \1eq{PPSTI} that are of first order in $q$. Evidently, the term $C_5$ does not contribute to this order. Then, the expansion leads to the appearance of derivatives of the gluon propagator, in analogy to \1eq{WIdis}, but also of the ghost-gluon kernel.  Specifically, we obtain for the WI of the three-gluon vertex and its displacement the expression
\be
 \Ls(r)= F(0)\left\{\widetilde{Z}_1 \frac{d\Delta^{-1}(r)}{dr^2} + \frac{\w(r)}{r^2} \Delta^{-1}(r)\right\}- \Cfat(r)\,.
\label{WIdis3g} 
\ee
In the above equation, $\Ls(r)$ is the form factor of the three-gluon vertex defined in \1eq{Lsg_def} and with lattice results shown in \fig{fig:Lsg}, while $\w(r)$ is a particular derivative of the ghost-gluon kernel, namely~\cite{Aguilar:2020yni,Aguilar:2021uwa}
\be 
\w(r) = - \frac{1}{3r^2}P^{\mu\nu}(r)\left[\frac{\partial H_{\nu\mu}(p,q,r)}{\partial q^\alpha } \right]_{q=0} \,. \label{HKtens}
\ee 
For the detailed derivation of \1eq{WIdis3g}, we refer to \cite{Aguilar:2021uwa,Papavassiliou:2022wrb}.

In the following section, we will use \1eq{WIdis3g} to determine the displacement amplitude $\Cfat(r)$ from lattice inputs. To this end, we must first pass to Euclidean space, where we note that
\be
\Cfat_{\s {\rm E}}(r^2_{\s {\rm E}}) = - \left. \Cfat(r)\right\vert_{r^2 = - r_{\s {\rm E}}^2} \,, \label{Cfat_Euc} 
\ee
with the extra sign originating from the fact that $\Cfat$ is a derivative [see \1eq{eq:theCs}].
Then, suppressing the indices ``${\rm E}$'' and solving for $\Cfat(r^2)$,  we obtain the central relation 
\be
\Cfat(r) = \Ls(r) - F(0)\left\{\frac{\w(r)}{r^2}\Delta^{-1}(r) + \widetilde{Z}_1 \frac{d\Delta^{-1}(r)}{dr^2} \right\} \,.
\label{centeuc}
\ee

For the determination of $\Cfat(r)$, 
we use lattice inputs for all the quantities that appear on the r.h.s. of \1eq{centeuc}, with the
exception of the function $\w(r)$, which will be computed from the SDE satisfied by the ghost-gluon kernel, derived and analyzed in the next section.

%%%%%%%%%%%%%%%%%%%%%%%%%%%%%%%%%%%%%%%%%%%%%%%%%%%%%%%%
\section{The ghost-gluon kernel contribution to the Ward identity}\label{sec:WSDE}
%%%%%%%%%%%%%%%%%%%%%%%%%%%%%%%%%%%%%%%%%%%%%%%%%%%%%%%%

In this section, we derive the SDE 
that determines the key function 
$\w(r)$; the resulting SDE will be solved using 
lattice inputs for the various quantities 
entering in it. In addition, 
the infrared behavior of $\w(r)$ will 
be analyzed in detail, following an analytic procedure.

\begin{figure}[ht!]
  \centering
  \includegraphics[width=\textwidth]{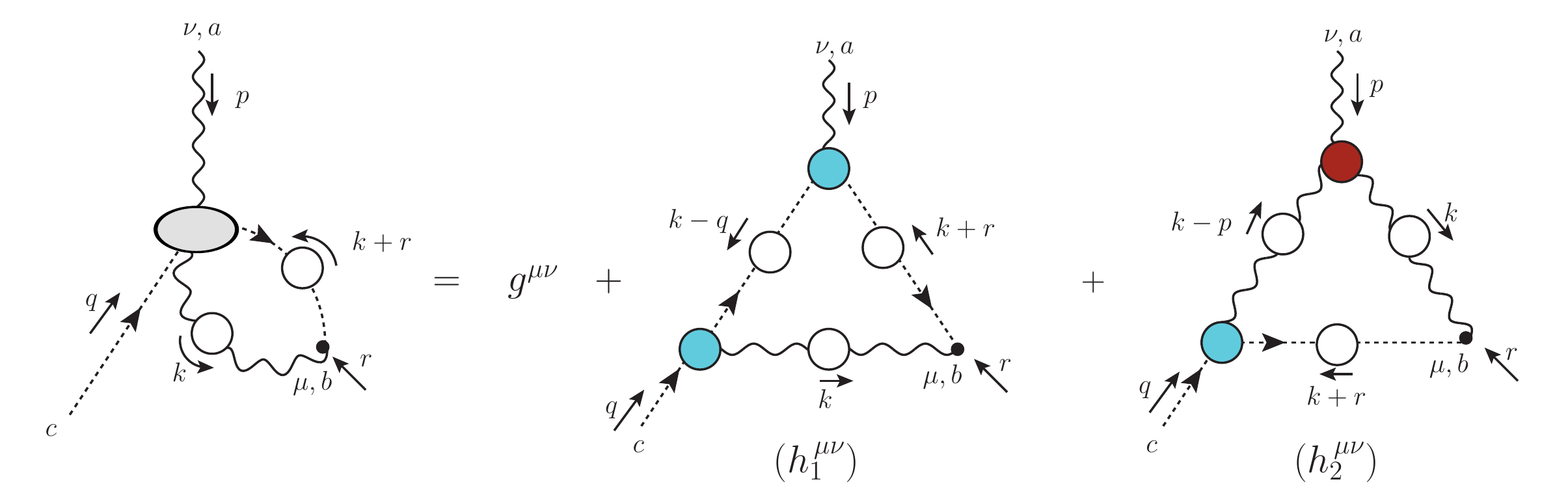}	
  \caption{ SDE for the ghost-gluon scattering kernel, $H_{\mu\nu}(r,q,p)$. We omit a diagram containing a 1PI four-point function. }
\label{fig:H_SDE}
\end{figure}

Our discussion starts with the SDE of the ghost-gluon kernel, $H_{\mu\nu}(r,q,p)$, shown diagrammatically in \fig{fig:H_SDE}, from which $\w(r)$ can be obtained using \1eq{HKtens}.

Note that the similarity between the diagrams 
shown in \fig{fig:H_SDE} and those in the 
bottom panel of \fig{fig:ghost_SDE}, depicting the  
SDE of the ghost-gluon vertex, is a simple  
reflection of the fundamental 
STI relating 
the ghost-gluon kernel with the ghost-gluon vertex,
\be 
\g_\nu(r,q,p) = r^\mu H_{\mu\nu}(r,q,p) \,. \label{H_to_Gamma}
\ee 
Specifically, \1eq{H_to_Gamma}
is preserved by the SDEs of $\g_\nu(r,q,p)$ and $H_{\mu\nu}(r,q,p)$; indeed, contraction of each diagram $(h_i^{\mu\nu})$ of \fig{fig:H_SDE} by $r^\mu$ yields the corresponding diagram $(g_i^\nu)$ of \fig{fig:ghost_SDE} (up to a shift of $k\to-k-r$ for $i = 1$, introduced to simplify certain expressions). 
Note that, in \fig{fig:H_SDE}, the diagram 
corresponding to the $(g_3)$ of \fig{fig:ghost_SDE}
has been omitted, for the reason explained in the item 
({\it i}) of Section~\ref{sec:ghost_dyn}.

%%%%%%%%%%%%%%%%%%%%%%%%%%%%%%%%%%
% Figure 19  - H_munu SDE
%%%%%%%%%%%%%%%%%%%%%%%%%%%%%%%%%%
%%%%%%%%%%%%%%%%%%%%%%%%%%%%%%%%%%

It is well known that, in the Landau gauge, 
the momentum $q$ of the ghost field in $H_{\mu\nu}(r,q,p)$ factors out of its quantum corrections~\cite{Marciano:1977su}, 
allowing us to write~\cite{Aguilar:2018csq,Aguilar:2020yni,Aguilar:2021uwa}
\be 
H_{\mu\nu}(r,q,p) = g_{\mu \nu} + q^\alpha K_{\mu\nu\alpha}(r,q,p) \,, \label{H_from_K}
\ee 
where no particular assumptions are made about the 
structure of the function $K_{\mu\nu\alpha}(r,q,p)$.  
Following \1eq{HKtens}, we 
differentiate the r.h.s. of \1eq{H_from_K} 
with respect to $q$ and subsequently set 
$q=0$, to obtain 
\be 
\w(r) = - \frac{1}{3}r^\alpha P^{\mu\nu}(r)K_{\mu\nu\alpha}(r,0,-r) \,. \label{w_from SDE}
\ee 

Lastly, the \emph{finite} renormalization of $\w$ proceeds through the use of \2eqs{Zs_def}{asym_def}, which leads to the appearance of an overall factor of ${\widetilde Z}_1$ in the equations.

The outcome of the above steps is that $\w(r)$ can be written as
\be
\w(r) = \w_1(r) + \w_2(r) \,, \label{W_conts}
\ee 
where the $\w_i(r)$ are the contributions originating from the diagrams $(h_i^{\mu\nu})$ of \fig{fig:H_SDE}, respectively, and read
\begin{eqnarray}
\w_1(r) &=& \frac{ \lambda {\widetilde Z}_1 }{3} \int_k \Delta(k) D(k) D(k+r) (r\cdot k ) f(k,r)B_1( k+r, - k , -r )B_1(k,0,-k) \,, \nonumber\\
\w_2(r) &=& \frac{ \lambda {\widetilde Z}_1 }{3} \int_k \Delta(k) \Delta(k+r) D(k+r) B_1(k+r,0,-k-r) \IW(-r, -k, k+r) \,, \label{W_diags}
\end{eqnarray}
where $f(k,r)$ is given by \1eq{fqk_def}, and we define the specific contribution of the three-gluon vertex to the kernel of $\w(r^2)$ as
\begin{align}
   \IW(q,r,p) &:=  \frac 1 2  (q-r)^\nu \overline\Gamma^\alpha_{\alpha\nu}(q,r,p) \,.
\label{eq:IWdef}
\end{align}  
Note that, from \1eq{eq:IWdef} and the Bose symmetry of the $\gbar_{\alpha\mu\nu}(q,r,p)$ under the exchange $\{q,\alpha\}\leftrightarrow \{r,\mu\}$, it follows that
\be 
\IW(q,r,p) = \IW(r,q,p) \,.
\ee 

At this point, by capitalizing on the planar degeneracy of $\gbar_{\alpha\mu\nu}(q,r,p)$ discussed in Sec.~\ref{sec:planar_deg}, we obtain a compact, and yet accurate, approximation for $\IW$. Specifically, using \1eq{eq:compact}, we find
\begin{align}
   \IW(q,r,p) &\approx \IW^0(q,r,p)\Ls(s)  \,,
\label{eq:IWcompact}
\end{align}  
where $\IW^0(q,r,p)$ is the tree-level value of $\IW$, given by
\be 
\IW^0(q,r,p) := \frac{2f(q,r)}{p^2}\left[ 2 q^2 r^2 - (q^2 + r^2)(q\cdot r) - (q\cdot r)^2\right] \,.
\ee 
We remark that the approximation given by \1eq{eq:IWcompact} becomes exact in the limit $p = 0$.

Using the above approximation for $\IW$, the contribution $\w_2(r)$ reads
\begin{align} 
\w_2(r) =& \frac{ 2\lambda {\widetilde Z}_1 }{3} \int_k \Delta(k) \frac{\Delta(k+r) D(k+r)}{(k+r)^2} B_1(k+r,0,-k-r) f(k,r)\nonumber\\
& \times \left[ 2 r^2 k^2 - (r^2 + k^2)(r\cdot k) - (r\cdot k)^2 \right]\Ls({\hat s}) \,, \label{W2_compact}
\end{align}
where we now have ${\hat s}^2 = r^2 + k^2 + (r\cdot k)$.

Lastly, we transform $\w_1$ of \1eq{W_diags} and $\w_2$ of \1eq{W2_compact} to Euclidean space to obtain the final expression to be used for the numerical determination of $\w$,
\begin{align}
\w_1(r) &= - \frac{r \alpha_s C_{\rm A}{\widetilde Z}_1}{12\pi^2}\!\int_0^\infty \! dk^2 k \Delta(k) F(k)B_1(k^2,k^2,\pi) \!\int_0^\pi \! d\phi s_\phi^4 c_\phi \frac{F(\sqrt{z})}{z}B_1(z,r^2,\chi) \,, \nonumber\\
\w_2(r) &= - \frac{ r \alpha_s C_{\rm A}{\widetilde Z}_1}{6\pi^2}\!\int_0^\infty \! dk^2\, k^3 \Delta(k) \!\int_0^\pi \! d\phi \, s_\phi^4 \Delta(\sqrt{z})B_1(z,z,\pi)\frac{F(\sqrt{z})}{z^2}\left[ k r(2 + c_\phi^2) - z c_\phi\right]\nonumber\\
& \quad \times \Ls\left(r^2 + k^2 + r k c_\phi\right) \,, \label{W_diags_euc}
\end{align}
where $z$ has been defined below \1eq{ab_euc} and
\be 
\chi := \cos^{-1}\left[-\frac{( r + k c_\phi)}{\sqrt{z}}\right] \,. \label{chi_def}
\ee 

We emphasize that we have used into the SDEs of both $B_1$ and $\w$, given by \2eqs{B1_SDE_euc}{W_diags_euc}, respectively, the same approximation for the three-gluon vertex, namely \1eq{eq:compact}. Therefore, our analyses of $B_1$ and $\w$ are self-consistent, in the sense that the STI in \1eq{H_to_Gamma} is strictly preserved.

Before embarking on the numerical determination of $\w(r)$ for the entire range of Euclidean momenta, we discuss the infrared behavior of this function, 
and demonstrate an important  
self-consistency proof involving $\Cfat(r)$.

Specifically, 
as discussed in Sec.~\ref{sec:ghost_loops}, the $\Ls(r)$ and $d\Delta^{-1}(r)/dr^2$ that appear in \1eq{centeuc} are infrared divergent, due to massless ghost loops present in their SDEs. Nevertheless, the BSE solutions for the amplitude $\Cfat(r)$ are all found to be finite at \mbox{$r = 0$}, (cf.~\fig{fig:C_gl_Cgh}) ~\cite{Aguilar:2011xe,Binosi:2017rwj,Aguilar:2017dco,Aguilar:2021uwa}. Therefore, in order for the WI displacement of \1eq{centeuc} to be consistent with the finite $\Cfat(0)$ obtained from BSE solutions, the infrared divergences of the ingredients appearing in \1eq{centeuc} must cancel against each other.

Indeed, a careful analysis of diagram $(e_2)$ of \fig{fig:supression} yields
\be  
\lim_{r\to 0} \Ls(r) = \left[ \frac{ \alpha_s C_{\rm A} {\widetilde Z}_1^3 F^3(0) }{96 \pi } \right] \ln \left( \frac{r^2}{\mu^2}\right) \,, \label{L_div}
\ee
up to infrared finite terms\footnote{We note that results identical to \2eqs{Delta_div}{L_div} for the infrared divergences of $d\Delta^{-1}(r)/dr^2$ and $\Ls(r)$, respectively, have been previously derived within the Curci-Ferrari model~\cite{Barrios:2022hzr}.}. Then, assuming that only $\Ls(r)$ and $d\Delta^{-1}/dr^2$ diverge and using the asymptotic form of $d\Delta(r)/dr^2$ given in \1eq{Delta_div} into \1eq{centeuc}, we find that the divergences do not fully cancel. Therefore, the finiteness of $\Cfat(0)$ demands that the term $\w(r)/r^2$ appearing in the WI must be infrared divergent.

Now, it is evident from \1eq{W_diags_euc} that $\w(r)$ vanishes as $r \to 0$. Nevertheless, the ratio $\w(r)/r^2$ is found to diverge at the origin. Specifically, expanding \1eq{W_diags_euc} around $r = 0$, it can be shown that $\w(r)/r^2$ has the asymptotic behavior
\be
\lim_{r\to 0} \frac{ \w(r) }{ r^2 } = - \left[ \frac{ \alpha_s C_{\rm A} {\widetilde Z}_1^3 \Delta(0) F^2(0) }{96 \pi } \right] \ln \left( \frac{r^2}{\mu^2}\right)\,. \label{W_div} 
\ee
Then, combining \3eqs{W_div}{L_div}{W_div} we find that the infrared divergences in \1eq{centeuc} cancel out exactly, leaving a finite $\Cfat(0)$, in full agreement with the BSE results.

%%%%%%%%%%%%%%%%%%%%%%%%%%%%%%%%%%
% Figure 20 - Cancelation of IR divergences
%%%%%%%%%%%%%%%%%%%%%%%%%%%%%%%%%%
\begin{figure}[h]
  \centering
\includegraphics[width=1\textwidth]{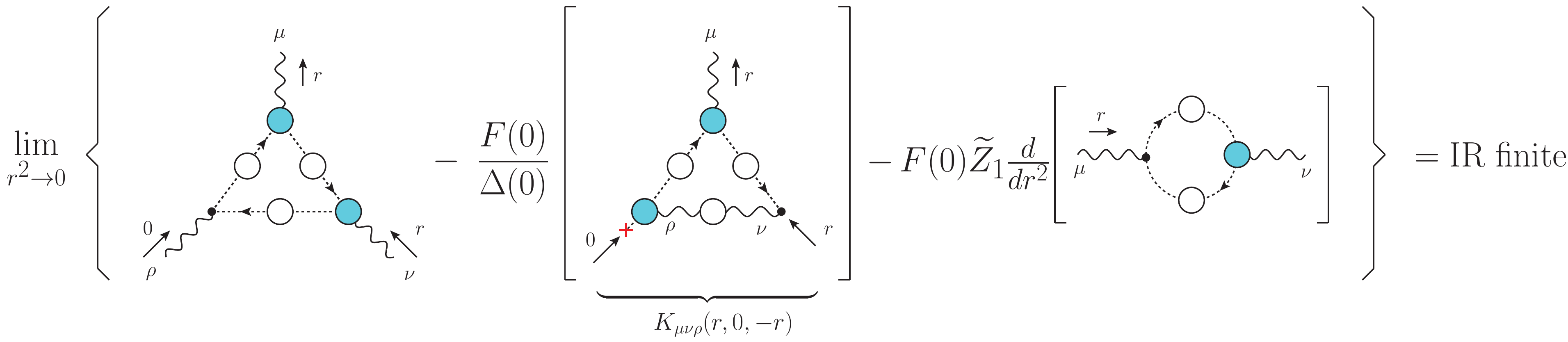}
\caption{ Diagrammatic representation of the cancellation of the infrared divergences originating from massless ghost loops in \1eq{centeuc} to yield a finite $\Cfat(0)$. The red cross indicates that the overall ghost momentum is factored out before being set to zero. }
\label{fig:divs_cancel}
\end{figure}
%%%%%%%%%%%%%%%%%%%%%%%%%%%%%%%%%%

We finish the discussion of the infrared finiteness of $\Cfat(0)$ with a remark. In the absence of the Schwinger mechanism, \ie for an identically zero $\Cfat(r)$, the infared divergences of $\Ls(r)$, $\w(r)/r^2$ and $d\Delta^{-1}(r)/dr^2$ must also cancel in \1eq{centeuc}. For instance, this cancellation can be explicitly verified at the one loop level\footnote{In the perturbative realization of \1eq{centeuc} $F(0)$ also diverges, participating in the overall cancellation of infrared divergences.}, where, evidently, $\Cfat(r) = 0$. In that case, however, the gluon propagator is also massless, causing the gluonic loops contributing to the functions entering \1eq{centeuc} to also diverge, such that the cancellation occurs among \emph{all} radiative diagrams. In contrast, in the presence of a gluon mass, the cancellation of the remaining infrared divergences takes place at the level of the ghost loops only, as illustrated diagrammatically in \fig{fig:divs_cancel}.

%%%%%%%%%%%%%%%%%%%%%%%%%%%%%%%%%%
% Figure 21 - W result
%%%%%%%%%%%%%%%%%%%%%%%%%%%%%%%%%%
\begin{figure}[h!]
\centering
\includegraphics[width=0.475\columnwidth]{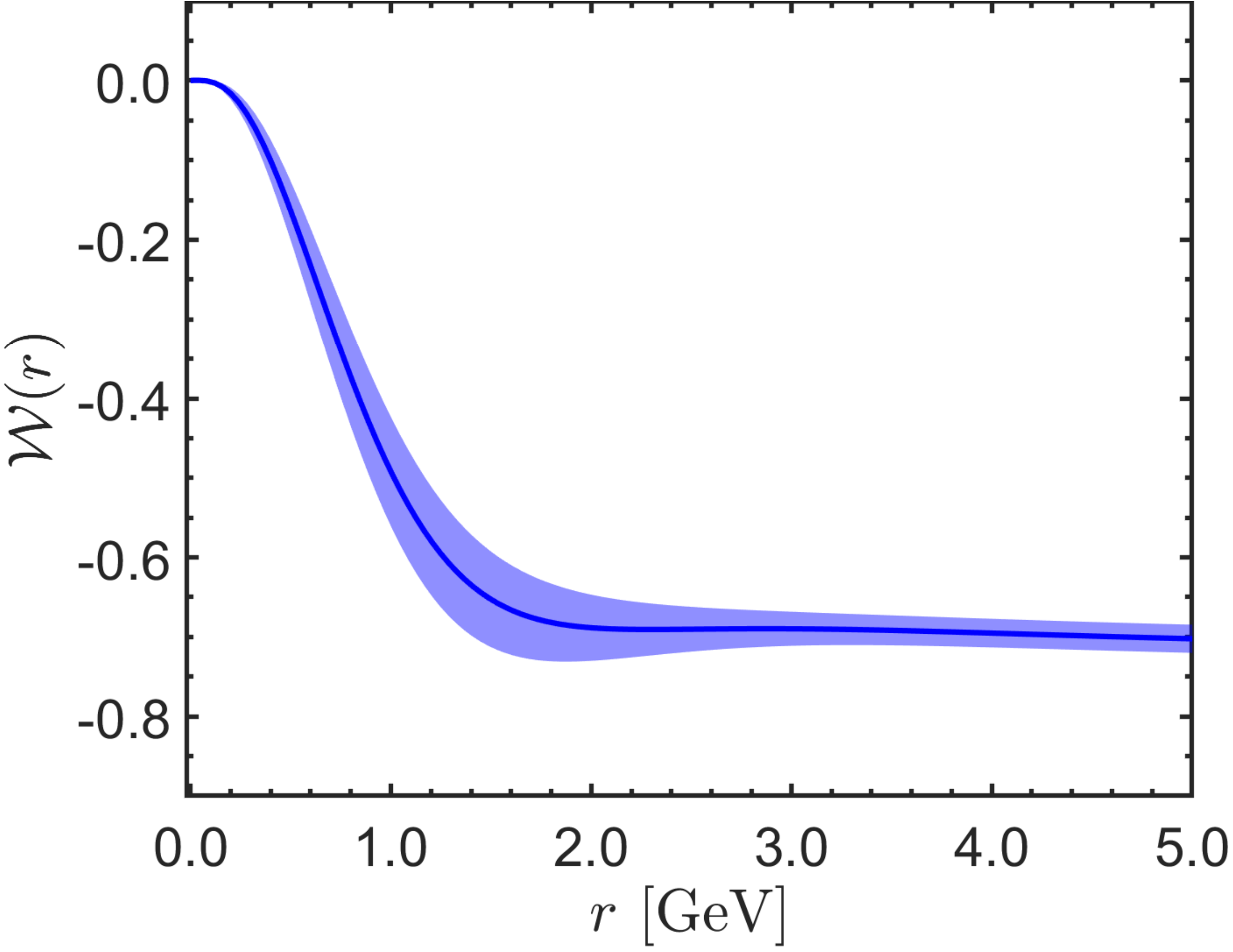} 
\caption{ $\w(r)$ obtained using the approximation \1eq{eq:IWcompact} based on the observed planar degeneracy of the three-gluon vertex in its kernel (blue solid curve) together with uncertainty estimate (blue band). }
\label{fig:W}
\end{figure}
%%%%%%%%%%%%%%%%%%%%%%%%%%%%%%%%%%

We now return to the numerical determination of $\w(r)$ 
from \1eq{W_diags_euc}. To this end, we 
employ the fits to lattice data of~\cite{Aguilar:2021lke} for $\Delta(q)$ and $\Ls(q)$ shown in Figs.~\ref{fig:gluon_prop} and \ref{fig:Lsg}, respectively, and the SDE solution for  $F(q)$ shown in the left panel of \fig{fig:B1_genkin}. All of the fits employed are constructed so as to reproduce the correct 
ultraviolet behavior 
of the respective Green's functions. For the value of the coupling in the asymmetric MOM scheme we employ $g^2 = 4\pi\alpha_s$, with $\alpha_s(4.3 \text{ GeV}) = 0.27$, as determined in the lattice study of~\cite{Boucaud:2017obn}. Lastly, for $B_1$ we use the SDE result of Sec.~\ref{sec:ghost_dyn}, shown in the right panel of \fig{fig:B1_genkin}, which reproduces accurately the available lattice data for the ghost-gluon vertex. 

Using the above ingredients into \1eq{W_diags_euc} we obtain the $\w(r)$ shown as the blue solid curve in \fig{fig:W}. 
The blue band in \fig{fig:W} represents the error estimate on our results; the procedures 
followed to obtain it are described in detail in~\cite{Aguilar:2022thg}.

%%%%%%%%%%%%%%%%%%%%%%%%%%%%%%%%%%%%%%%%%%%%%%%%%%%%%%%%
\section{Displacement function from lattice inputs}\label{sec:wilat}
%%%%%%%%%%%%%%%%%%%%%%%%%%%%%%%%%%%%%%%%%%%%%%%%%%%%%%%%

In this section we determine $\Cfat(r)$ from the 
main relation given in \1eq{centeuc}.

\begin{figure}[h]
\centering
\includegraphics[width=0.475\columnwidth]{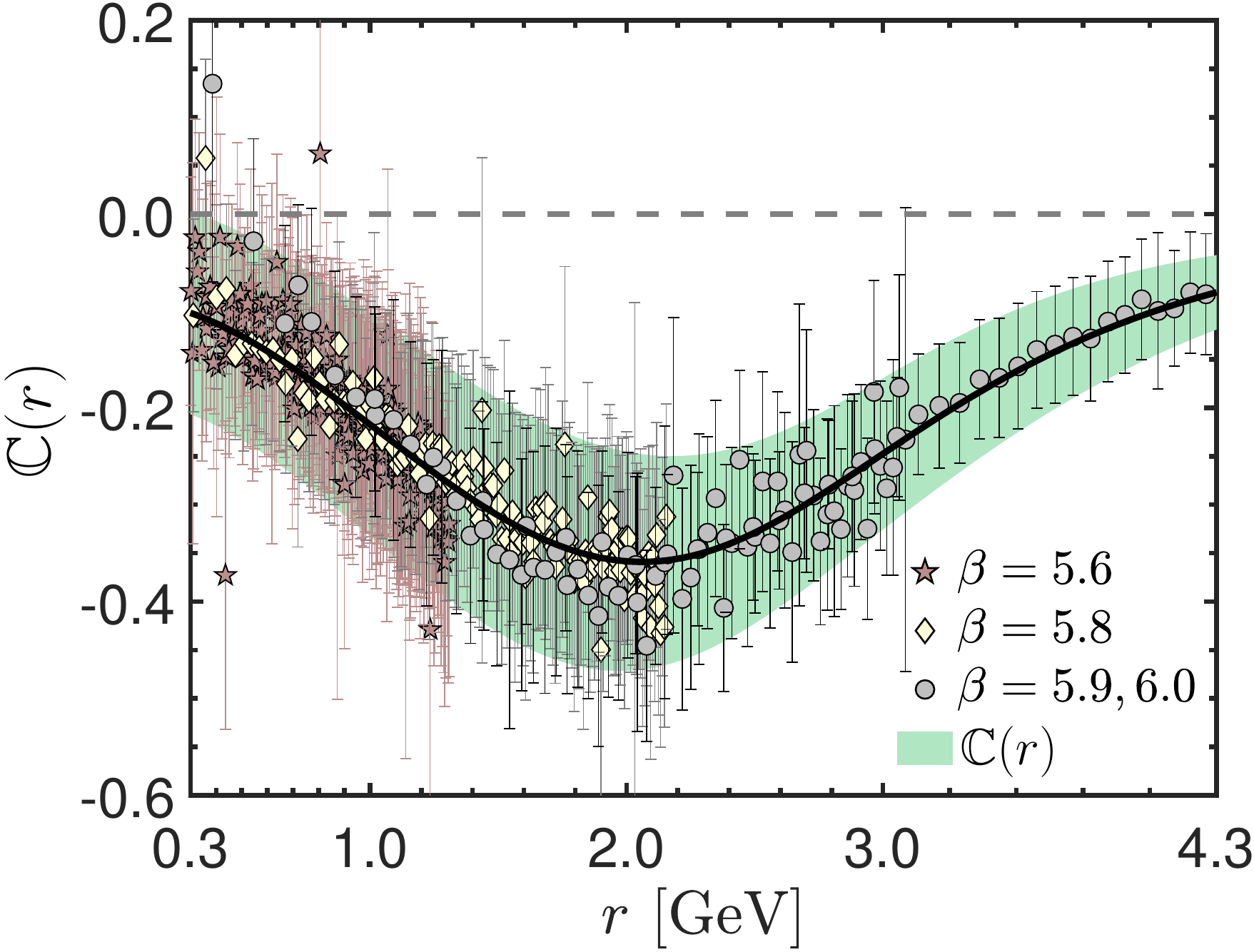}\hfil\includegraphics[width=0.475\columnwidth]{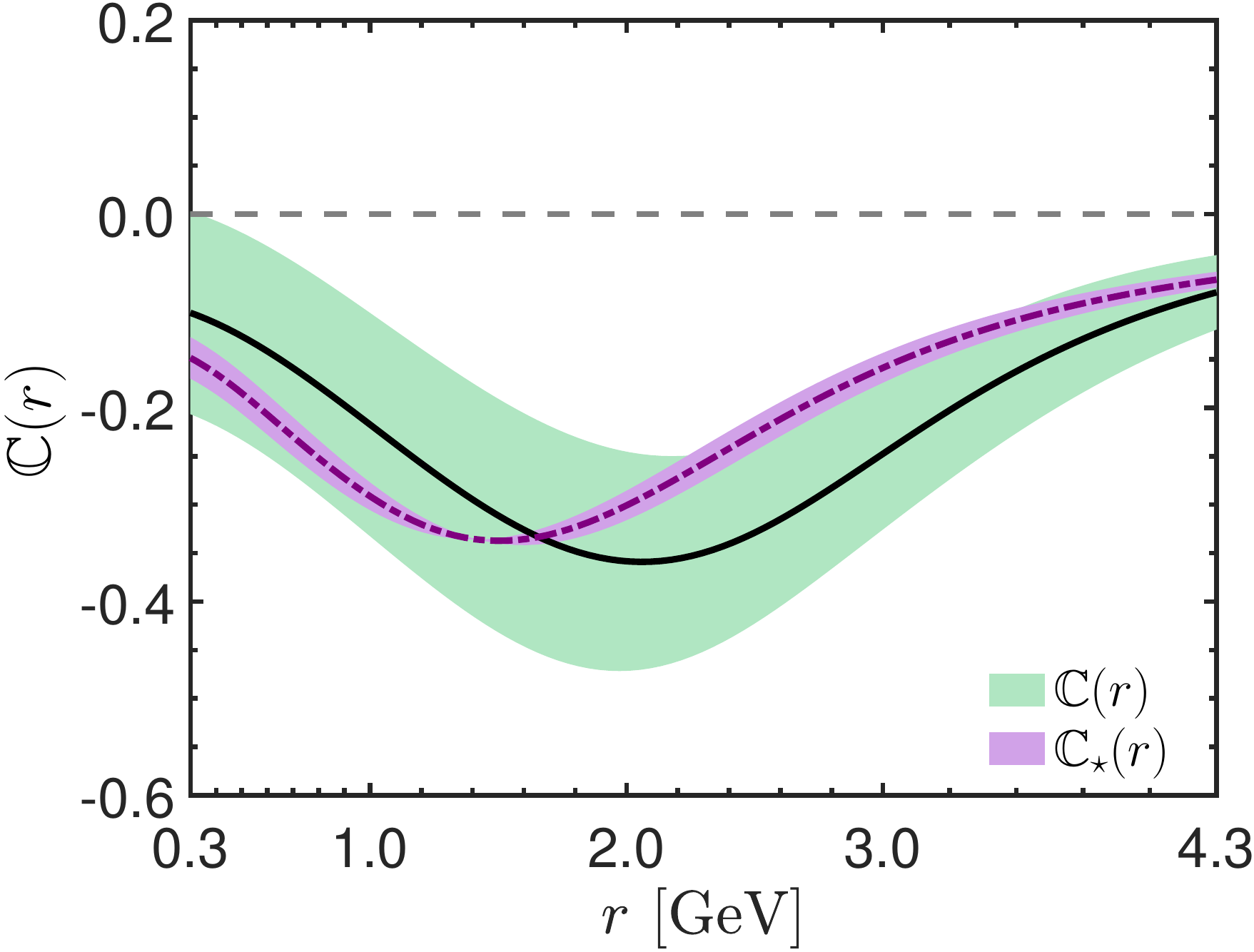}

\caption{ Left: Result for $\Cfat(r)$ (black continuous line) obtained from \1eq{centeuc} using the $\w(r)$ shown in \fig{fig:W}, the fits to lattice data for $\Delta(r)$ and $\Ls(r)$ shown in Figs.~\ref{fig:gluon_prop} and \ref{fig:gluon_der_and_Lsg}, respectively, and the SDE solution for $F(r)$ shown in \fig{fig:B1_genkin}. The points are obtained using for $\Ls(r)$ the data in Ref.~\cite{Aguilar:2021lke}, with error bars denoting the error propagated from $\Ls$ and $\w$.  The green band is obtained by fitting the upper and lower bounds of the points and guide the eye to the typical error associated with $\Cfat(r)$. Right: The $\Cfat(r)$ of the left panel is compared to the BSE prediction $\Cfat_\star(r)$ (purple dot-dashed and error band) of \fig{fig:C_gl_Cgh}.
}
\label{fig:Cfat}
\end{figure}

For $\w(r)$ we use the result shown in \fig{fig:W}, together with the curves for $\Ls(r)$ from \fig{fig:Lsg}, $\Delta(r)$ and $d\Delta^{-1}(r)/dr^2$ from Figs.~\ref{fig:gluon_prop} and \ref{fig:gluon_der_and_Lsg}, respectively, and the $F(r)$ of \fig{fig:B1_genkin}. 
The $\Cfat(r)$ obtained is shown as a black solid curve in the left panel of \fig{fig:Cfat}. In the same panel, we show as points the estimates of $\Cfat(r)$ obtained by using into \1eq{centeuc} the lattice data points of Ref.~\cite{Aguilar:2021lke} \emph{directly}, rather than a fit. To estimate the uncertainty in the resulting $\Cfat(r)$, we combine the error estimate of $\w(r)$, represented by the blue band in \fig{fig:W}, with the statistical error of the lattice data points for $\Ls(r)$ of~\cite{Aguilar:2021lke}, and neglect the error in the gluon propagator, which is much smaller than the errors in $\Ls$ and $\w$. Then, a conservative error propagation analysis produces the error bars shown in \fig{fig:Cfat}.

%%%%%%%%%%%%%%%%%%%%%%%%%%%%%%%%%%
% Figure 22 - C lattice result
%%%%%%%%%%%%%%%%%%%%%%%%%%%%%%%%%%
%%%%%%%%%%%%%%%%%%%%%%%%%%%%%%%%%%

At this point, we quantify the significance of the $\Cfat(r)$ obtained above, in comparison to the null hypothesis result; evidently, in the absence of the 
Schwinger mechanism, this latter quantity, to be denoted by $\Cfat_0$ in what follows, 
vanishes identically, 
namely $\Cfat_0 =0$.
To this end, we first compute the $\chi^2$ of our points as
\be
\chi^2 = \sum_{i} \frac{\left[\Cfat(r_i)-\Cfat_0(r_i)\right]^2}{\epsilon_{\s{\Cfat(r_i)}}^2} \,, \label{chi2_def}
\ee 
\ie the null hypothesis is taken as the estimator for our data. The sum runs over the $n_r = 515$ indices $i$ such that $r_i \in [0.3, 4.3]$ GeV, the interval of momenta for which the systematic error in our calculation of $\w(r)$ is best known, and $\epsilon_{\s{\Cfat(r_i)}}$ denotes the error estimate of $\Cfat(r_i)$. Then we obtain $\chi^2 = 2\,630$, corresponding to $\chi^2_{\rm d.o.f.} = 5.11$. The probability $P_{\Cfat_0}$ that our result for $\Cfat$ is
consistent with the null hypothesis is 
vanishingly small, given by the formula
\be 
P_{\Cfat_0} = \int_{\chi^2=2\,630}^\infty \chi_{\rm \s{PDF}}^2(515,x) dx = \left.\frac{\Gamma(n_r/2,\chi^2/2)}{\Gamma(n_r/2)}\right\vert_{n_r = 515}^{\chi^2=2\,630} = 7.3\times10^{-280}\,.
\ee 
In fact, even if the errors were $0.95\%$ larger, \ie nearly doubled, we could still discard $\Cfat_0$ at the $5\sigma$ confidence level.

In the right panel of \fig{fig:Cfat} we compare $\Cfat(r)$ to the BSE prediction, $\Cfat_\star(r)$, of \fig{fig:C_gl_Cgh}, shown as a purple dot-dashed curve and corresponding error band. In that panel, we observe an excellent qualitative agreement between the two results. The most noticeable quantitative difference is in the position of the minimum. Specifically, $\Cfat$ reaches the minimum value of $-0.36\pm 0.11$ at $r = 1.93\substack{+0.09 \\ -0.06}$ GeV, while the minimum of $\Cfat_\star$ is $-0.341\pm0.003$ at $r=1.5\pm 0.1$.

Nevertheless, it is clear in the right panel of \fig{fig:Cfat} that the BSE prediction lies within the error estimate of the lattice-derived $\Cfat(r)$. In fact, defining a $\chi^2$ measure for the discrepancy between $\Cfat$ and $\Cfat_\star$ as
\be
\chi^2_\star = \sum_{i} \frac{\left[\Cfat(r_i)-\Cfat_\star(r_i)\right]^2}{\epsilon_{\s{\Cfat(r_i)}}^2} \,, \label{chi2_star_def}
\ee 
we obtain $\chi^2_\star = 258.5$, which is smaller than the number of degrees of freedom. Then, this value of $\chi^2_\star$ amounts to a probability of
\be 
P_{\Cfat_\star} = \left.\frac{\Gamma(n_r/2,\chi^2_\star/2)}{\Gamma(n_r/2)}\right\vert_{n_r = 515}^{\chi^2_\star=258.5} = 1 - 2.0\times 10^{-23} \,,
\ee 
showing that $\Cfat_\star$ is statistically compatible with the lattice derived $\Cfat$, with probability extremely near unit.

%%%%%%%%%%%%%%%%%%%%%%%%%%%%%%%%%%%%%%%%%%%%%%%%%%%%%%%%
\section{Conclusions}\label{sec:conc}
%%%%%%%%%%%%%%%%%%%%%%%%%%%%%%%%%%%%%%%%%%%%%%%%%%%%%%%%

The gauge sector of QCD is
host to a wide array of subtle mechanisms that
are of vital importance for the self-consistency and infrared
stability of the theory. In the present work, we have
offered a comprehensive review of the intricate 
dynamics that account for some of the most prominent infrared 
phenomena, such as the generation of a gluon mass through the action
of the Schwinger mechanism, the nonperturbative masslessness of the 
ghost, and the characteristic features induced by this particular mass pattern to the form factors of the three-gluon vertex.

The SDEs, supplemented by the judicious use of certain key results from lattice QCD, 
provide a robust continuum framework for carrying out
such a demanding investigation. In fact, 
the results obtained from the SDEs are increasingly reliable,
passing successfully all sorts of tests imposed on them.
A particularly impressive, and certainly not isolated, case of such
a success has been outlined in detail in Section~\ref{sec:dhat}.

Symmetry and dynamics are tightly interwoven; therefore, 
the information encoded in the STIs and WIs of the theory is 
particularly decisive for unraveling basic dynamical patterns. 
A striking manifestation of the profound connection between symmetry and dynamics is provided by the dual role played by the function 
$\Cfat(r)$: it is both 
the BS amplitude of the massless states composed by a pair of gluons, 
and the quantity that embodies
the displacement induced to the WIs by the presence of 
these states.

In our opinion, 
the determination of $\Cfat(r)$ described in Section~\ref{sec:wilat}  represents a major success of the entire set of concepts and techniques 
surrounding the generation of a gluon mass 
through the action of the Schwinger mechanism.
Thus, fifty years after the genesis of QCD, 
we seem to be closing in on the mechanism that the theory uses  
for curing  
the infrared instabilities endemic to perturbation theory.
We hope to be able to report further progress in this direction in the near future.

%%%%%%%%%%%%%%%%%%%%%%%%%%%%%%%%%%%%%%%%%%
\vspace{6pt} 

%%%%%%%%%%%%%%%%%%%%%%%%%%%%%%%%%%%%%%%%%%

%%%%%%%%%%%%%%%%%%%%%%%%%%%%%%%%%%%%%%%%%%
\funding{The authors are supported by the Spanish MICINN grant PID2020-113334GB-I00. M.~N.~F. acknowledges financial support from Generalitat Valenciana through contract CIAPOS/2021/74. J.~P. also acknowledges  
funding from the regional Prometeo/2019/087 from the Generalitat Valenciana.}

\dataavailability{Not applicable.}
%In this section, please provide details regarding where data supporting reported results can be found, including links to publicly archived datasets analyzed or generated during the study. Please refer to suggested Data Availability Statements in section ``MDPI Research Data Policies'' at \url{https://www.mdpi.com/ethics}. If the study did not report any data, you might add ``Not applicable'' here.}

\acknowledgments{The authors thank A.C.~Aguilar, D.~Binosi, D.~Ib\'a\~nez, J.~Pawlowski, C.D.~Roberts, and
J.~Rodr\'iguez-Quintero for several collaborations.}

\conflictsofinterest{The authors declare no conflict of interest.}

\abbreviations{Abbreviations}{
The following abbreviations are used in this work:\\

\noindent
\begin{tabular}{@{}ll}
BFM & background field method \\
BQI & background-quantum identity \\
BRST & Becchi-Rouet-Stora-Tyutin \\
BS & Bethe-Salpeter \\
BSE & Bethe-Salpeter equation \\
EHM & emergent hadron mass \\
MOM & momentum subtraction (renormalization schemes)\\
PT & pinch technique \\
QCD & Quantum Chromodynamics \\
QED & Quantum Electrodynamics \\
RGI & renormalization group invariant \\
SDE & Schwinger-Dyson equation \\
STI & Slavnov-Taylor identity \\
WI & Ward identity
\end{tabular}

}

%%%%%%%%%%%%%%%%%%%%%%%%%%%%%%%%%%%%%%%%%%
%% Optional
\appendixtitles{yes} % Leave argument "no" if all appendix headings stay EMPTY (then no dot is printed after "Appendix A"). If the appendix sections contain a heading then change the argument to "yes".
\appendixstart
\appendix

%%%%%%%%%%%%%%%%%%%%%%%%%%%%%%%%%%%%%%%%%%%%%%%%%%%%%%%%
\section[\appendixname~\thesection]{BQIs for the BSE amplitudes}\label{app:poleBQI}
%%%%%%%%%%%%%%%%%%%%%%%%%%%%%%%%%%%%%%%%%%%%%%%%%%%%%%%%

In this Appendix, we use two particular BQIs in order to relate the displacement functions 
$\C$ and $\Cfat$ with their BFM counterparts $\Ctilde$ and 
$\Cfattilde$, respectively. 

\begin{figure}[ht]
    \begin{center}
 	\includegraphics[width=0.445\textwidth]{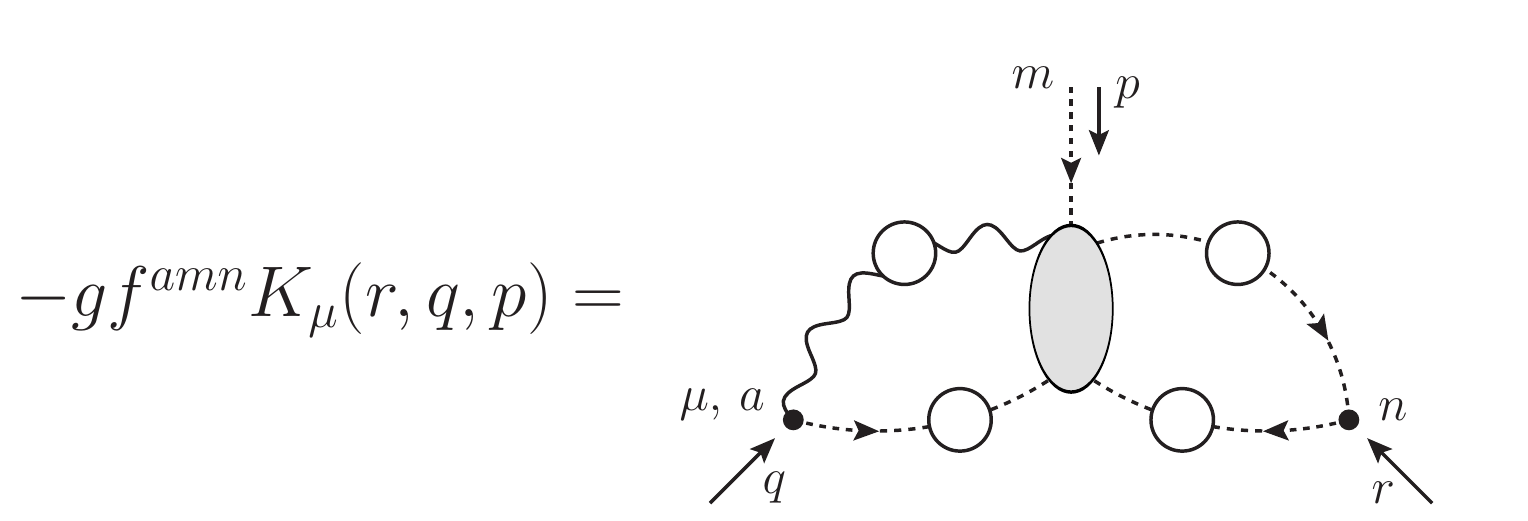}\hfill
	\includegraphics[width=0.535\textwidth]{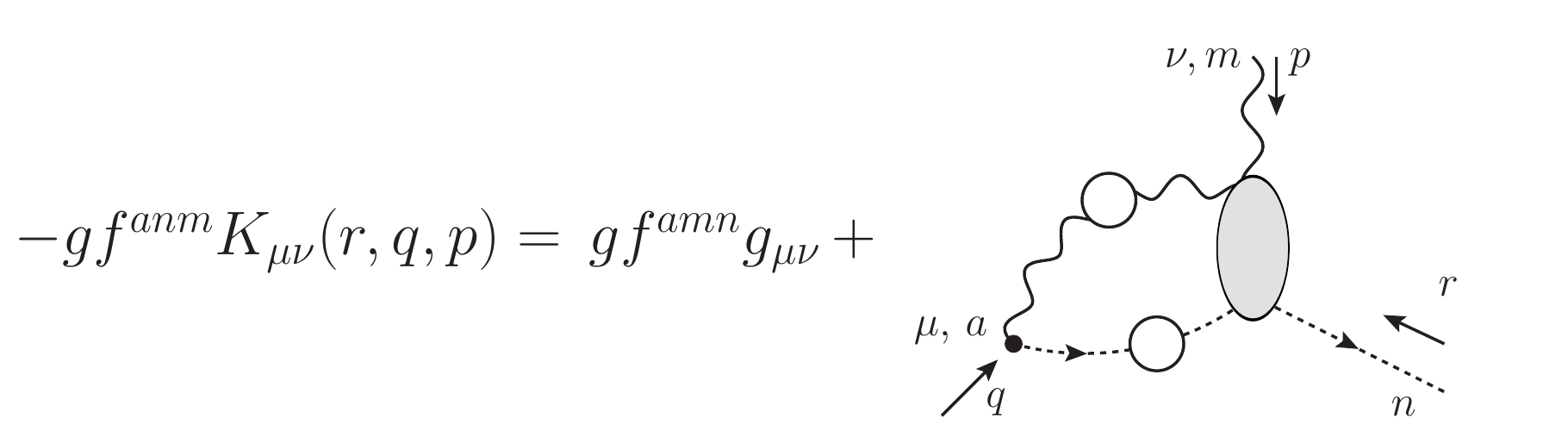}	
    \end{center}
    \caption{ The auxiliary functions $K_\mu(q,r,p)$ and $K_{\mu\nu}(q,r,p)$, appearing in the BQI of \1eq{BQI_Bcc}. }\label{fig:BQI_Ks}
\end{figure}

The ghost-gluon vertices $\fatg_\mu(r,p,q)$ and 
$\fatgt_\mu(r,p,q)$ are related by a BQI~\cite{Binosi:2009qm}, which reads
\begin{eqnarray}
\fatgt_\mu(r,p,q) &=& \left\lbrace \left[ 1 + G(q) \right] g^\nu_\mu + L(q) \frac{q_\mu q^\nu}{q^2} \right\rbrace \fatg_\nu(r,p,q) \nonumber\\
&& + F^{-1}(p) p^\nu K_{\mu\nu}(r,q,p) + r^2 F^{-1}(r)K_\mu(r,q,p) \,, \label{BQI_Bcc}
\end{eqnarray}
where $K_{\mu}$ and $K_{\mu\nu}$ are two auxiliary functions, shown diagrammatically in \fig{fig:BQI_Ks}, while  $G(q)$ and $L(q)$ are the form factors of $\Lambda_{\mu\nu}(q)$, defined in \1eq{Lambda_GL}.

%%%%%%%%%%%%%%%%%%%%%%%%%%%%%%%%%%
% Figure A1 - BQI kernels
%%%%%%%%%%%%%%%%%%%%%%%%%%%%%%%%%%
%%%%%%%%%%%%%%%%%%%%%%%%%%%%%%%%%%

Next, we decompose the $\fatgt_\mu(r,p,q)$ and $\fatg_\mu(r,p,q)$ in \1eq{BQI_Bcc} into their regular and pole parts, using \2eqs{fullgh}{ghsm}, respectively.
Note that the second and third terms in \1eq{BQI_Bcc} do not contain poles in $q^2$; this is 
so because 
$K_{\mu\nu}(r,q,p)$ can contain (longitudinally coupled) poles only in the $p_\nu$ channel, whereas $K_\mu(r,q,p)$ has no external gluon legs,
and hence no poles.

Then, multiplying \1eq{BQI_Bcc} by $q^2$ we obtain
\be 
q_\mu {\widetilde C}(r,p,q) = q_\mu \left[ 1 + G(q) + L(q) \right] \Cgh(r,p,q) + {\cal O}(q^2) \,. \label{BQI_Bcc_2}
\ee

Setting $q = 0$ in \1eq{BQI_Bcc_2} and using \1eq{F0_G0}, we find
\be 
\Cgh(r,-r,0) = Z_1 F(0){\widetilde C}(r,-r,0) \,. \label{Cgh_0}
\ee
Hence, using \1eq{Cant}, we obtain the result in \1eq{C1_0}.

Then, expanding \1eq{BQI_Bcc_2} to first order in $q$, using \1eq{eq:theCs} for $C(r,p,q)$ and \1eq{lhssm2} for ${\widetilde C}(r,p,q)$,
entails
\be 
\C(r) = Z_1 F(0)\Ctilde(r) \,, \label{BQI_Cgh}
\ee
which is one of the main results of this Appendix.

A relation identical to \1eq{BQI_Cgh} can be obtained for $\Cfat(r)$ and its BFM analog, $\Cfattilde(r)$. The starting point of the derivation is the BQI~\cite{Binosi:2009qm}
\begin{eqnarray} 
\fatgt_{\alpha\mu\nu}(q,r,p) &=& \left\lbrace \left[ 1 + G(q) \right] g^\rho_\alpha + L(q) \frac{q_\alpha q^\rho}{q^2} \right\rbrace \fatg_{\rho\mu\nu}(q,r,p) \label{BQI_BQQ}\\
&& + K_{\rho\nu\alpha}(r,q,p)P^\rho_\mu(r)\Delta^{-1}(r) - K_{\rho\mu\alpha}(p,q,r)P^\rho_\nu(p)\Delta^{-1}(p) \,, \nonumber
\end{eqnarray}
where $K_{\mu\nu\alpha}(r,q,p)$ is the function defined in \1eq{H_from_K}.

Then, we note that the only longitudinal poles at $q = 0$ present in \1eq{BQI_BQQ} are those contained in the $\fatg_{\alpha\mu\nu}(q,r,p)$ and $\fatgt_{\alpha\mu\nu}(q,r,p)$ vertices, with the auxiliary functions $K_{\alpha\nu\rho}(q,p,r)$ having poles only in the $r_\mu$ and $p_\nu$ channels. As such, isolating the $q_\alpha g_{\mu\nu}/q^2$ pole and expanding around $q = 0$, one eventually finds
\be 
{\widetilde C}_1(0,r,-r) = Z_1^{-1} F^{-1}(0)C_1(0,r,-r) = 0\,, \label{BQI_Cgl_0}
\ee
and 
\be 
\Cfat(r) = Z_1 F(0)\Cfattilde(r) \,, \label{BQI_Cgl} \ee 
where ${\widetilde C}_1(q,r,p)$ and $\Cfattilde(r^2)$ are defined in analogy to the \2eqs{eq:Cdec}{eq:theCs}, and we used \1eq{C1_0}.

%%%%%%%%%%%%%%%%%%%%%%%%%%%%%%%%%%%%%%%%%%
%\begin{adjustwidth}{-\extralength}{0cm}

\reftitle{References}


%\end{adjustwidth}
\end{document}